\documentclass[journal,10pt]{IEEEtran}
\ifCLASSINFOpdf
  \usepackage[pdftex]{graphicx}
  \usepackage{amsthm,amsmath,amssymb}
\usepackage{makecell}
\usepackage{mathrsfs}
  \usepackage{float}
  \usepackage{algorithm}
\usepackage{algpseudocode}
\usepackage{amsmath}
\usepackage{bm}
\usepackage{color}
\usepackage{comment}
\usepackage{booktabs}
 
\else
\fi
\usepackage{leftidx}
\usepackage{epstopdf}
\usepackage{amsmath}
\usepackage{subfigure}
\usepackage{makecell}
\usepackage{multirow}
\usepackage{hyperref}
\usepackage{amssymb}
\begin{document}

\title{Multi-Modal Intelligent Channel Modeling: A New Modeling Paradigm via Synesthesia of Machines}

\author{Lu~Bai,~\IEEEmembership{Member,~IEEE}, Ziwei~Huang,~\IEEEmembership{Member,~IEEE}, Mingran~Sun,~\IEEEmembership{Graduate Student Member,~IEEE}, Xiang~Cheng,~\IEEEmembership{Fellow,~IEEE}, and Lizhen~Cui,~\IEEEmembership{Senior Member,~IEEE}

\thanks{The authors would like to thank Mengyuan Lu, Zengrui Han, and Huijuan Qiao for their help in the figure creation and multi-modal information processing.}
\thanks{L.~Bai is with the Joint SDU-NTU Centre for Artificial Intelligence Research (C-FAIR), Shandong University, Jinan, 250101, P. R. China (e-mail: lubai@sdu.edu.cn).}
\thanks{Z.~Huang, M.~Sun, and X.~Cheng are with the State Key Laboratory of Advanced Optical Communication Systems and Networks, School of Electronics, Peking University, Beijing, 100871, P. R. China (e-mail: ziweihuang@pku.edu.cn; mingransun@stu.pku.edu.cn;  xiangcheng@pku.edu.cn).}	
\thanks{L.~Cui is with the School of Software, Shandong University, Jinan 250101, P. R. China, and also with the Joint SDU-NTU Centre for Artificial Intelligence Research (C-FAIR), Shandong University, Jinan 250101, P. R. China (e-mail: clz@sdu.edu.cn).}	
}

\markboth{}
{Zeng \MakeLowercase{\textit{et al.}}: Bare Demo of IEEEtran.cls for IEEE Journals}

\maketitle

\markboth{IEEE Communications Surveys \& Tutorials, vol. xx, no. xx, XX 2024}
{Submitted paper}
		\maketitle

\begin{abstract}
In the future sixth-generation (6G) era, to support accurate localization sensing and efficient communication link establishment for intelligent agents, a comprehensive understanding of the surrounding environment and proper channel modeling are indispensable. The existing method, which solely exploits radio frequency (RF) communication information, is difficult to accomplish accurate channel modeling, and hence cannot support the aforementioned application associated with intelligent agents. Fortunately, multi-modal devices are deployed on intelligent agents to obtain environmental features, which could further assist in channel modeling. Currently, some research efforts have been devoted to utilizing multi-modal information to facilitate channel modeling, while still lack a comprehensive review. To fill this gap, we embark on an initial endeavor with the goal of reviewing multi-modal intelligent channel modeling (MMICM) via Synesthesia of Machines (SoM), which refers to intelligent multi-modal sensing-communication integrated technology. In the MMICM, the multi-modal information from communication devices and various sensors is intelligently processed to explore the mapping relationship between communications and sensing. Compared to channel modeling approaches that solely utilize RF communication information, the utilization of multi-modal information can provide a more in-depth understanding of the propagation environment around
the transceiver, thus facilitating more accurate channel modeling.
First, this paper introduces existing channel modeling approaches from the perspective of the channel modeling evolution, i.e., starting from conventional channel modeling, the channel modeling
approach moves on to the RF-only intelligent channel
modeling and finally lands on MMICM. Then, we have elaborated and investigated recent advances in the topic of capturing typical channel characteristics and features, i.e., channel non-stationarity and consistency, via conventional channel modeling, RF-only intelligent channel modeling, and MMICM by characterizing the mathematical, spatial, coupling, and mapping relationships. In addition, applications that can be supported by MMICM are summarized and analyzed. To corroborate the superiority of MMICM via SoM, we give the simulation result and analysis. Finally, some open issues and potential directions for the MMICM are outlined from the perspectives of measurements, modeling, and applications.

\end{abstract}

\IEEEpeerreviewmaketitle

\section{Introduction}
\IEEEPARstart {T}{h}{e} fifth-generation (5G) wireless communication network has entered the commercial deployment phase and is changing various facets of life and industrial structures \cite{TANGPAN11}. Based upon the International Telecommunication Union Radiocommunication Sector (ITU-R) classification, there are three typical application scenarios, such as enhanced mobile broadband (eMBB), massive machine type communications (mMTC), and ultra-reliable and low latency communications (uRLLC) \cite{TAO11}. In the future sixth-generation (6G) era, three application scenarios in 5G will be expanded and deepened by integrating communications, sensing, computing, artificial intelligence (AI), and other capabilities to establish six application scenarios, including immersive
communication, integrated sensing and communication (ISAC), massive communication, ubiquitous
connectivity, hyper reliable and low-latency communication, and integrated AI and communication \cite{6G11}--\cite{6G12}. Compared to 5G, 6G has further put forward tenfold to even hundredfold more strict network requirements for key system metrics, such as reliability, peak data rate, over-the-air latency, connection density, and security \cite{TAO22}--\cite{6G14}.  This poses higher demands for the accuracy, intelligence, and involvement in system design of 6G channel modeling compared to 5G channel modeling.
		\begin{figure*}[!t]
		\centering	\includegraphics[width=0.9\textwidth]{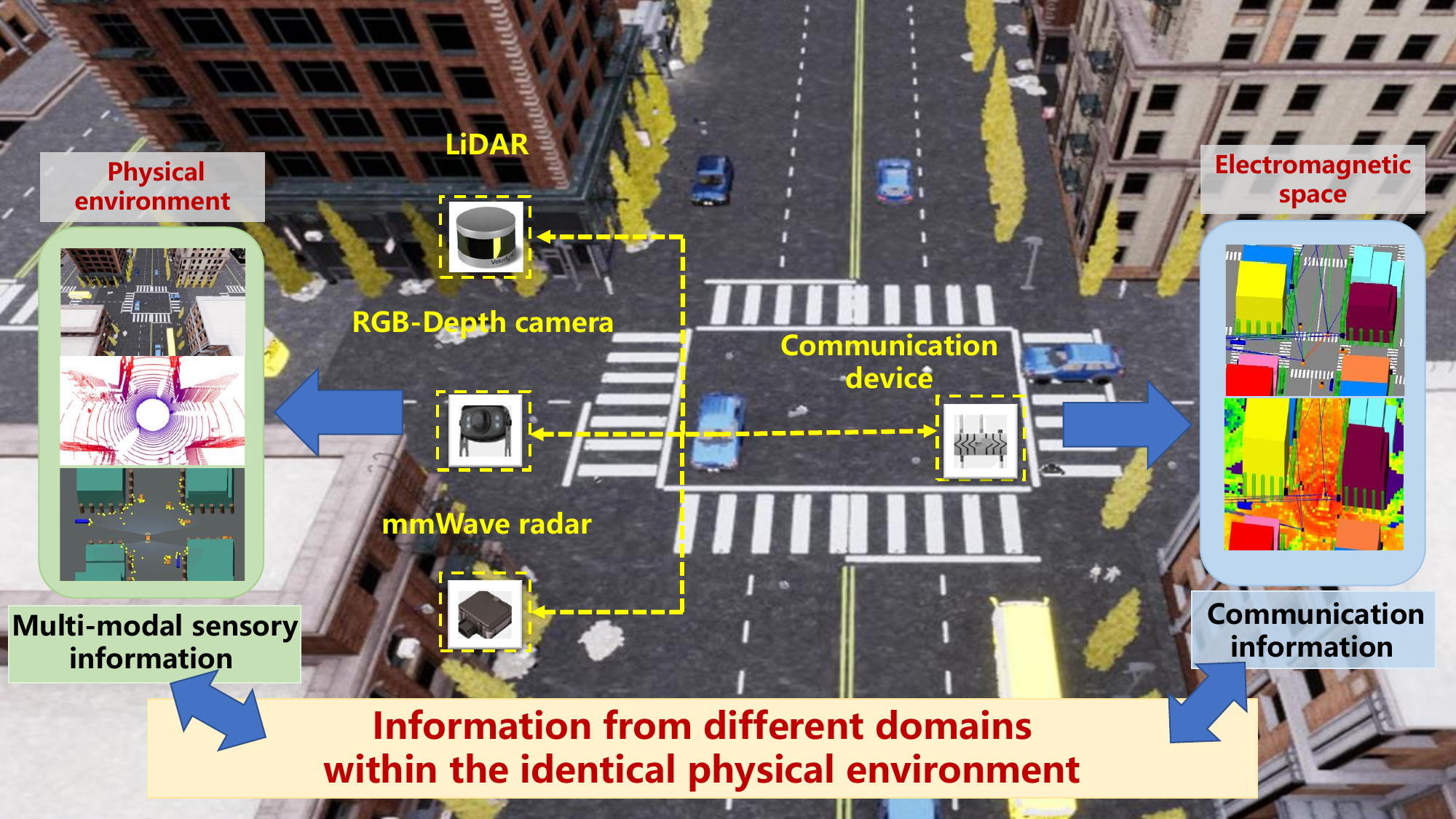}
	\caption{Correlations between multi-modal sensory information and communication information.}
	\label{Correlation}
	\end{figure*}

As a typical application in 6G, artificial intelligence of things (AIoT) has received extensive attention \cite{AIoT}. 
In AIoT, networked intelligent agents, such as autonomous vehicles, unmanned aerial vehicles (UAVs), and robotics, are naturally  equipped with various multi-modal sensors and communication devices \cite{SoM,wang112}. Meanwhile, thanks to its innate ability to process nonlinear and complex models, the AI technology will become an essential native component of AIoT in 6G. By intelligent processing the collected multi-modal sensory information and communication information via the AI technology, the intelligent agent implements robust environmental perception together with efficient link establishment \cite{Example1,Example2}. Therefore, it is essential to investigate the intelligent integration and mutually beneficial mechanism between communication and multi-modal sensing based on the AI technology. Currently,  the widely investigated ISAC cannot adequately support the aforementioned task associated with communications and multi-modal sensing. This is because that ISAC focuses on the integration of radio frequency (RF) radar sensing and RF communications. To fill this gap, inspired by human synesthesia, we proposed the concept of Synesthesia of Machines (SoM) \cite{SoM}. SoM refers to the intelligent multi-modal sensing-communication integration, which aims to achieve mutual benefit between communications and sensing via machine learning (ML) technology. As a result, SoM is a more generalized concept than ISAC and ISAC can be regarded as a special case of SoM. Furthermore, similar to how humans sense the surrounding environment through multiple organs, communication devices and multi-modal sensors can also acquire environmental information, and thus are referred to as machine senses. As the foundation of SoM research, it is necessary to explore the complex SoM mechanism, i.e., mapping relationship, between communication information and multi-modal sensory information, and further conduct high-precision and intelligent channel modeling \cite{SoM,mingran11}. Certainly, channel modeling is the cornerstone of any system design and algorithm development \cite{UAV-B-1}--\cite{V2V-H-1}. This also holds true for SoM system design and algorithm development. In the SoM system, multi-modal sensors can sense physical environment information and channel modeling aims to accurately characterize the electromagnetic environment via the communication information collected by communication device. Fortunately, multi-modal sensory and communication information are data collected from different domains originating from the same environment, and thus exhibit strong correlations, as shown in Fig.~\ref{Correlation}. Based on the strong correlation, the mapping relationship between multi-modal sensing and communications is significant to be explored. By exploring the mapping relationship, an in-depth understanding of the surrounding physical environment at the transmitter (Tx) and  receiver (Rx) can be obtained during channel modeling. As a result, the high-precision and multi-modal intelligent channel
modeling (MMICM) can be conducted which has higher involvement in the system design of 6G and SoM. Furthermore, the multi-modal information obtained by communication devices and various sensors is intelligently processed in MMICM to explore the mapping relationship between communications and sensing, which can facilitate the channel modeling. Compared to channel modeling approaches that solely rely on RF communication information, the utilization of multi-modal information can provide a more in-depth understanding of the propagation environment around the transceiver, resulting in a more accurate and less complex modeling of the electromagnetic space.
Therefore, to support the SoM research, the MMICM needs to be investigated urgently. Towards this objective, we conduct a comprehensive survey on recent advances in the MMICM in this paper.

\subsection{Related Work}
In the area of channel modeling, there have been many surveys focusing primarily on reviewing existing channel models under a certain application scenario, e.g., vehicle-to-vehicle (V2V) \cite{V2Vsur1, V2Vsur1122}, high-speed train (HST) \cite{HSTsur1,HSTsur2}, UAV \cite{UAVsur1}--\cite{UAVsur3}, millimeter wave (mmWave) \cite{7109864,7762981}, or massive multiple-input multiple-output (MIMO) \cite{8570036,mMIMOsur}. Since the vision of wireless communication networks is the so-termed space-air-ground-sea integrated network (SAGSIN), some surveys in \cite{9237116}--\cite{COMSTmy} presented a comprehensive review on the 6G channel modeling under diverse application scenarios, such as  V2V, HST, UAV, satellite, maritime, mmWave, and massive MIMO communications. Although the aforementioned surveys in \cite{V2Vsur1}--\cite{COMSTmy} provided a clear bird’s-eye view of  channel modeling approach, while do not emphasize on the combination of a promising  technology, i.e., AI, and channel modeling. To overcome this limitation, the authors in \cite{huangchen1} presented a comprehensive survey to introduce recent advances in the AI enabled channel modeling, i.e., intelligent channel modeling. Furthermore, the authors in \cite{yanngmi11} provided a systematic overview on intelligent channel modeling, and further developed an intelligent channel modeling architecture to construct complex models between environmental features and channel characteristics. Nonetheless, the survey in \cite{huangchen1,yanngmi11} focused on utilizing ML to process communication information to reveal channel characteristics and facilitate channel modeling, while neglecting the sensory information.

Currently, ISAC, which can enhance the physical system through the combination of wireless communications and sensing capability, has received extensive attention \cite{Liu11}. As the foundation of ISAC system design, an in-depth understanding of channel characteristics and proper channel modeling are essential. The authors in \cite{Zhang11} given a nice summary of research progress
of the ISAC channel measurements, characteristics, and modeling. Liu \emph{et al.} \cite{LiuT11} gave a state-of-the-art review of the channel model utilized in the ISAC application, where the challenge encountered in the ISAC modeling approach was also analyzed. Yang \emph{et al.} \cite{YangW11} thoroughly reviewed the ISAC channel measurement methodology, and further proposed a novel architecture of general ISAC channel models to describe the sensing and communication channel. However, as previously mentioned, the ISAC technology cannot support the task of intelligent multi-modal sensing-communication integration, i.e., SoM. Furthermore, since the ISAC technology focuses on the integration of RF communications and sensing, its understanding of the surrounding environment is limited. 
Conversely, by exploring the nonlinear mapping relationship between communications and RF/non-RF sensing via SoM, various features of the propagation environment around the transceiver are obtained, which can support the construction of high-precision MMICM. Compared to channel modeling approaches that solely rely on RF communication information, the utilization of multi-modal information from communication devices and various sensors can provide a more in-depth understanding of the propagation environment around the transceiver, leading to a more precise channel modeling.
Currently, a comprehensive survey of the MMICM via SoM is still lacking. Furthermore, the existing survey ignored the analysis of the evolution of channel modeling from the conventional channel modeling approach to the intelligent channel modeling approach with the help of uni-modal and multi-modal information. Meanwhile, the existing survey also ignored the introduction of channel non-stationarity and consistency based on different channel modeling approaches by capturing different relationships.
Therefore, a comprehensive survey to provide recent advances in the MMICM and analyze its applications and potential directions is still lacking in the existing literature.

\subsection{Contributions and Organization of This Paper}

This survey aims to fill this gap, and its major contributions are outlined below.

\begin{enumerate}
\item According to whether ML-based models and uni/multi-modal information are utilized, recent channel modeling approaches are classified and their evolution is further elaborated for the first time. Specifically, starting from conventional channel modeling, the channel modeling approach moves on to the RF-only intelligent channel modeling and finally lands on  MMICM. Furthermore, 
the conventional channel modeling, RF-only intelligent channel modeling, and MMICM are adequately compared and analyzed.

\item As the typical channel characteristic and feature, the investigation of channel non-stationarity and consistency is of paramount importance. To capture channel non-stationarity and consistency, current methods based on the conventional channel modeling, RF-only intelligent channel modeling, and MMICM are reviewed. By delineating mathematical, spatial, coupling, and mapping relationships of scatterers/clusters in the modeling of channel non-stationarity and consistency, the utilization of the environmental features increases sequentially in the three aforementioned channel modeling approaches. This results in a more precise capturing of channel non-stationarity and consistency.
\item To further support and enhance the involvement of channel modeling in system design and technological advancements, the huge facilitation of the MMICM on communication networks and networked intelligent applications, including communication system transceiver design, network optimization, localization sensing, and intelligent agent cognitive intelligence, is discussed for the first time.  Due to the in-depth understanding of physical environment via multi-modal information and the exploration of mapping relationship, the MMICM has a more significant facilitation on communication networks and networked intelligent  applications in comparison with conventional and RF-only intelligent channel modeling. 
\item Channel statistical properties and  path loss prediction results of the conventional channel modeling, RF-only intelligent channel modeling, and MMICM are compared and analyzed thoroughly. Simulation results demonstrate  the necessities of utilizing the multi-modal information and exploring the mapping relationship between communications and sensing in the channel modeling.
\item  For the MMICM, a list of future challenges and research directions were presented from three perspectives, including measurements, modeling, and applications. These include: i) the construction of real-world and synthetic datasets with communication and multi-modal sensory information; (ii) the exploration the mapping relationship in various conditions under multi-vehicle/UAV communications and multi-modal sensing scenarios; (iii) more comprehensive MMICM for digital twin (DT), embodied intelligence, and networked intelligence.

\end{enumerate}

\begin{figure}[!t]
		\centering	\includegraphics[width=0.45\textwidth]{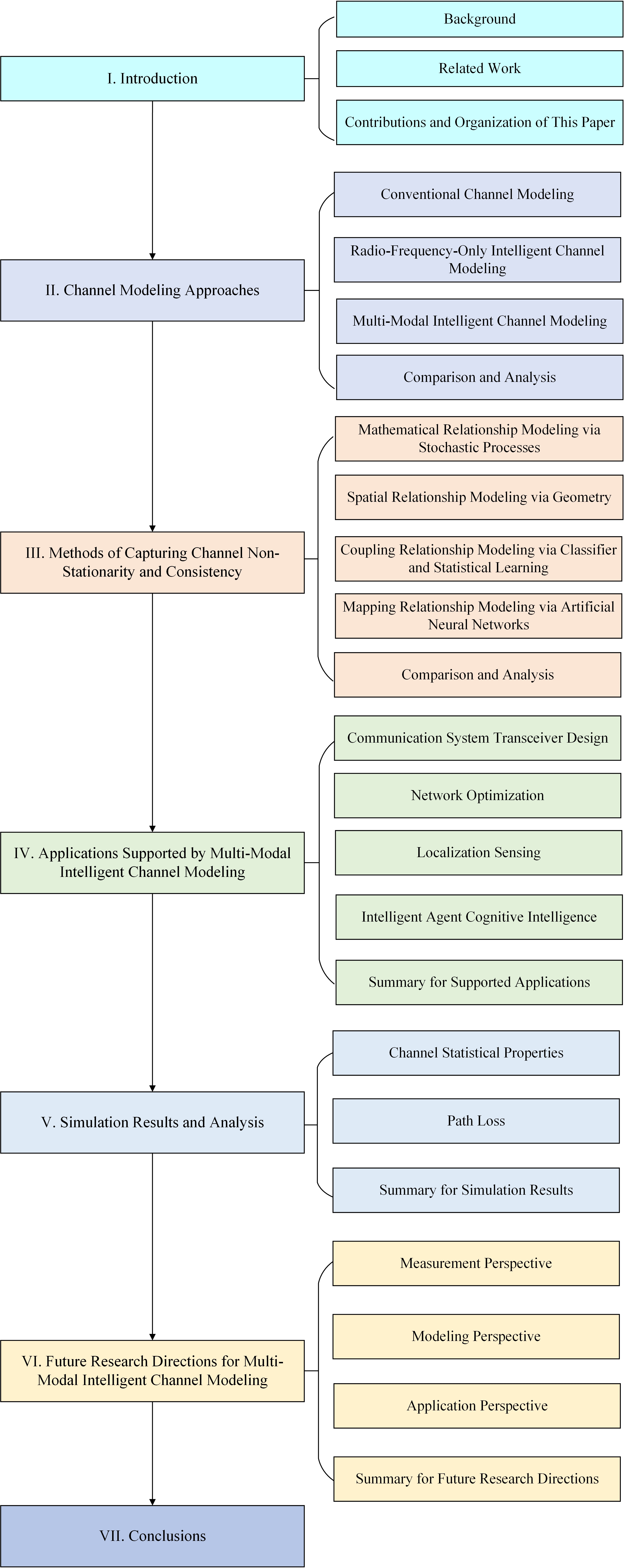}
	\caption{The organization of this paper.}
	\label{organization}
	\end{figure}

The organization of this paper is presented in Fig.~\ref{organization} and is analyzed as follows. In Section~II, existing channel modeling approaches and their evolution are presented. Section~III gives the method of capturing the typical channel characteristic and feature, i.e., channel non-stationarity and consistency. In Section~IV, the applications, which can be facilitated by MMICM, are elaborated. Section~V presents the simulation result and analysis concerning the conventional channel modeling, RF-only intelligent channel modeling, and MMICM. In Section~VI, open issues of future research directions for the MMICM are outlined. Finally, the conclusions are drawn in Section~VII.

\section{Channel Modeling Approaches}

In this section, according to whether ML-based models and uni/multi-modal information are utilized, existing channel modeling approaches are classified and their evolution is further given. To be specific, starting from conventional channel modeling, the channel modeling approach moves on to the RF-only intelligent channel modeling and finally lands on MMICM. For the conventional channel modeling, the deterministic and statistical methods are exploited, while ML-based models are ignored. Considering the huge advantage of ML-based models, the RF-only intelligent channel modeling utilizes ML-based models to intelligently process uni-modal communication information. To further enhance the understanding of physical environment, the MMICM intelligently processes multi-modal information from communication devices and sensors via ML-based models. In terms of the number of sensory modality, MMICM can be further categorized into  MMICM based on uni-modal sensory information and MMICM based on multi-modal sensory information.
In addition, we adequately compare and analyze the aforementioned three approaches.  

\subsection{Conventional Channel Modeling}

\begin{figure*}[!t]
		\centering	\includegraphics[width=0.95\textwidth]{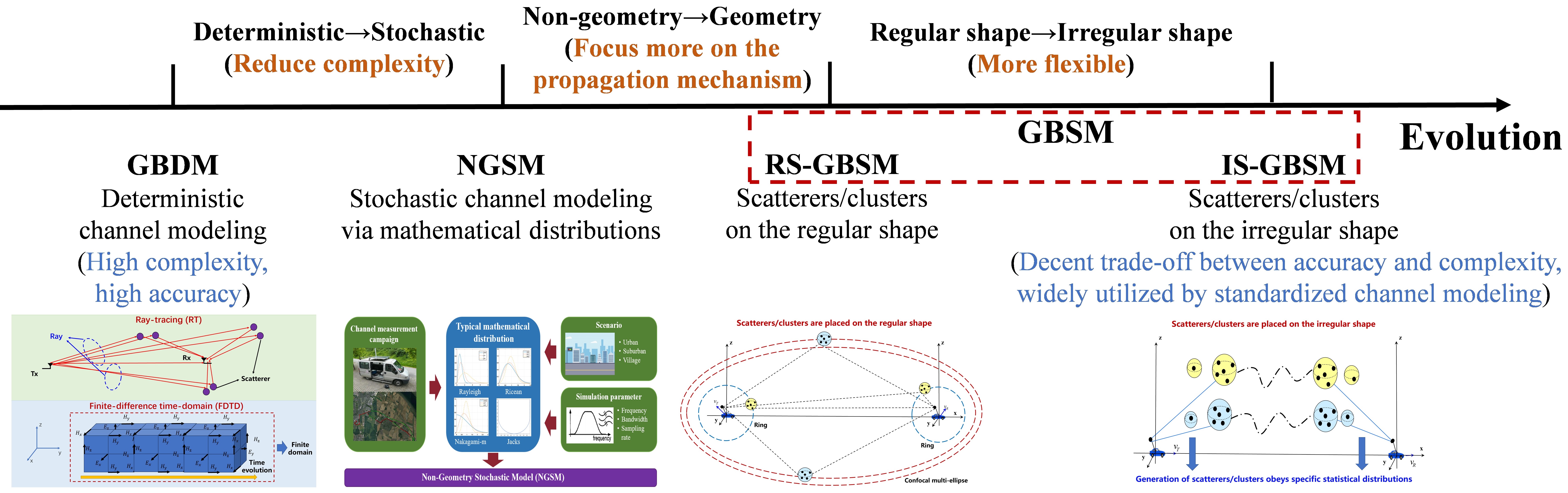}
	\caption{The evolution of conventional channel modeling approaches.}
	\label{conventional_evolution}
	\end{figure*}

In general, the existing conventional channel modeling can be classified into the geometry-based deterministic modeling (GBDM), non-geometry stochastic modeling (NGSM), and geometry-based stochastic modeling (GBSM) \cite{hengtai}--\cite{yuxiangzhang}.  At the beginning, in the GBDM, the channel model is constructed in a deterministic manner for a site-specific scenario, where the procedure of physical radio propagation can be reproduced. Nonetheless, the time-consuming and precise representation of the site-specific scenario leads to high complexity of the GBDM. To reduce the channel modeling complexity, the NGSM determines channel parameters in a stochastic manner, where probability density functions (PDFs), e.g., Rayleigh, Ricean, and Nakagami distributions,  of empirical parameters are utilized based on channel measurement results. The main limitation of NGSMs is the ignorance of geometrical scattering mechanism in the propagation environment, where the channel is regarded as a black box. To overcome this limitation, the concept of rays  is introduced to intuitively mimic the propagation procedure of multipath components (MPCs). In addition, the concept of scatterers that generate MPCs is introduced to model the interaction between radio waves and objects, where scatterers with  similar delays and azimuth/elevation directions are further grouped into clusters. Based on the concept of scatterers/clusters, the GBSM can be derived from the predefined stochastic distribution of  scatterers/clusters in the propagation environment. GBSMs can be further divided into regular-shaped GBSMs (RS-GBSMs) and irregular-shaped GBSMs (IS-GBSMs), which depend on whether scatterers/clusters are placed on the regular shape, e.g., cylinders, ellipsoids, and spheres, or the irregular shape.  The RS-GBSM leverages a regular geometry to imitate the distribution of the scatterer/cluster with low complexity, and thus has been widely leveraged in the theoretical research on channel modeling. Compared to the RS-GBSM, by assuming that locations of scatterers/clusters obey a specific statistical distribution rather than being limited to a regular shape, the IS-GBSM is more flexible and more suitable for high-mobility communication scenarios. Owing to the decent trade-off between accuracy and complexity, the IS-GBSM has been extensively utilized in the existing standardized channel modeling, e.g., European COoperation in the field of Scientific and Technical research (COST) 2100 channel model \cite{SRM4-zw}, Third Generation Partnership Project (3GPP) TR38.901 channel model \cite{standardized-zr}, Mobile and wireless communications Enablers for the Twenty-twenty Information Society (METIS) channel model \cite{METIS-zw}, IMT-2020 channel model  \cite{IMT-zw}, and QUAsi Deterministic RadIo channel GernerAtor (QuaDRiGa) channel model  \cite{QuaDRiGa-zw}. For clarity, Fig.~\ref{conventional_evolution} intuitively gives the evolution of conventional channel modeling approaches, including the GBDM, NGSM, RS-GBSM, and IS-GBSM.

\begin{figure}[!t]
		\centering	\includegraphics[width=0.49\textwidth]{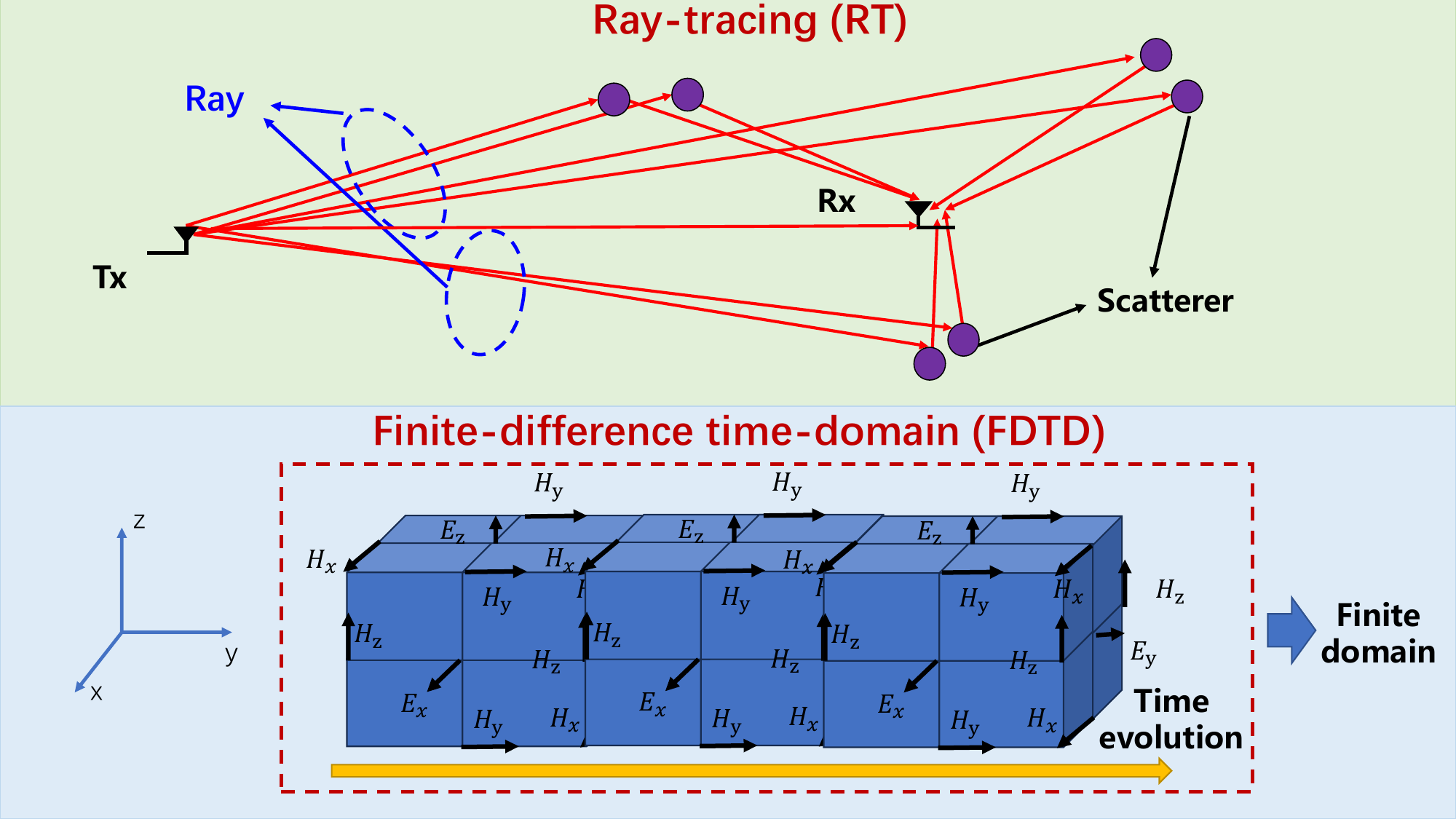}
	\caption{Two typical modeling approaches, i.e., RT and FDTD, in the GBDM.}
	\label{GBDM}
	\end{figure}

\subsubsection{GBDM}
GBDM aims to reproduce the procedure of physical radio propagation in site-specific scenarios. According to the electromagnetic field theory, parameters of the GBDM are defined in a deterministic way, including ray-tracing (RT) \cite{danping,guanke} and finite-difference time-domain (FDTD) \cite{FDTD11}, as shown in Fig.~\ref{GBDM}. In \cite{mengyuan-EI-my}, the authors developed a UAV-to-ground GBDM under sub-6 gigahertz (GHz), 15 GHz, and 28 GHz frequency bands in urban and suburban scenarios based on the RT technology. Different vehicular traffic densities (VTDs) and UAV heights were taken into account to depict the propagation characteristics of UAV-to-ground communication channels. In addition to the sub-6 GHz and mmWave frequency bands, the authors in \cite{gbdm-zhangjianhua-my} further developed a RT-based GBDM in the  terahertz (THz) frequency band. Specifically, the authors conducted the measurement in two typical indoor scenarios, i.e., empty room and spacious hall, where a close agreement between simulation results based on the developed GBDM and measurement data was achieved. However, the GBDMs, e.g., \cite{mengyuan-EI-my,gbdm-zhangjianhua-my}, require an accurate electromagnetic space information to determine the parameter, resulting in the high complexity. Furthermore, there is currently no standardized channel model based on the GBDM modeling approach, which limits the applicability and development of GBDMs.

\subsubsection{NGSM} 
\begin{figure}[!t]
		\centering	\includegraphics[width=0.49\textwidth]{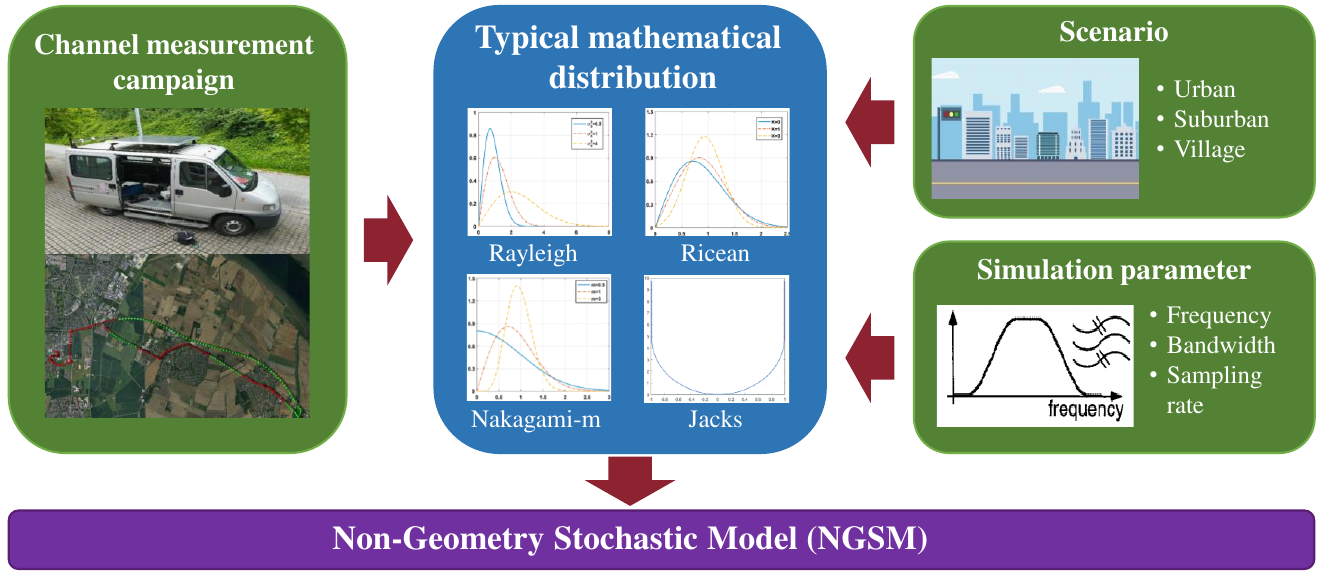}
	\caption{Construction of NGSMs based on PDFs of empirical parameters according to channel measurement campaigns.}
	\label{NGSM_fig}
	\end{figure}

Different from the GBDM, the stochastic channel model reduces dependence on the physical environment and utilizes statistical methods to determine the channel-related parameter \cite{ztecom-zw}. In the NGSM, the parameter of channels are determined in a stochastic manner, where the construction of NGSMs depends generally on channel statistical properties together with PDFs of empirical 
parameters based on measurement campaigns, as shown in Fig.~\ref{NGSM_fig}. Due to the low complexity of the NGSM, it is utilized in the standardized COST 207 and COST 259 channel models \cite{cost207}--\cite{cost2592}. In \cite{MRM10-zw}, the authors proposed a wideband non-stationary NGSM for  vehicular channels. A non-uniformly distributed tap phase was generated to include the line-of-sight (LoS) component. By modifying the auto-correlation function, variable types of Doppler spectra for different delays was modeled. The authors in \cite{MRM10-zw} further derived significant statistical properties, including power delay profiles (PDPs), tap correlation coefficient matrices, and Doppler power spectral densities (DPSDs).
In \cite{NGSM2-my}, the authors utilized measurement data at 28 GHz and 73 GHz in New York to propose a non-stationary NGSM. Based on the proposed NGSM in \cite{NGSM2-my}, key channel parameters and characteristics, including number of spatial clusters, and angular dispersion, were derived and investigated. The accuracy of the proposed NGSM in \cite{NGSM2-my} was further verified in various non-line-of-sight (NLoS) environments. Since channel-related parameters in NGSMs can be derived statistically and the scattering geometry is not taken into account for simplicity, the computational complexity of NGSMs is low. However, the accuracy of the NGSM is lower than that of the GBDM as the underlying scattering geometry in the propagation environment is neglected.

\begin{figure}[!t]
		\centering	\includegraphics[width=0.49\textwidth]{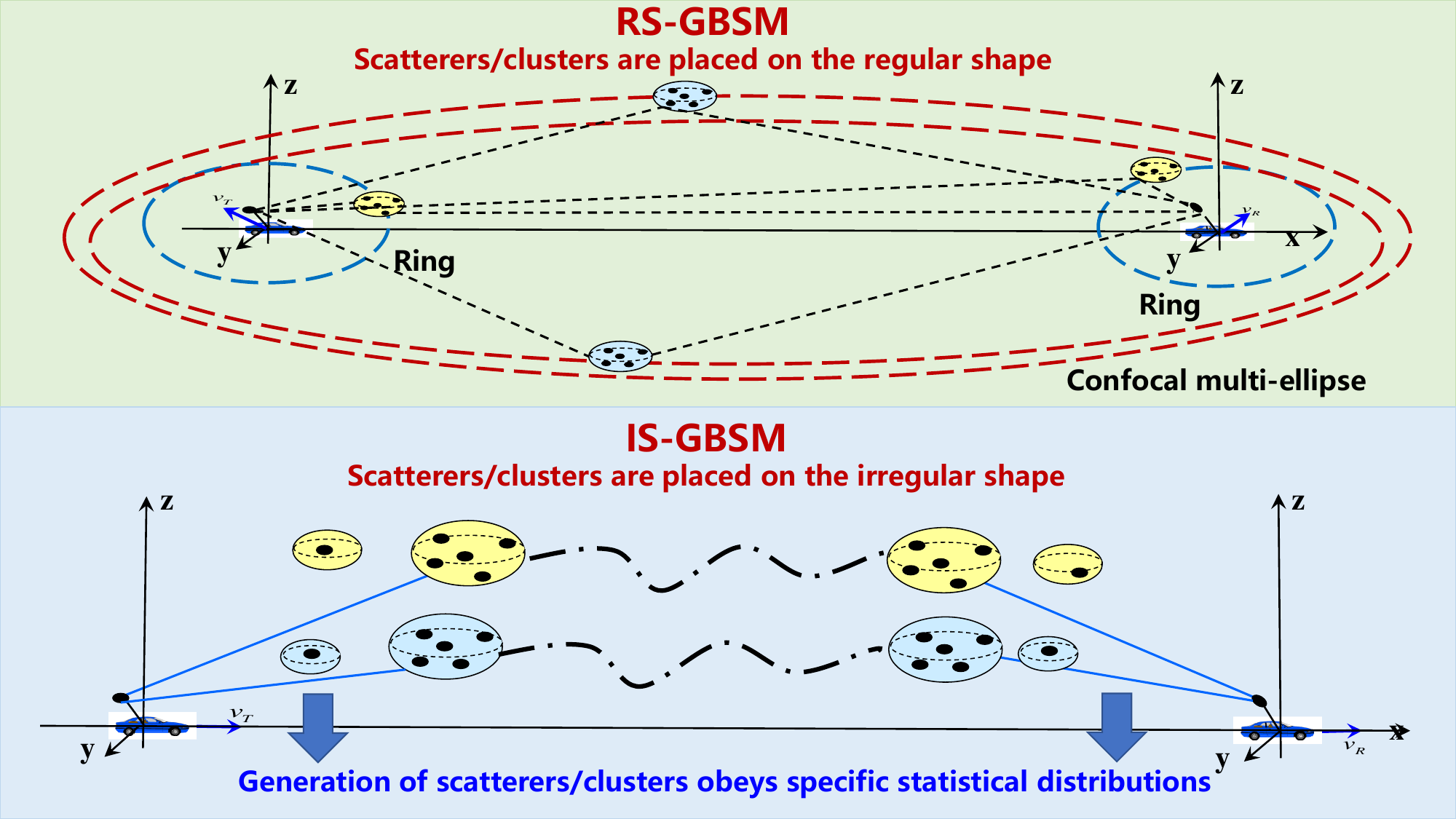}
	\caption{Two categories of GBSMs, i.e., RS-GBSM and IS-GBSM, based on locations of scatterers/clusters. }
	\label{GBSM}
	\end{figure}



\subsubsection{GBSM}
To overcome the limitation of low accuracy of NGSMs caused by neglecting scattering geometry, extensive GBSMs have been proposed. Unlike the NGSM, the GBSM is proposed from the predefined stochastic distribution of scatterers/clusters in the geometrical propagation environment. As shown in Figs.~\ref{GBSM}, GBSMs can be further
divided into RS-GBSMs and IS-GBSMs, which depend on whether the scatterer/cluster is set on the regular shape or the irregular shape. In \cite{RS-GBSM-1-zw}, the authors developed a  two-ring RS-GBSM, where the single- and double-bounced components were modeled.  To leverage the advantage of scatterers/clusters located on the ellipse having the same delay to the transceiver, a ellipse-based RS-GBSM for the dynamic  channel was proposed in \cite{RS-GBSM-2-zw}, which modeled the time non-stationarity. However, the proposed RS-GBSM in \cite{RS-GBSM-2-zw} was limited to narrowband communication systems. For a wideband communication system under dynamic scenarios, the authors in  \cite{RS-GBSM-3-zw} developed a time non-stationary RS-GBSM with multi-confocal ellipses.  However, the developed RS-GBSM in  \cite{RS-GBSM-3-zw} ignored the distinction between dynamic and static scatterers. To overcome this drawback, the authors in \cite{RS-GBSM-4-zw} proposed a combined two-ring model and confocal ellipse RS-GBSM, which also modeled the time non-stationarity. However, the aforementioned RS-GBSMs in \cite{RS-GBSM-1-zw}--\cite{RS-GBSM-4-zw} are limited to two-dimensional (2D) propagation environment. To overcome this limitation, the authors in \cite{RS-GBSM-cylinder-my} proposed a three-dimensional (3D) cylinder RS-GBSM for air-to-air (A2A) channels in UAV communication scenarios. The proposed RS-GBSM in \cite{RS-GBSM-cylinder-my} utilized a 3D Markov mobility model to characterize the movement of the UAV in horizontal and vertical directions. In addition, statistical properties, e.g., time-frequency correlation function (TF-CF) and DPSD, were analyzed.  To further divide clusters into dynamic clusters and static clusters, the authors in \cite{newiot-zw} developed a 3D vehicular RS-GBSM, which combined 3D multi-confocal semi-ellipsoids and semi-spheres. To reduce the complexity and enhance the accuracy, channel parameters were calculated through 3D vectors in the proposed RS-GBSM \cite{newiot-zw}. Important statistical properties, e.g., space-time correlation function (ST-CF), space cross-correlation function (SCCF), time auto-correlation function (TACF), and DPSD, were further derived and investigated.


Different from RS-GBSMs, IS-GBSMs assume that locations of scatterers/clusters in the propagation environment obey a specific statistical distribution.   In comparison of the RS-GBSM,   the IS-GBSM assumes that scatterers/clusters are generated according to  the statistical distribution rather than being limited to the regular shape, and thus is more flexible and more suitable for high-mobility communication scenarios, such as vehicular scenario, UAV scenario, and HST scenario. Meanwhile, based on the IS-GBSM, the trajectory of scatterers/clusters can be mimicked and adjusted flexibly.
Since the IS-GBSM approach achieves the  decent trade-off between accuracy and complexity, it has been widely adopted in the existing standardized channel models, including COST 2100, 3GPP TR38.901, METIS, IMT-2020, and QuaDRiGa channel models. Based on the IS-GBSM, 
the authors in \cite{CCC-zw} proposed a general 3D channel model, which modeled space-time non-stationarity in massive MIMO dynamic channels via the birth-death process approach. To further model and investigate channel characteristics under THz bands,  the authors proposed a general 3D space-time-frequency non-stationary massive MIMO THz IS-GBSM  \cite{ISGBSM2-my}. The proposed IS-GBSM in \cite{ISGBSM2-my} investigated different channel characteristics in diverse massive MIMO THz scenarios, e.g., indoor scenarios, device-to-device (D2D) communications, and long traveling paths of users. 
However, various trajectories of transceivers were ignored in the proposed IS-GBSM \cite{ISGBSM2-my}. To overcome this limitation, the authors in \cite{Bailu-1} proposed a space-time-frequency non-stationary IS-GBSM for 6G massive MIMO mmWave UAV channels. The proposed model in \cite{Bailu-1} considered  diverse 3D trajectories of transceivers/clusters and self-rotations of UAVs. Furthermore, the influence of UAV-related parameters, e.g., the UAV’s moving direction, height, and speed, on channel statistical properties was explored in \cite{Bailu-1}.   However, the proposed IS-GBSM in \cite{Bailu-1} cannot capture single-bouncing transmissions generated by single-clusters. To overcome this drawback, the authors in \cite{hengtai11-zw,liuyu11-zw}  conducted a preliminary attempt and classified clusters into single-clusters and twin-clusters in UAV channels. Nonetheless, the relationship between angle of arrivals (AoAs) and angle of departures (AoDs) of single-clusters was neglected, and the influence of the ratio of single-clusters and twin-clusters on channel statistical properties was not investigated. To fill this gap, the authors in \cite{Bailu-22} proposed a mixed-bouncing based UAV IS-GBSM, which simultaneously modeled single-bouncing transmissions and multi-bouncing transmissions. The proposed IS-GBSM in \cite{Bailu-22} characterized the complex relationship between AoAs and AoDs. Furthermore, a novel parameter, i.e., cluster density index, to model the ratio of single-cluster number and twin-cluster number, was introduced and its effect on the channel channel statistical properties was further explored. 
Overall, since the channel-related parameter is determined in a stochastic manner and the underlying scattering geometry in the propagation environment is adequately considered, the GBSM achieves a decent trade-off between accuracy and complexity. 

In summary, the conventional channel modeling approach includes GBDM, NGSM, and GBSM. The GBDM has the capability to reproduce the procedure of physical radio propagation in certain scenarios. For the NGSM,  the construction depends on the channel statistical property as well as the PDF according to measurement campaigns. The GBSM is constructed from the predefined stochastic distribution of scatterers/clusters in the geometrical propagation environment. The GBSM can be further classified into RS-GBSM and IS-GBSM, depending upon
whether scatterers/clusters are located in regular shapes, e.g., cylinders,
ellipsoids, and spheres, or the irregular shape. 

\subsection{Radio-Frequency-Only Intelligent Channel Modeling}
\begin{figure}[!t]
\centering	
\includegraphics[width=0.49\textwidth]{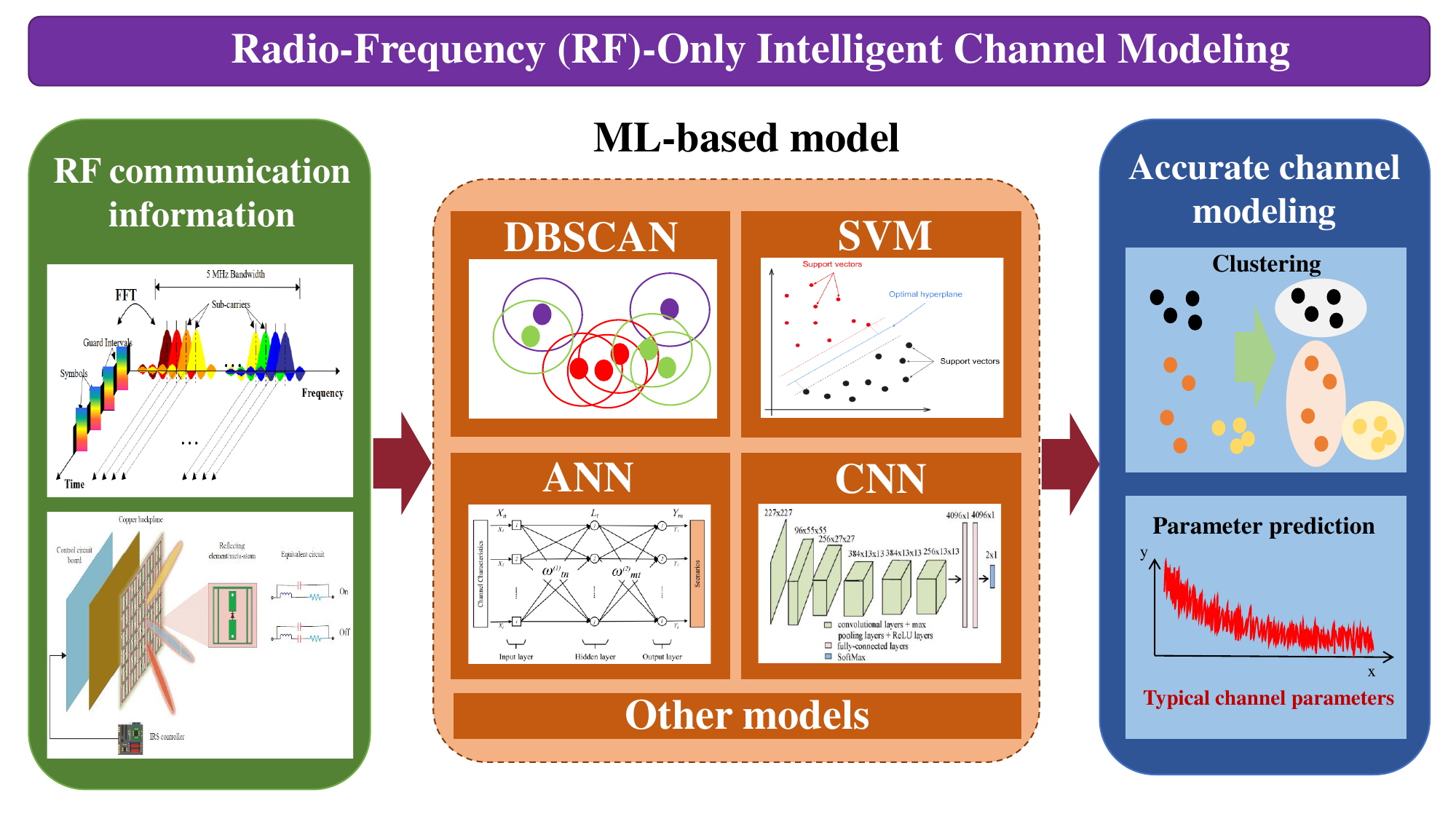}
\caption{RF-only intelligent channel modeling by intelligently processing RF communication information.}
\label{RF-only}
\end{figure}

Different from the conventional channel modeling, RF-only intelligent channel modeling utilizes the ML-based model to intelligently process uni-modal communication information, as shown in Fig.~\ref{RF-only}.
Due to the utilization of ML-based models, the RF-only intelligent channel modeling approach can effectively extract channel information from RF communication data, thus augmenting the capability and adaptability of models.
Considering the aforementioned advantage, extensive work related to RF-only intelligent channel modeling has been conducted. According to the purpose and output of ML-based models, the existing work can be divided into RF-only intelligent channel modeling for clustering and RF-only intelligent channel modeling for parameter prediction.

\subsubsection{RF-Only Intelligent Channel Modeling for Clustering} 
The RF-only intelligent channel modeling for clustering aims to cluster the MPC component. At the beginning, the authors in \cite{cluster11-zw} conducted the RF-only intelligent channel modeling and utilized the $K$-Means clustering algorithm to cluster MPC components in indoor scenarios, where the number of clusters was obtained. To obtain more comprehensive cluster parameters and extend indoor scenarios to outdoor scenarios, Gan et al. \cite{cluster22-zw} proposed a cluster identification approach in vehicular scenarios based on the  density-based spatial clustering of applications with noise (DBSCAN) algorithm. The delay information and Doppler information were intelligently processed via the DBSCAN algorithm to obtain cluster number and lifetime by tracking the cluster centroid. To further investigate the power information of clusters, a novel power weighted MPC clustering method was developed in \cite{cluster33-zw} based on the group tracking, which was widely utilized in the image processing. In addition, a non-stationary RF-only intelligent channel model based on the measurement data was proposed to validate the accuracy of the developed clustering and tracking method. However, the size, position, and shape features information were ignored during the MPC clustering and tracking in \cite{cluster33-zw}. To overcome this limitation,  a power-angle-spectrum-based clustering and tracking method was developed in  \cite{cluster44-zw} based on the Kuhn-Munkres (K-M) algorithm. With the help of the developed method,  the number, appearance/disappearance behavior, and size of clusters were adequately explored. The comparison result demonstrated that the developed method achieved the excellent accuracy on the MPC clustering and tracking with lower complexity compared to the conventional method. Nevertheless, the main drawback of the developed method in  \cite{cluster44-zw} was the ignorance of trajectories in the MPC clustering. To overcome this drawback, the authors in \cite{huangchenscatterer-zr} proposed a trajectory-joint clustering method in vehicular scenarios. In the developed method \cite{huangchenscatterer-zr}, a novel distance measurement function based on the clustering algorithm was developed to measure the distance between trajectories in the time domain, which identified the dynamic cluster evolution in high-mobility vehicular channels. Therefore, by intelligently processing the uni-modal communication information via ML-based models, the MPC component can be properly clustered and tracked in the  RF-only
intelligent channel modeling.


\subsubsection{RF-Only Intelligent Channel Modeling for Parameter Prediction} 
    The RF-only intelligent channel modeling for parameter prediction aims to model and predict key channel parameters. Owing to the paramount importance of angular parameters in the channel modeling, the authors in \cite{fastaoa-zr} proposed a fast ML-based AoA recognition method, which contained offline training and online prediction processes. In the offline training, a prediction model was developed through a support vector machine (SVM) based on a measurement channel dataset under sub-6 GHz frequency bands. In the online prediction, the proposed model in \cite{fastaoa-zr} was utilized to realize the fast AoA recognition for single-input multiple-output (SIMO) channels. However, the proposed method in \cite{fastaoa-zr} ignored the utilization of artificial neural network (ANN) in the RF-only intelligent channel modeling for parameter prediction. To overcome this limitation, an ANN was utilized to model and predict the channel excess attenuation of satellite channels \cite{baisatellite-zr}. The input was weather data and the propagation measurement at Q-bands, i.e., 39.402 GHz, the output was channel excess attenuation in  satellite communications, where the loss function was the mean square error (MSE). Specifically, the proposed method in \cite{baisatellite-zr} conducted channel fading predictions by utilizing low-cost weather sensors and feedback on the channel state information (CSI) from the return link. However, the proposed method in \cite{baisatellite-zr} was limited to the modeling and prediction of one single channel parameter.  To model and predict more channel parameters, the authors in \cite{zhaox1-zr} proposed a framework for SIMO channel modeling and prediction based on ANNs under mmWave frequency bands. In the proposed framework \cite{zhaox1-zr}, the input was the measured channel impulse response (CIR) processed by space-alternating generalized expectation- maximization (SAGE) and the output contained channel small-scale and large-scale fading characteristics, such as direction of arrival (DoA) of MPCs, delay spread, angular spread, and path loss. To further model and predict temporal/spatial-varying channel parameters, the authors in \cite{zhaox2-zr} proposed a channel modeling and prediction framework based on ANNs, which was implemented to accurately playback the measured SIMO channel under mmWave frequency bands. In addition to the measured CIR processed by SAGE, the input in \cite{zhaox2-zr} also contained 3D transceiver coordinates. Meanwhile, the output also contained  channel small-scale and large-scale fading characteristics. Different from ANNs, the authors in \cite{bai11-zw} considered the advantage of convolution neural networks (CNNs), which can facilitate the modeling and prediction of more diverse channel parameters. To be specific, the input was the 3D location information of transceiver antennas and the output contained path loss, delay spread,  delay mean, angular parameters, etc, where the MSE  was utilized as the loss function. To  further select and condense more efficient input parameters, the authors in \cite{bai22-zw} utilized the least absolute shrinkage and selection operator (LASSO) algorithm to model and predict Q-band channel attenuation. Specifically, seven atmospheric parameters of capturing satellite channel attenuation were selected by the LASSO algorithm from fourteen commonly utilized atmospheric parameters. As a result, the input contained air temperature, rainfall rate, relative humidity, thickness of rainfall amount, visibility, wind speed, and average particle speed.  Furthermore, considering the utility of long short-term memory (LSTM) for handling time-series problems, it is strongly applicable for satellite channel fading modeling and prediction based on time-series atmospheric parameters. Therefore, with the help of LSTM and the aforementioned  seven atmospheric parameters selected by LASSO, the output was the Q-band channel attenuation, which can be modeled and predicted in real-time \cite{bai22-zw}. Therefore, the uni-modal communication information can be intelligently processed to conduct the RF-only intelligent channel modeling for parameter prediction.

In summary, the RF-only intelligent channel modeling approach extracts channel information from RF communication data and augments the capability and adaptability of channel models.
Currently, based on the ML-based model, extensive RF-only intelligent channel modeling has been conducted for specific static/dynamic scenarios. In terms of  the purpose and output of ML-based models, the existing work can be classified into RF-only intelligent channel modeling for clustering and RF-only intelligent channel modeling for parameter prediction. The RF-only intelligent channel modeling for clustering aims to cluster and track the MPC component. The RF-only intelligent channel modeling for parameter prediction aims to model and predict  parameters related to channel large-scale fading characteristics and/or channel small-scale fading characteristics.

\subsection{Multi-Modal Intelligent Channel Modeling}
\begin{figure}[!t]
		\centering	\includegraphics[width=0.49\textwidth]{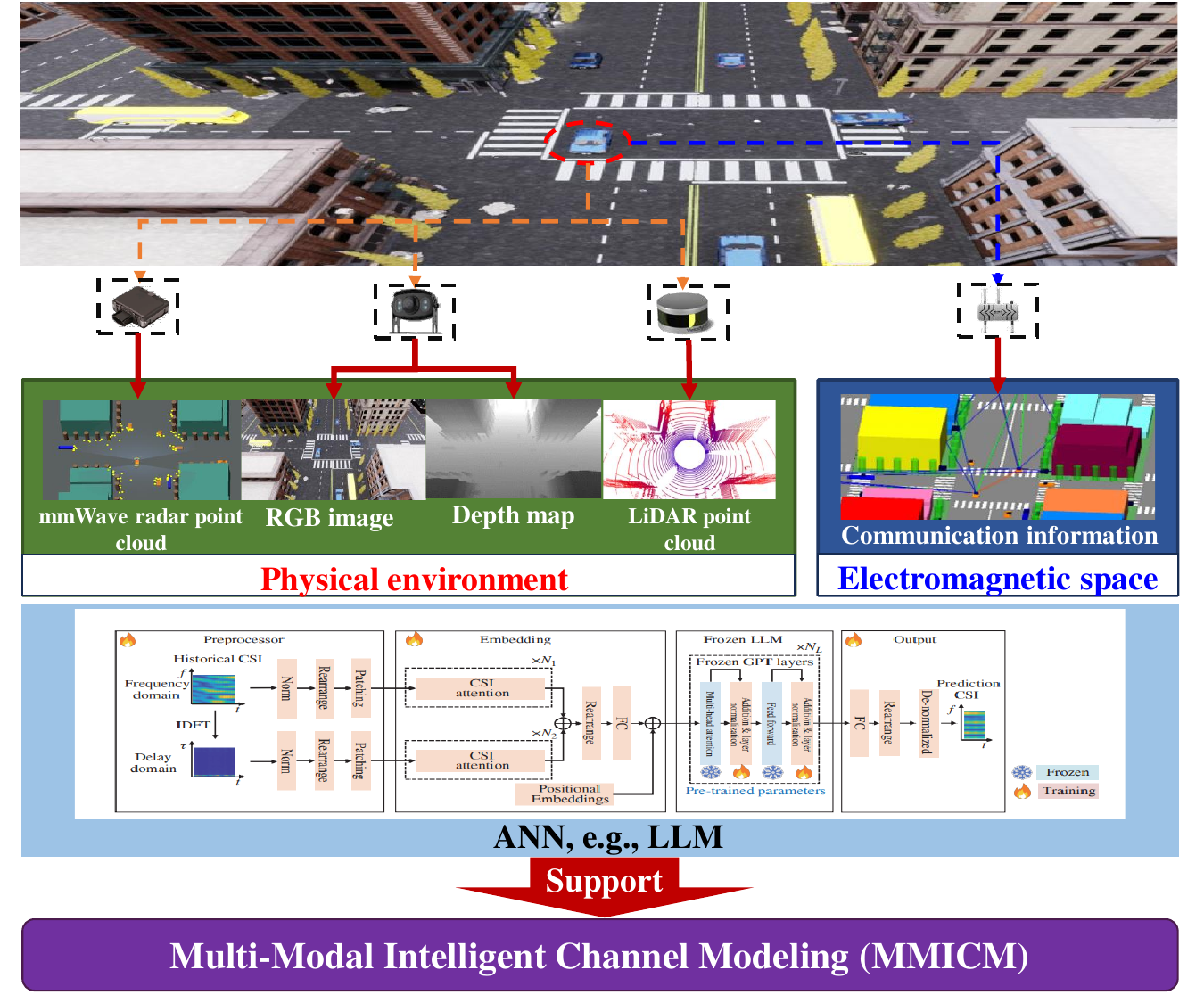}
	\caption{MMICM with the exploration of mapping relationship between sensory information in the physical environment and communication information in the electromagnetic space.}
	\label{MMICM}
	\end{figure}

 Different from the conventional and RF-only intelligent channel modeling that solely utilize communication information, the MMICM utilizes the multi-modal information from communication devices and various sensors to achieve intelligent channel modeling. By exploring the nonlinear mapping relationship
among communications information and various sensory information via SoM,
the propagation environment feature around the
transceiver is obtained, which can facilitate more accurate, less complex, and more general channel modeling, as shown in Fig.~\ref{MMICM}. Compared with the aforementioned channel modeling
approaches that solely rely on RF communication information,
the utilization of multi-modal information from communication devices and various sensors can provide a more in-depth understanding of the propagation environment, leading to a more accurate channel modeling. According to the number of modalities of sensory information utilized in the MMICM, existing MMICMs can be classified into those based on uni-modal sensory information and those based on multi-modal sensory information. Note that, for the MMICM based on the uni-modal sensory information, although solely uni-modal sensory information is exploited, due to the introduction of communication information, it still belongs to multi-modal intelligent modeling.

\subsubsection{MMICM Based on Uni-Modal Sensory Information} As previously mentioned, sensory information is introduced in the MMICM to improve the accuracy of channel modeling for the first time. For the MMICM based on uni-modal sensory information, only one modal of the sensory information is utilized in the channel modeling. The developed channel model in \cite{satellite1-mr} utilized 2D satellite maps to predict path loss exponent and shadowing factor in urban scenarios under sub-6 GHz frequency band. To explore the mapping relationship between 2D satellite maps in the physical environment and channel large-scale fading characteristics, i.e., path loss, in the electromagnetic space, the authors in \cite{satellite1-mr} utilized two networks, i.e., VGG-16 and ResNet-50, respectively. To describe the electromagnetic space more precisely, the authors in \cite{satellite2-mr}  transformed the classification task in \cite{satellite1-mr} into a regression task. Specifically, the proposed MMICM in \cite{satellite2-mr} utilized 2D satellite maps to predict path loss distribution within the specific city coverage in urban scenarios under sub-6 GHz frequency band. In \cite{satellite2-mr}, a VGG-16 neural network was trained to explore the mapping relationship between 2D satellite maps in the physical environment and path loss distribution in the electromagnetic space. To extract environmental features from satellite images more accurately, the proposed MMICM in \cite{satellite3-mr} utilized residual structures and attention mechanisms to predict path loss values in urban scenarios under sub-6 GHz frequency band. However,  2D satellite images only contain 2D information of the physical environment, lacking of 3D information of the buildings in the physical environment. To acquire more comprehensive 3D information of physical environment, the 3D city map was utilized in \cite{citymap1-mr,citymap2-mr} to predict path loss maps in urban scenarios under sub-6 GHz frequency bands. In \cite{citymap1-mr,citymap2-mr}, the architectures of UNet and CNN were utilized to process the 3D city map data. However, map data, including satellite maps and city maps, solely contains global information of the physical environment, lacking of detailed information of the physical environment near the transceivers. To overcome this limitation, sensory data acquired by sensing devices was introduced in the path loss prediction. Light detection and ranging (LiDAR) point clouds with 3D information of the physical environment were utilized in the proposed MMICM \cite{lidar-mr}. To be specific, the authors in \cite{lidar-mr} utilized an encoder-decoder architecture of CNNs to process LiDAR point clouds and predict path loss values in urban scenarios under mmWave frequency bands. However, LiDAR point clouds have a limited sensing range and cannot capture the broad range of physical environmental information. To overcome this drawback, 3D information with locations of buildings and transceivers was utilized in \cite{location-mr} to predict path loss values in urban scenarios under mmWave frequency bands. To explore the mapping relationship, a neural network, named SegNet, was utilized in \cite{location-mr}  to process the structured location information. However, for the aforementioned MMICMs based on uni-modal sensory information \cite{satellite1-mr}-\cite{location-mr}, the understanding of the physical environment was limited, resulting in a decrease in the accuracy of channel modeling.

\subsubsection{MMICM Based on Multi-Modal Sensory Information} 
Compared to the MMICM based on uni-modal sensory information, the MMICM based on multi-modal sensory information utilizes at least two modals of sensory information, including RGB images, depth maps, LiDAR point clouds, and mmWave radar point clouds, to achieve intelligent channel modeling. Visual sensory data, including RGB images and depth maps, facilitates easier object recognition. However, the visual sensory data is easily influenced by lighting conditions. To overcome this drawback, radar sensory data, including LiDAR and mmWave radar, is introduced in sensory data acquisition. Compared to visual sensory data, radar data contains more comprehensive 3D information and is generally unaffected by external environmental conditions. When the multi-modal sensory information is introduced in the MMICM, the understanding of the physical environment is more comprehensive, thus improving the accuracy of channel modeling. The proposed MMICM in \cite{mingran11} utilized RGB images and depth maps to predict path loss distribution in urban scenarios under mmWave frequency band. The authors in \cite{mingran11} constructed a new mixed multi-modal sensing-communication integration dataset in the UAV-to-ground scenario. Based on the constructed dataset, the multi-modal sensory information-based real-time path loss prediction scheme was developed in \cite{mingran11}. To capture more comprehensive physical environmental information, the authors in \cite{XGD-mr} utilized RGB images, depth maps, and LiDAR point clouds to predict path loss distribution in urban scenarios under sub-6 GHz and mmWave frequency bands. By leveraging a generative adversarial network (GAN) for parameter initialization coupled with fine-tuning through supervised learning, the mapping relationship between multi-modal sensory information and path loss distribution was explored  \cite{XGD-mr}. However, the aforementioned MMICMs in \cite{satellite1-mr}-\cite{XGD-mr} were limited in modeling channel large-scale fading characteristics, i.e., path loss. To model channel small-scale  fading characteristics, the authors in \cite{Daitou-zw} proposed a novel vehicular MMICM by scatterer recognition from LiDAR point clouds. By using LiDAR point clouds to implement scatterer recognition, multipath effect, i.e., channel small-scale fading characteristics, were mimicked closely associated with  the environment in \cite{Daitou-zw}.

In summary, the MMICM utilizes the multi-modal information from communication devices and various sensors to achieve intelligent channel modeling. The MMICM can be classified into uni-modal sensory information-based and multi-modal sensory information-based according to the number of sensory information modalities utilized. The MMICM based on uni-modal sensory information 
 utilizes communication information and one kind of sensory information to achieve intelligent channel modeling. To implement more comprehensive understanding of physical environment and improve the accuracy of channel modeling, the MMICM based on multi-modal sensory information utilizes communication information and at least two kinds of sensory information to achieve intelligent channel modeling.

\subsection{Comparison and Analysis}

\begin{figure*}[!t]
		\centering	\includegraphics[width=0.95\textwidth]{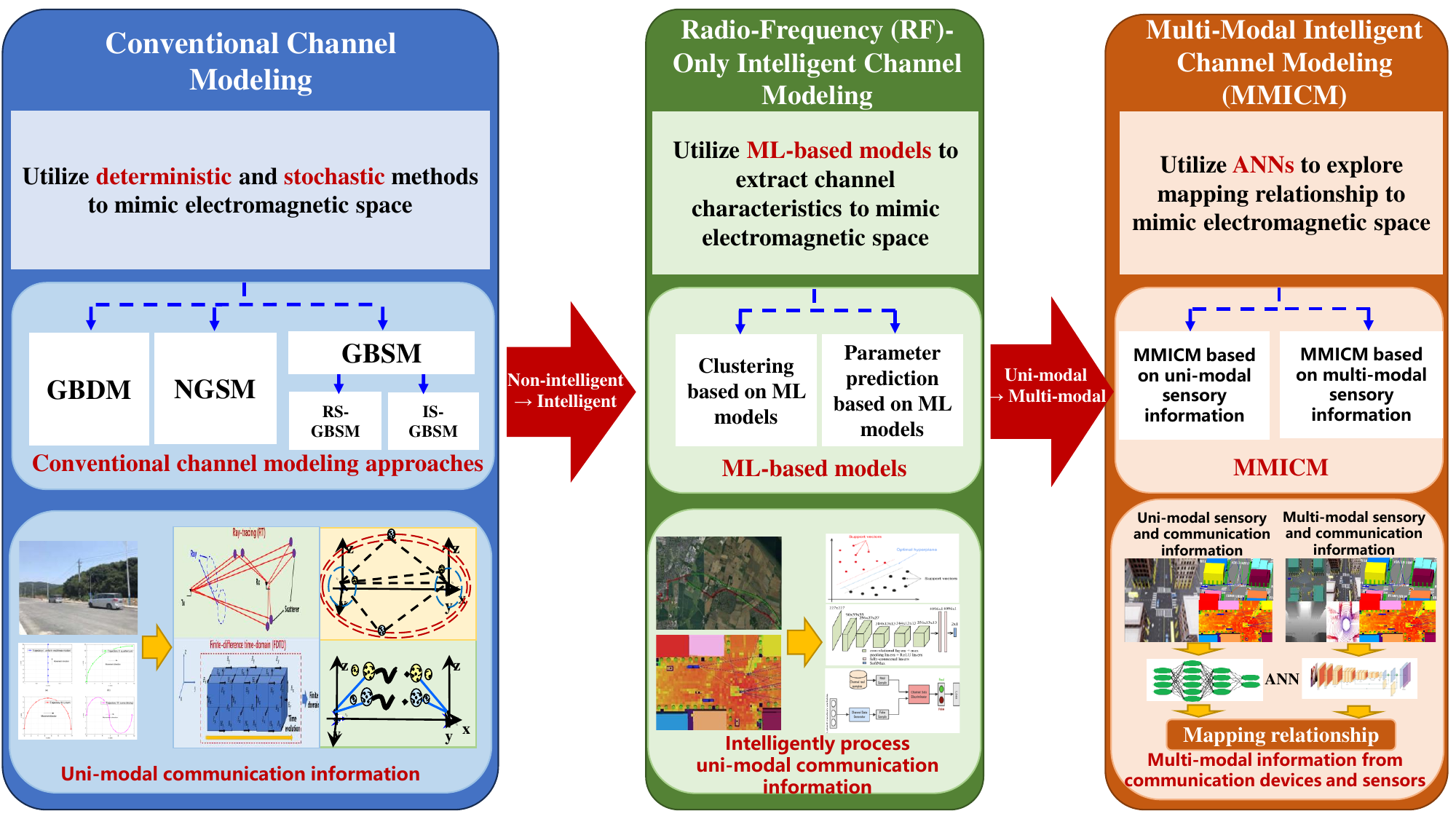}
	\caption{Evolution of channel modeling approaches, including conventional channel modeling, RF-only intelligent channel modeling, and MMICM.}
	\label{model_evolution}
	\end{figure*}

In this section, according to whether ML-based models and uni/multi-modal information are utilized, current channel modeling approaches are classified and their evolution is also presented. The evolution is that, starting from conventional channel modeling, the channel modeling approach moves on to the RF-only intelligent channel modeling and finally lands
on MMICM. In Table~\ref{Modeling_approaches}, different channel modeling approaches are compared analyzed. For clarity, Fig.~\ref{model_evolution} further presents the evolution of channel modeling from the
conventional channel modeling approach to the intelligent
channel modeling approach with the help of uni-modal and
multi-modal information.

\begin{table*}[!t]
\centering
    \begin{small}
\caption{Comparisons of Channel Modeling Approaches.}
\renewcommand\arraystretch{2.3}
\begin{tabular}{|c|c|c|c|c|}
    \hline			
    \textbf{\multirow{1}{*}{\makecell[c]{Channel modeling \\approach}}} &\multicolumn{2}{c|}{\textbf{\multirow{1}{*}{\makecell[c]{Classification}}}}&\textbf{\multirow{1}{*}{\makecell[c]{Data modality}}}
     &\textbf{\multirow{1}{*}{\makecell[c]{Description}}}\\
    \hline
    \multirow{4}{*}{\makecell[c]{Conventional \\channel modeling}} & \multicolumn{2}{c|}{\multirow{1.1}{*}{GBDM}}                      & \multirow{4}{*}{\makecell[c]{Uni-modal\\ communication \\information}} & \multirow{4}{*}{\makecell[c]{Utilize deterministic  \\and statistical methods \\to  mimic electromagnetic space}} \\ 
    \cline{2-3}
    & \multicolumn{2}{c|}{\multirow{1.1}{*}{NGSM}}&  &   \\
    \cline{2-3}
    & \multicolumn{1}{l|}{\multirow{2}{*}{GBSM}} & \multirow{1.1}{*}{\makecell[c]{RS-GBSM}} &   &  \\ \cline{3-3}
    & \multicolumn{1}{l|}{}  & \multirow{1.1}{*}{\makecell[c]{IS-GBSM}} &     &    \\ 
    \hline
   
    \multicolumn{1}{|c|}{\multirow{2.5}{*}{\makecell[c]{RF-only intelligent \\channel modeling}}} & \multicolumn{2}{c|}{\makecell[c]{RF-only intelligent \\channel modeling \\for clustering}} & \multicolumn{1}{c|}{\multirow{2.5}{*}{\makecell[c]{Uni-modal\\communication \\information }}}  & \multicolumn{1}{c|}{\multirow{2.6}{*}{\makecell[c]{Utilize ML-based models to \\extract channel characteristics \\from RF communication information, \\thus achieving intelligent \\ channel modeling}}} \\ 
    \cline{2-3}
    \multicolumn{1}{|c|}{}& \multicolumn{2}{c|}{\makecell[c]{RF-only intelligent \\channel modeling \\for parameter prediction}} & \multicolumn{1}{c|}{} & \multicolumn{1}{c|}{}\\
    \hline

    \multicolumn{1}{|c|}{\multirow{3}{*}{MMICM}} & \multicolumn{2}{c|}{\makecell[c]{MMICM based on \\uni-modal sensory\\ information}} & \multicolumn{1}{c|}{\makecell[c]{Communication information \\and one kind of \\sensory information }}  & \multicolumn{1}{c|}{\multirow{2.5}{*}{\makecell[c]{Utilize ANNs to explore the mapping \\relationship between physical \\environment and electromagnetic space, \\thus improving the accuracy \\of intelligent channel modeling}}} \\ 
    \cline{2-4}
    \multicolumn{1}{|c|}{}& \multicolumn{2}{c|}{\makecell[c]{MMICM based on \\multi-modal sensory \\information}} & \multicolumn{1}{c|}{\makecell[c]{Communication Information \\and at least two kinds \\of sensory information}} & \multicolumn{1}{c|}{}\\
    \hline
\end{tabular} 
\label{Modeling_approaches}
    \end{small}
\end{table*}

The conventional channel modeling aims to properly characterize and represent the electromagnetic space. Currently, the conventional channel modeling contains the  GBDM, NGSM, and GBSM. In the GBDM, the procedure of physical radio propagation in a specific scenario can be characterized. In the NGSM, the construction
depends upon the channel statistical property and the
PDF according to measurement campaigns. In the GBSM, the construction is based on the predefined stochastic distribution of 
scatterers/clusters in the geometrical propagation environment. In general, the GBSM can be further classified into RS-GBSM and IS-GBSM.

Different from the conventional channel modeling,  the RF-only intelligent channel modeling intelligently processes the RF communication information via the ML-based method to enhance the capability and adaptability of channel models. Currently, the RF-only intelligent channel modeling is conducted for specific static/dynamic scenarios due to the utilization of data from specific scenarios that cannot represent the environment. Furthermore, aiming at adapting to 6G complicated communication scenarios, intelligent channel models which incorporate adequate channel characteristics under various 6G scenarios, is being formulated.

Unlike the conventional and RF-only intelligent channel modeling that solely utilize communication information, the MMICM utilizes the multi-modal information from communication devices and various sensors to achieve intelligent channel modeling. By exploring the nonlinear mapping relationship among communications and multi-modal sensing via SoM, the propagation environment feature around the transceiver is obtained, which can facilitate more accurate, less complex, and more general channel modeling.  The MMICM can be classified into uni-modal sensory information-based and multi-modal sensory information-based according to the number of sensory information modalities used. For the MMICM based on uni-modal sensory information, it exploits communication information and one kind of sensory information. For the MMICM based on
multi-modal sensory information, it exploits communication information and at least two kinds of sensory information. Note that, different modalities of sensory information have their own advantages and disadvantages. The simultaneous utilization of multi-modal sensory information can compensate for their respective disadvantages. However, whether different modalities of sensory information can be simultaneously utilized depends on the specific application scenario. Therefore, MMICMs based on
multi-modal sensory information have their own application scenarios.

\section{Methods of Capturing Channel Non-Stationarity and Consistency}

\begin{figure*}[!t]
	\centering
	\subfigure[]{\includegraphics[width=0.49\textwidth]{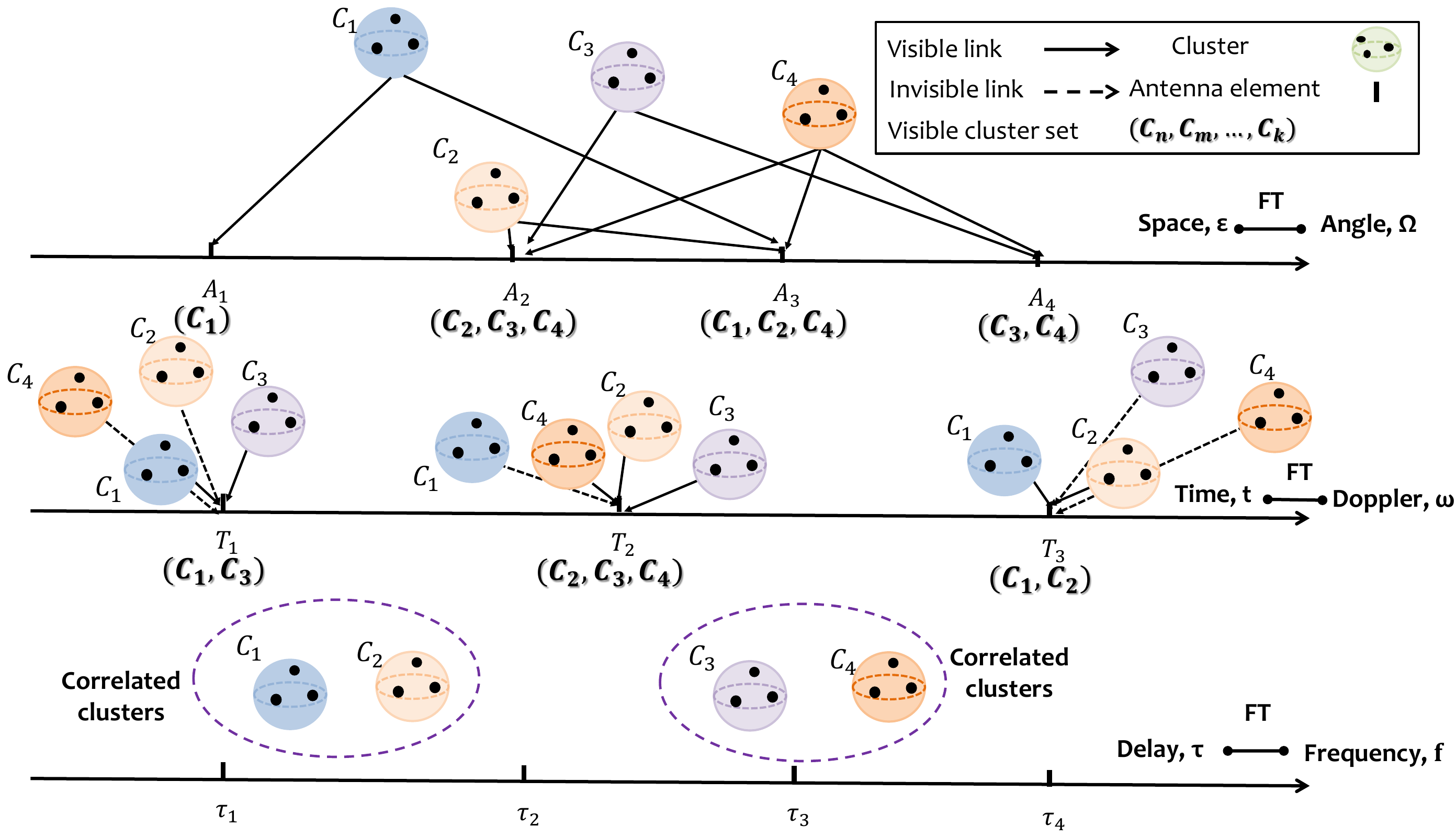}}
	\subfigure[]{\includegraphics[width=0.49\textwidth]{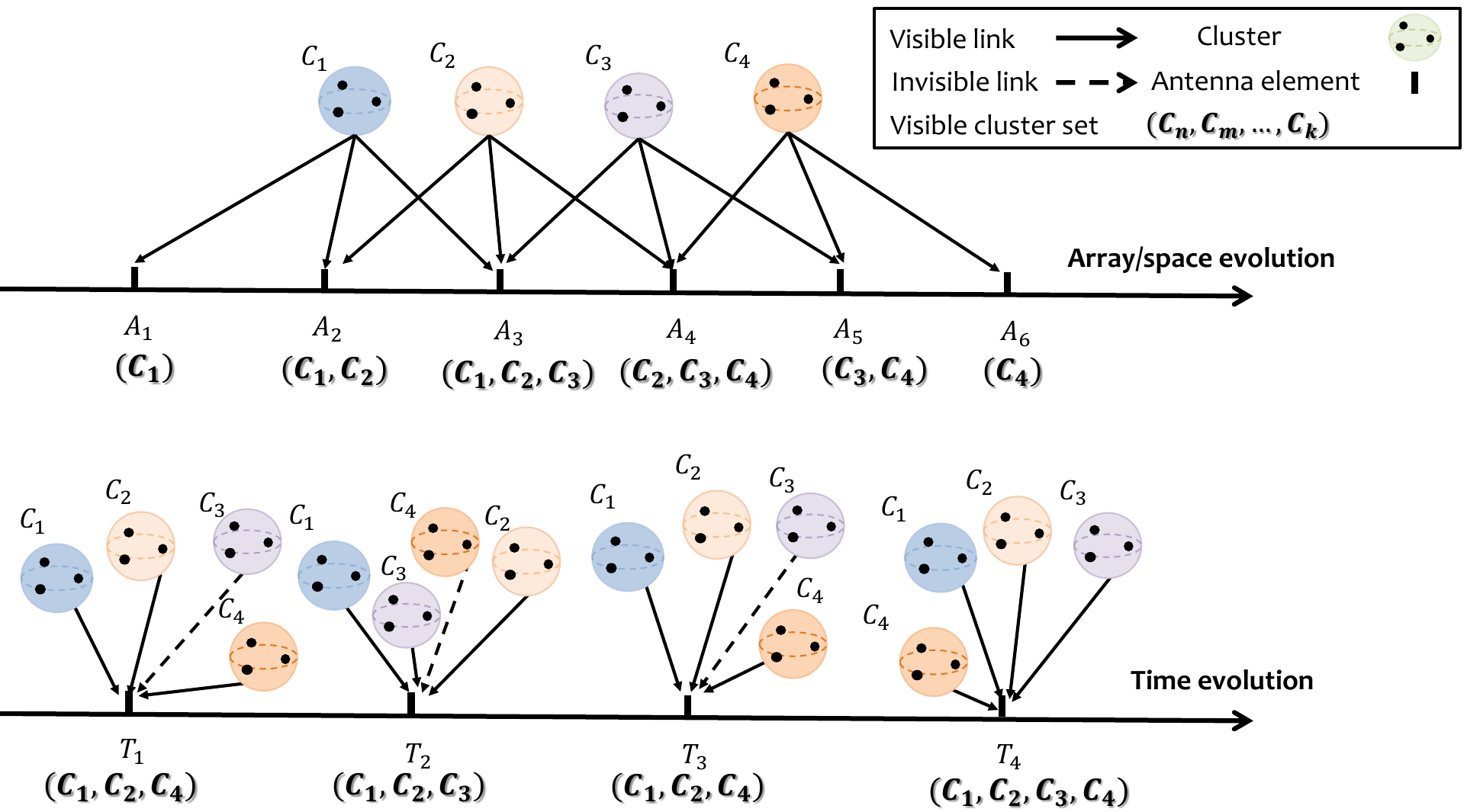}} 
	\caption{Physical mechanism underlying channel non-stationarity and consistency. (a) Channel non-stationarity. (b) Channel consistency.}
  \label{discription_non-stationarity_consistency}
\end{figure*}

Since channel non-stationarity is a typical channel characteristic and channel consistency is a physical channel feature, an in-depth understanding and accurate capturing of channel non-stationarity and consistency are of paramount importance. In the 6G era, more dynamic propagation environments, higher frequency bands, larger-scale antenna arrays further increase the necessity of investigating channel non-stationarity and consistency \cite{COMSTmy}. 
For the channel non-stationarity is an important channel characteristic, i.e., the channel exhibits non-stationarity in different domains. Channel non-stationarity in a specific domain, such as space/time/frequency, means that channel statistical properties vary in this domain. To be specific, in the high-mobility scenarios, the appearance and disappearance  of scatterers/clusters are rapid over time, leading to time non-stationarity. In massive MIMO scenarios, the appearance and disappearance of scatterers/clusters can be observed in the antenna array, leading to space non-stationarity. In mmWave communications, different frequency components exhibit different transmission coefficients in the same signal, leading to frequency non-stationarity. For clarity, the physical mechanism underlying channel non-stationarity is shown in Fig.~\ref{discription_non-stationarity_consistency}(a).
Unlike channel non-stationarity, channel consistency is a channel physical feature and inherently exists in each channel. Channel consistency in the space/time domain means that channels change smoothly and consistently as the array/time evolves. In the high-mobility mmWave communication, channels possess significantly small coherence time, resulting in the frequent channel update \cite{add11-zw}. As a consequence, adjacent moments sharing similar propagation environments have many identical scatterers/clusters, resulting in time consistency. Also, short carrier wavelength in mmWave communications results in the small adjacent antenna spacing in the massive MIMO antenna array. As a result, adjacent antenna elements sharing similar positions have many identical scatterers/clusters, resulting in  space consistency.  For clarity, the physical mechanism underlying channel consistency is given in Fig.~\ref{discription_non-stationarity_consistency}(b).
To support the precise channel modeling, it is exceedingly necessary to review the recent advance in capturing channel non-stationarity and consistency. Towards this objective, in this section, based on the conventional channel modeling, RF-only intelligent channel modeling, and MMICM, typical methods of capturing channel non-stationarity and consistency by modeling mathematical, spatial, coupling, and mapping relationships are presented in turn. Different channel modeling approaches have varying abilities to model different types of relationships. Finally, these methods of capturing channel non-stationarity and consistency are adequately compared and analyzed.

\subsection{Mathematical Relationship Modeling via Stochastic Processes}

 		\begin{figure*}[!t]
		\centering	\includegraphics[width=0.99\textwidth]{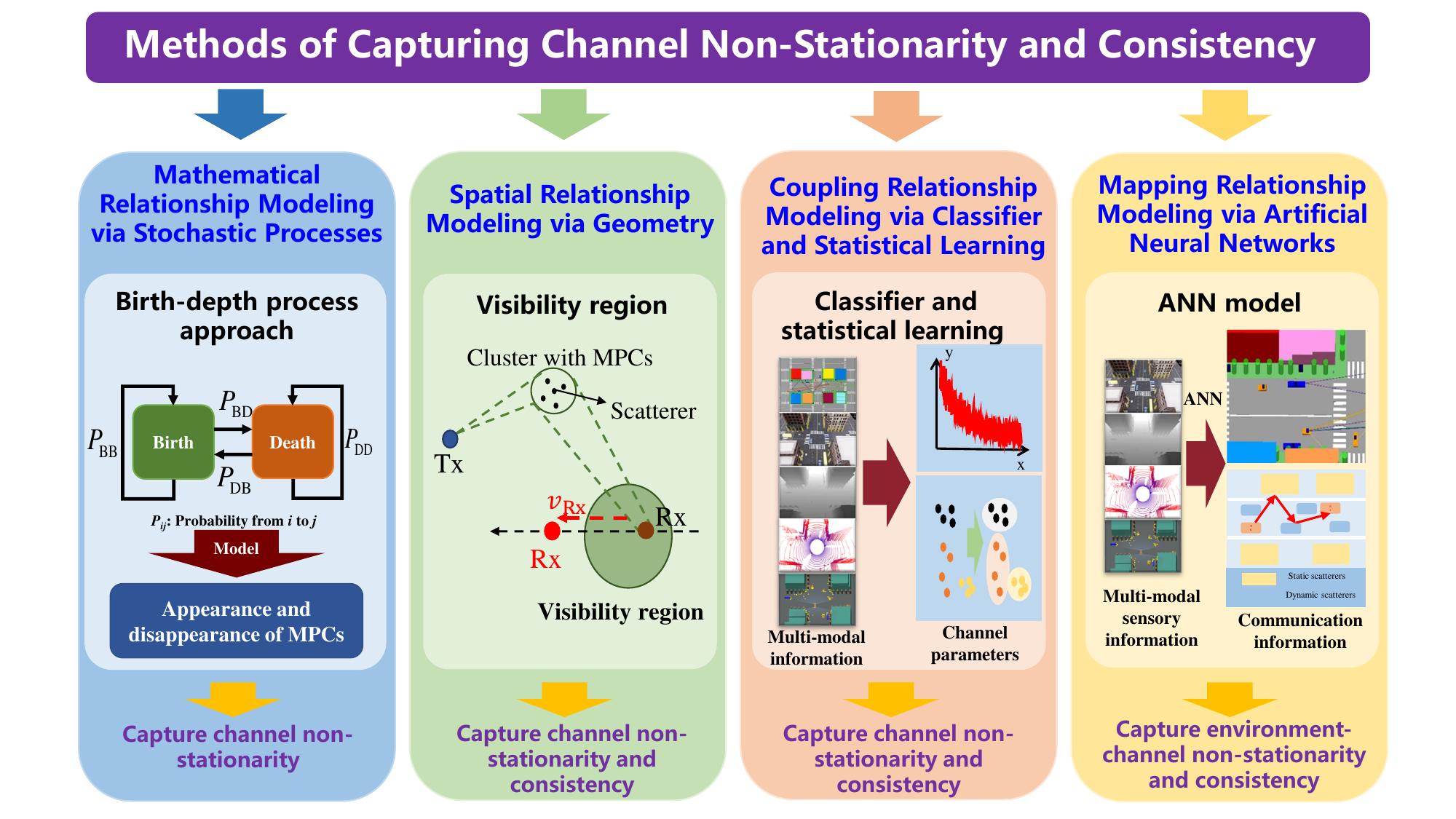}
	\caption{Typical methods of capturing channel non-stationarity and consistency by modeling the mathematical relationship, the spatial relationship, the coupling relationship, and the mapping relationship.}
	\label{non-stationarity-consistency_four_methods}
	\end{figure*}


According to channel measurement campaigns \cite{nonstationarity1-zw}--\cite{nonstationarity7-zw},
channel non-stationarity is an important channel characteristic. The mathematical definition of channel non-stationarity in a specific domain, such as space, time, or frequency domain, is that the channel statistical property varies in this domain.  The corresponding physical definition of the channel non-stationarity in a specific domain, i.e., space, time, or frequency domain, is that the appearance and disappearance of MPCs exhibit in this domain.  The birth-death process approach is a special case of continuous Markov process and state transitions contain two categories, including birth state and death state. In the birth-death process approach, the birth or death state of MPCs in a certain domain determines whether the MPCs exist or disappear in this domain, respectively. By utilizing the birth-death process approach to properly characterize the appearance and disappearance of MPCs, the channel non-stationarity can be modeled.  Because of the excellent trade-off between complexity and accuracy of the birth-death process approach, it has been extensively leveraged  to characterize the mathematical relationship for the variation of the MPC number, as shown Fig.~\ref{non-stationarity-consistency_four_methods}.

In second-generation (2G), narrowband wireless communication systems were widely exploited to highly dynamic application scenarios. In this case, channel non-stationarity in the time domain, i.e., time non-stationarity, needs to be modeled \cite{book1-zw,nonstationarity2-zw}. Towards this objective, the authors in \cite{MRM1-zw} developed a single-input single-out (SISO) NGSM at the sub-6 GHz frequency band. In the developed NGSM \cite{MRM1-zw}, the appearance and disappearance of MPCs, i.e., the variation of MPC number, over time were characterized by the birth-death process approach via the transition probability and steady-state probability. Similarly, the developed NGSMs in \cite{MRM2-zw}--\cite{MRM7-zw} also utilized the birth-death process approach to mimic the appearance and disappearance of MPCs over time, where MPCs with varying numbers at different time instances were modeled. 

As the communication bandwidth increases, channel  non-stationarity in the frequency domain, i.e., frequency non-stationarity, needs to be captured. When wideband communications are utilized to high-mobility scenarios in third-generation (3G) and fourth-generation (4G), the channel simultaneously exhibits time non-stationarity and frequency non-stationarity \cite{book2-zw}. Aiming at modeling time-frequency non-stationarity, two wideband SISO  NGSMs with tapped delay line (TDL) structure were proposed in \cite{MRM8-zw,MRM9-zw}, where the birth-death process approach was utilized to model the correlated appearance and disappearance of taps with different delays over time. Furthermore, the authors in \cite{MRM10-zw,MRM11-zw} proposed two TDL-based NGSMs and also modeled the complex time evolution of taps by using the birth-death process approach.

In 5G, the utilization of massive MIMO technology in high-mobility scenarios, the channel exhibits non-stationarity in both the space and time domains, and thus space-time  non-stationarity needs to be modeled. Two 5G standardized 3D channel models, i.e., 5G Channel Model (5GCM) in \cite{5GCM} and mm-wave based Mobile radio Access network for fifth Generation Integrated Communications (mmMAGIC) channel model in  \cite{mmMAGIC} based on the GBSM approach were developed. Aiming at capturing space-time non-stationarity, the 5GCM in \cite{5GCM} and the mmMAGIC channel model in \cite{mmMAGIC} exploited the birth-death process approach to characterize the appearance and disappearance of clusters in the time and space/array domain. As a result, different time instants and different antennas have different numbers of effective clusters in the propagation environment. 

In the future 6G era, it can be predictable that the massive MIMO technology and mmWave communications will be simultaneously exploited to high-mobility scenarios. In such a condition, the channel will exhibit space, time, and frequency non-stationarity, and thus space-time-frequency non-stationarity needs to be modeled \cite{TWCMY-zw}. To this aim, the authors in \cite{MRM12-zw} proposed a general IS-GBSM, where the  birth-death process approach was simultaneously utilized to model the appearance and disappearance of clusters in the space, time, and frequency domains. To further characterize the correlated scattering in the propagation environment, our previous work in \cite{TITS-huang-zw} obtained correlated clusters, which can adequately characterize the correlated attenuation and phase shift of rays with different delays. Based on the correlated cluster, by utilizing the birth-death process approach, the complex appearance and disappearance of correlated clusters in the time and space domains can be mimicked, which jointly modeled space, time, and frequency non-stationarity.

In summary, the birth-death process approach aims to mimic the mathematical relationship for the variation of MPCs in the space, time, and frequency domains, thus modeling channel non-stationarity in the space, time, and  frequency domains. Currently, the birth-death process approach has been widely utilized in the conventional channel modeling.

\subsection{Spatial Relationship Modeling via Geometry}
 
The channel measurement and analysis in \cite{COMSTmy}, \cite{SRM1-zw}--\cite{SRM3-zw} indicate that channel consistency is a typical channel physical feature, which inherently exists in each channel.  The mathematical definition of channel consistency in a specific domain, such as space  domain or time domain, is that the channel changes in this domain in a consistent and smooth  manner. In such a condition, the channel is spatially consistent and smoothly temporal-evolving, where the MPC parameter also varies in the space or time domain continuously and smoothly. From the perspective of actual propagation mechanism, the physical mechanism underlying channel consistency in a specific domain, such as space domain or time domain, is that adjacent time instants or adjacent antenna elements have similar MPC sets, which share many identical MPCs in the propagation environment. To capture channel consistency, it is necessary to characterize the spatial feature of MPCs by considering the geometry of the propagation environment. Since the geometry of propagation environment is properly considered in the visibility region approach, it has been widely exploited to characterize the spatial relationship for the smooth evolution of MPCs via geometry. In the visibility region approach, a portion of the propagation environment is defined as the visibility region, where MPCs are effective when they are located within this region, thus contributing to the CIR, as shown in Fig.~\ref{non-stationarity-consistency_four_methods}. With the movement of transceivers and MPCs in the propagation environment, effective MPCs are smoothly and continuously change as time and antenna evolves, thus capturing channel non-stationarity and consistency.

In the highly dynamic communication scenario, channels experience severe Doppler shift and have significantly short coherence time, and thus need to be updated  frequently. In such a condition, a continuous and consistent CIR over time needs to be generated, which can be ensured via the characterization of channel consistency in the time domain, i.e., time consistency \cite{SRMziwei-zw}. In the standardized COST 2100 channel method \cite{SRM4-zw}, a visibility region approach was developed  to model time non-stationarity and consistency for the first time. In the developed visibility region approach, scatterers/clusters are assigned to visibility regions. Once the terminal is within a visibility region, the scatterer/cluster assigned to this visibility region is an effective scatterer/cluster, which contributes to the CIR and channel realization. As mobile stations (MSs) and scatterers/clusters continuously move, MSs smoothly enter and leave visibility regions assigned by scatterers/clusters, leading to effective and ineffective scatterers/clusters over time, respectively. The effectiveness and ineffectiveness of scatterers/clusters can be smoothly mimicked via the visibility region approach. Similar to the standardized COST 2100 in \cite{SRM4-zw}, the authors in \cite{SRM41-zw}--\cite{SRM43-zw} also developed visibility region approach to characterize smooth time evolution of MPCs and time non-stationarity and consistency.
Therefore, the spatial relationship for the smooth evolution of MPCs in the time domain can be characterized by considering the geometry of the propagation environment, which can model time non-stationarity and consistency. 

As the massive MIMO technology and mmWave communications are adopted to the high-mobility application scenario in 6G, space-time non-stationarity of channels with time-space consistency needs to be mimicked \cite{COMSTmy}. The authors in \cite{SRMbai-zw} proposed a UAV IS-GBSM for massive MIMO mmWave channels. In the proposed model \cite{SRMbai-zw}, each Tx antenna and each Rx antenna were assigned to a visibility region, which is a sphere with the center of the Tx antenna and the Rx antenna. In this case, due to the movement of transceivers and scatterers/clusters, there are different visibility regions for different antennas and time instants, thus modeling space-time non-stationarity. In addition, as the movement of transceivers and scatterers/clusters is continuous and consistent, there are similar visibility regions at adjacent antennas and time instants, thus capturing space-time consistency. To further model channel non-stationarity and consistency in multi-UAV cooperative channels, the authors in \cite{SRMbai2-zw} proposed an IS-GBSM with the utilization of the visibility region approach. The visibility region of a specific UAV was modeled as a sphere with the center of the UAV. As a consequence, in the multi-UAV communication, each UAV has its own visibility region, where effective MPCs of different UAVs are different. Furthermore, adjacent UAVs with similar positions have similar effective MPCs. In this case, channel non-stationarity and consistency in multi-UAV cooperative channels were captured in \cite{SRMbai2-zw}.  

In summary, the visibility region approach aims to characterize the spatial relationship for the smooth evolution of MPCs in the space and time domains, thus capturing channel non-stationarity and consistency in the space and time domains.  Currently, the visibility region approach has been widely utilized in the GBSM.

\subsection{Coupling Relationship Modeling via Classifier and Statistical Learning}

To further leverage geometrical information of the propagation environment, classifier and statistical learning are extensively employed \cite{classifer11-zw}--\cite{classifer33-zw}. The utilization of classifier and statistical learning can intelligently process communication information and/or sensory information collected from physical environment to accurately explore the complex evolution of scatterers/clusters and channel characteristics in the electromagnetic space. In this case, the coupling relationship between physical environment and electromagnetic space can be intelligently modeled. As a result, through the intelligent modeling of the coupling relationship, the comprehensive scatterer/cluster evolution and channel characteristics can be accurately revealed, thus precisely capturing channel non-stationarity and consistency, as shown in Fig.~\ref{non-stationarity-consistency_four_methods}. Note that the complex mapping relationship between scatterers/clusters and objects is not explored in the coupling relationship modeling. This is essentially different from the mapping relationship modeling, which will be described in Section III-D.


In \cite{cr11-zw}, the authors developed a clustering and tracking algorithm based on power-angle-spectrum (PAS). First, PAS was extracted from the measurement data via a Bartlett beamformer. For each PAS, potential targets were selected from the background and were divided into multiple clusters. Additionally, a minimum-cost tracking method based on the K-M algorithm was proposed to accurately identify clusters in channels, thus modeling time non-stationarity. To further consider the  time evolution of clusters, the authors in \cite{cr22-zw} introduced the power spectrum-based sequential tracker, which identified clusters through three-stage power spectrum processing. In terms of tracking, a Kalman filter was implemented to  predict the candidate range of tracked clusters in consecutive snapshots. In addition, a gradient-based histogram of power method was proposed to determine the time evolution of clusters, thus modeling time non-stationarity. However, the aforementioned methods in \cite{cr11-zw,cr22-zw} cannot capture time consistency due to the ignorance of temporal correlation between channels at different moments. To capture time consistency, the authors in \cite{cr33-zw} proposed a novel distance measurement function to measure the distance between trajectories in the time domain, which  considered differences in trajectories and positions. 
By modeling the variation of cluster trajectories and time evolution of clusters, time non-stationarity of channels with time consistency were captured. However, the aforementioned work in \cite{cr11-zw}--\cite{cr33-zw} solely utilized the uni-modal communication information to capture time non-stationarity and/or consistency.

In addition to solely utilizing the communication information, the authors in \cite{cr44-zw} developed a cluster-based stochastic model for ISAC channels. By utilizing the RF sensory information, a new channel modeling framework in \cite{cr44-zw} was proposed based on the 3GPP standard, which contained  communication clusters and sensing clusters. The clustering and tracking algorithms were utilized to extract and analyze ISAC channel characteristics, which can model time non-stationarity of channels with time consistency. To further utilize the non-RF sensory information, the authors in \cite{LA-GBSM-zw} exploited the sensory information, i.e., LiDAR point clouds, to detect  dynamic and static objects via DBSCAN, which can distinguish the dynamic and static scatterers. In addition, statistical distributions of parameters of dynamic and static scatterers can be explored in high, medium, and low VTDs, thus precisely revealing mmWave vehicular channel characteristics. Based on the revealed channel characteristics, a novel mmWave vehicular MMICM was proposed \cite{LA-GBSM-zw}, where channel non-stationarity and consistency were accurately modeled according to explored statistical distributions. Therefore, through the intelligent processing of LiDAR point clouds and communication information, the coupling relationship between physical environment and electromagnetic space can be modeled to facilitate the capturing of  channel non-stationarity and consistency.

In summary, by intelligently processing communication information and/or sensory information, the coupling relationship between physical environment and electromagnetic space can be modeled via the classifier and statistical learning. The coupling relationship modeling results in the accurate revelation of channel characteristics, which can facilitate the capturing of channel non-stationarity and consistency. Currently, the coupling relationship modeling via the classifier and statistical learning has been widely utilized in the RF-only intelligent channel modeling, while has been preliminarily utilized in the MMICM.

\subsection{Mapping Relationship Modeling via Artificial Neural Networks}

To conduct tasks related to intelligent sensing-communication integration, networked intelligent agents in 6G AIoT, including autonomous vehicles, UAVs, and robotics, are naturally equipped with diverse multi-modal sensors and communication devices. These multi-modal sensors and communication devices can collect extensive multi-modal sensory information and communication information. To achieve mutual facilitation between communications and multi-modal sensing, the nonlinear mapping relationship between communication information and multi-modal sensory  information needs to be explored. Fortunately, communication information and multi-modal sensory information are physically correlated collected from different domains within the identical physical environment. However, there are significant differences between communication and multi-modal sensing in terms of data structures, operation frequency bands, and application orientation, leading to the fact that the exploration of mapping relationship is of high challenges. According to data structures, the wireless communication channel parameter is one-dimensional (1D) discrete multipath channel information. For the multi-modal sensory information, RGB images are  2D dense visual information, depth images are 2D dense depth information, and LiDAR point clouds and mmWave radar point clouds are 3D sparse point cloud information. According to operation frequency bands, the communication information and multi-modal sensory information differ by more than 4 orders of magnitude, resulting in the fact that non-RF propagation more directional compared to RF propagation while with weaker diffraction and scattering capabilities. According to application orientation, communication information aims to establish a comprehensive RF propagation environment to support efficient and reliable wireless transmissions, whereas multi-modal sensory information aims to reconstruct, understand, and perceive the physical environment. Therefore, the exploration of nonlinear mapping relationship between communication information and multi-modal sensory information is of paramount importance and huge challenge. Thanks to the innate ability to process nonlinear and complex models \cite{ANN11-zw}--\cite{ANN33-zw}, the ANN will become an essential native component for the exploration of mapping relationship. Based on the mapping relationship modeling via ANNs, the correspondence and complex evolution between scatterers/clusters with MPCs and objects can be accurately modeled, thus capturing the tight interplay between physical environment and channel non-stationarity/consistency, as shown in Fig.~\ref{non-stationarity-consistency_four_methods}.


For the capturing of channel non-stationarity and consistency, considering the importance and challenge of the mapping relationship modeling, a preliminary work has been carried out. In \cite{Daitou-zw}, the authors explored the complicated mapping relationship between the scatterer in electromagnetic space and the LiDAR point cloud in physical environment, was explored through
multilayer perceptron (MLP) by considering electromagnetic propagation effect. Based on the explored mapping relationship, a mmWave MMICM for vehicular scenarios was proposed. In the proposed MMICM \cite{Daitou-zw}, based on the scatterer recognition from LiDAR point clouds
and the modeling of tight interplay between physical environment and electromagnetic space, the scatterer smoothly varied with the LiDAR point cloud. 
Therefore, by modeling the mapping relationship between LiDAR point clouds and communication information via MLP, channel non-stationarity and consistency
were modeled closely combined with the environment,
 i.e., environment-channel non-stationarity and consistency were modeled.  
However, due to the huge challenge of mapping relationship modeling, the capturing of environment-channel non-stationarity and consistency is still in its infancy.

In summary, the complicated mapping relationship between multi-modal sensory information and communication information can be modeled via ANNs, which can capture channel non-stationarity and consistency closely combined with the environment, i.e., environment-channel non-stationarity and consistency. Currently, the mapping relationship modeling via ANNs has been preliminarily utilized in the MMICM.

\subsection{Comparison and Analysis}

Channel non-stationarity is an important channel characteristic. Channel consistency is a  typical channel physical feature, which inherently exists in each channel. In this section, based on the conventional channel modeling, RF-only intelligent channel
modeling, and MMICM, we present typical methods of capturing channel non-stationarity and consistency by modeling mathematical, spatial, coupling, and mapping relationships. For clarity, these methods of capturing channel non-stationarity and consistency are listed and compared in  Table~\ref{consistency}.

\begin{table*}[!t]
\centering
    \begin{small}
\caption{Comparisons of Channel Non-Stationarity and Consistency Capturing Methods.}
\renewcommand\arraystretch{2.5}
\begin{tabular}{|c|c|c|c|c|c|}
    \hline			
    \textbf{\makecell[c]{Relationship}} & \textbf{\makecell[c]{Modeling \\method}} &\textbf{\makecell[c]{Channel \\non-stationarity}}&\textbf{\makecell[c]{Channel \\consistency}}
    &\textbf{\makecell[c]{Channel modeling \\approach}} &\textbf{\makecell[c]{Description}}\\
    \hline
    \makecell[c]{Mathematical \\relationship}& \makecell[c]{Stochastic \\process}	& \makecell[c]{\checkmark} & \makecell[c]{×}   & \makecell[c]{Widely utilized  \\in NGSM} & \makecell[c]{Model the mathematical \\relationship for the \\variation of MPCs}\\	
    \hline
    \makecell[c]{Spatial \\relationship}& \makecell[c]{Geometry}	& \makecell[c]{\checkmark} & \makecell[c]{\checkmark}   & \makecell[c]{Widely utilized  \\in GBSM} & \makecell[c]{Model spatial relationship \\for the smooth \\evolution of MPCs}\\	
    \hline
    \makecell[c]{Coupling \\relationship}& \makecell[c]{Classifier and \\statistical \\learning}	& \makecell[c]{\checkmark} & \makecell[c]{\checkmark}   & \makecell[c]{RF-only \\intelligent channel \\modeling \& MMICM} & \makecell[c]{Model coupling relationship \\between physical environment \\and electromagnetic space \\to reveal channel characteristics}\\	
    \hline
    \multicolumn{1}{|c|}{\multirow{2}{*}{\makecell[c]{Mapping \\relationship}}} & \multicolumn{1}{c|}{\multirow{2}{*}{ANN}} & \multicolumn{1}{c|}{\checkmark} & \multicolumn{1}{c|}{\checkmark} & \multicolumn{1}{c|}{\multirow{2}{*}{MMICM}}& \multicolumn{1}{c|}{\multirow{2}{*}{\makecell[c]{Model mapping relationship \\between multi-modal \\sensing and communications \\to capture environment-channel \\non-stationarity and consistency}}} \\ 
    \cline{3-4}
    \multicolumn{1}{|c|}{}& \multicolumn{1}{c|}{}                  & \multicolumn{2}{c|}{\makecell[c]{Closely combined \\with the environment}}& \multicolumn{1}{c|}{} &\multicolumn{1}{c|}{} \\ 
    \hline
\end{tabular} 
\label{consistency}
    \end{small}
\end{table*} 

For the mathematical relationship modeling via stochastic processes, the mathematical relationship for the appearance and disappearance of MPCs in the space, time, and frequency domains can be captured by the birth-death process approach.  As a consequence, channel non-stationarity in the space, time, and frequency domains can be modeled. Currently, the birth-death process approach has been extensively exploited in the NGSM.

For the spatial relationship modeling via geometry,  the spatial relationship for the smooth MPC evolution in the space and time domains can be modeled by the visibility region approach. As a result, channel non-stationarity and consistency in the space and time domains can be modeled. 
 Currently, the visibility region approach has been
extensively utilized in the GBSM.

For the coupling relationship modeling via classifier and statistical learning, the communication information and/or sensory information can be intelligently processed to model the coupling relationship between physical environment and electromagnetic space. As a result, the channel characteristics can be accurately revealed, which can support the capturing of channel non-stationarity and consistency. Currently, the coupling relationship modeling via the classifier and statistical learning has been extensively exploited in the RF-only intelligent channel modeling, while has been preliminarily exploited in the MMICM.

For the mapping relationship modeling via ANNs, the nonlinear mapping relationship between multi-modal sensory information and communication information can be explored. As a consequence, channel non-stationarity and consistency can be modeled closely combined with the environment, i.e., environment-channel non-stationarity and consistency. Currently, the mapping relationship modeling via ANNs has been preliminarily exploited in the MMICM. 

Finally, it is noteworthy that, by delineating mathematical relationships, spatial relationships, coupling relationships, and mapping relationships of scatterers/clusters, the utilization of the environmental features increases sequentially in the three aforementioned channel modeling approaches. This leads to a more accurate capturing of channel non-stationarity and consistency.

\section{Applications Supported by Multi-Modal Intelligent Channel Modeling}

As previously mentioned, the MMICM utilizes the multi-modal information from communication devices and various sensors to conduct intelligent channel modeling. To achieve more accurate channel modeling, the propagation environment feature around the transceiver is obtained by exploring the nonlinear mapping relationship between communications and RF/non-RF sensing via SoM. 
Due to the utilization of multi-modal information and the exploration of mapping relationship, various 6G applications can be supported.
In this section, 6G potential applications supported by the MMICM are analyzed from four perspectives, including communication system transceiver design, network optimization, localization sensing, and intelligent agent cognitive intelligence. For clarity, the aforementioned application supported by the MMICM is further summarized. 

\subsection{Communication System Transceiver Design}
Through the MMICM, the application related to communication system transceiver design, such as channel estimation, channel prediction, and beam prediction, can be supported. 
\begin{figure*}[!t]
    \centering
    \subfigure[]{
    \includegraphics[width=0.32\textwidth]{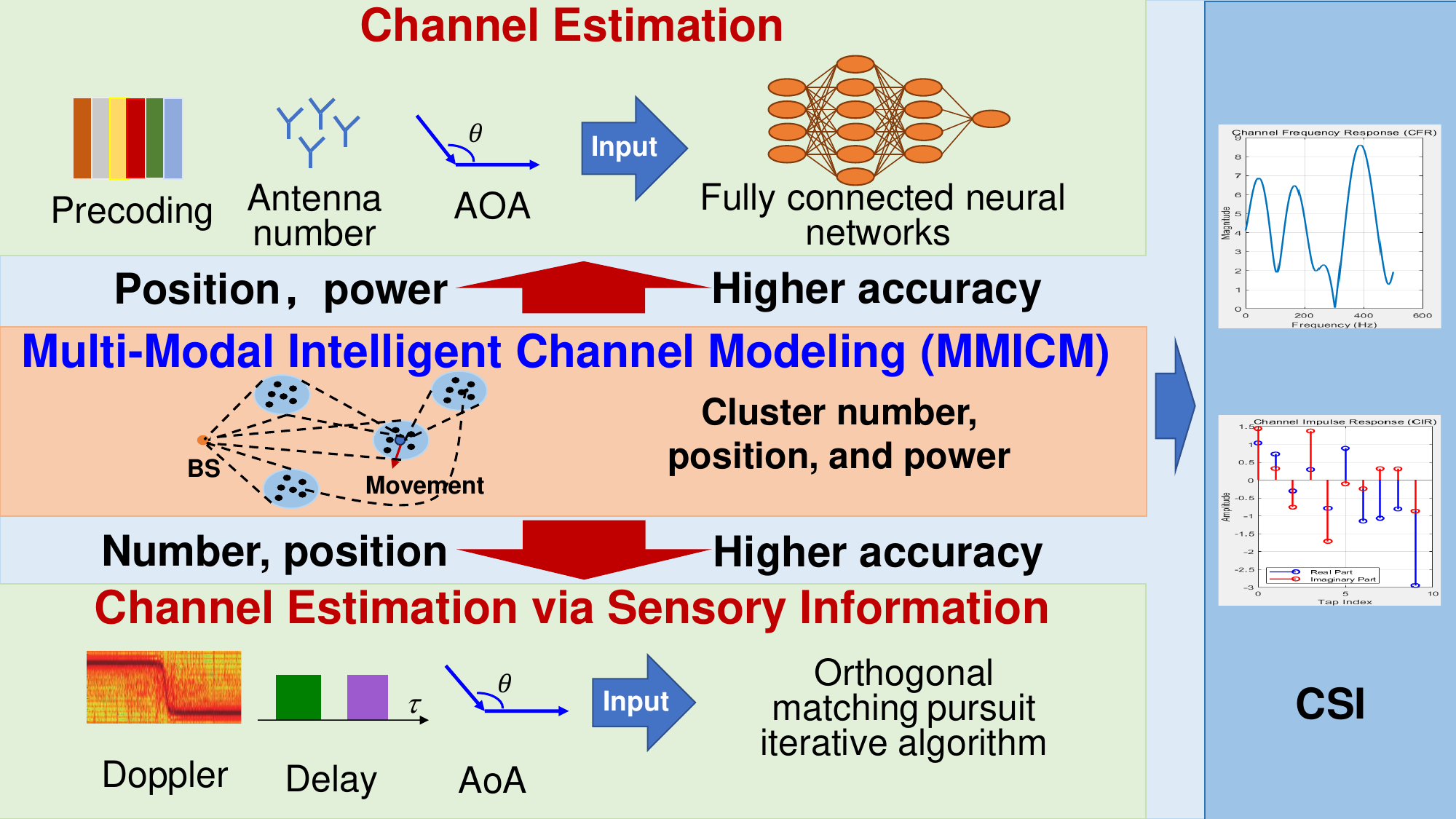}}
    \hfill
     \subfigure[]{
    \includegraphics[width=0.32\textwidth]{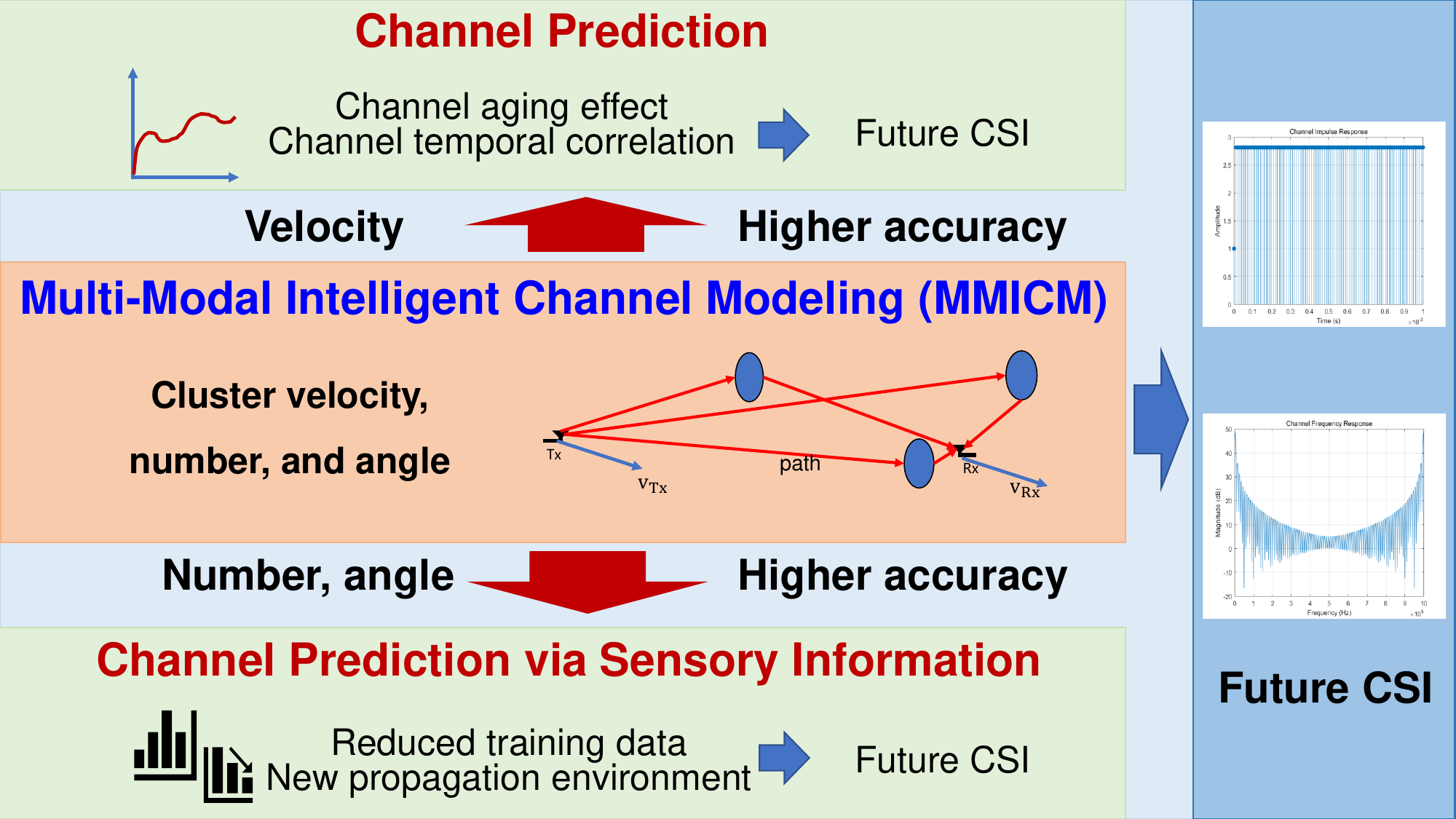}}
    \hfill
    \subfigure[]{
    \includegraphics[width=0.32\textwidth]{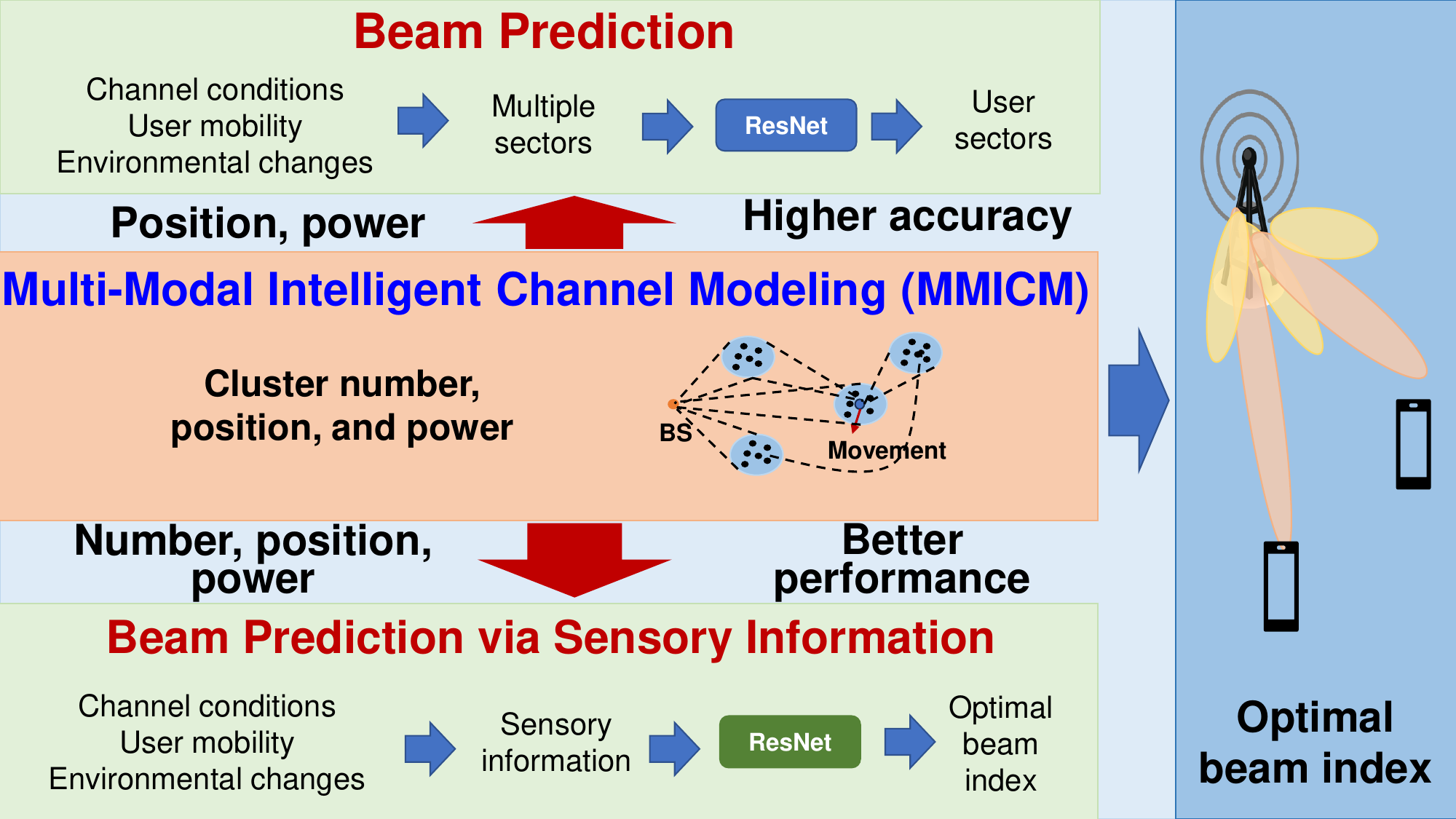}}
    \caption{Applications related to communication system transceiver design through the MMICM. (a) Channel estimation. (b) Channel prediction. (c) Beam prediction.}
    \label{Communication System Transceiver Design}
\end{figure*}

\subsubsection{Channel Estimation}
Channel estimation refers to the process of estimating CSI, including the channel frequency response (CFR), CIR, and other channel-related parameters, which is important for efficiently decoding transmitted signals in the wireless communication system \cite{CE11, sunshu}. In \cite{CE12}, the authors developed a deep learning compressed sensing channel estimation method based on fully connected neural networks under the mmWave frequency band. In the developed method \cite{CE12}, the input was the analog/digital precoding, number of base station (BS) antennas, AoA, and the output was the channel amplitude, which was further utilized to reconstruct channels. Certainly, the position and power of clusters can be obtained through the MMICM while not through the conventional channel modeling and RF-only intelligent channel modeling. Based on the developed method in  \cite{CE12}, apart from analog/digital precoding, BS antenna number, and AoA, the obtained cluster position and power can be further utilized as input to the network  to accurately predict the channel amplitude. Due to the high correlation between the cluster information and channel amplitude, the accuracy of the predicted channel amplitude can be improved. Different from the developed method in \cite{CE12} that leveraged the communication information to conduct channel estimation under the 6 GHz frequency band, the authors in \cite{CE13} exploited the sensory information, i.e., mmWave radar. The Doppler, delay and AoA of objects were obtained via mmWave radar as the initial value of the channel orthogonal matching pursuit (OMP) iterative algorithm, where the number of iterations was proportional to the number of paths detected by the mmWave radar. The main drawback of the developed method in \cite{CE13} is that the object detected by the mmWave radar is not necessarily a cluster in the channel, leading to a decrease in the accuracy of channel estimation. In this case, based on the MMICM \cite{Daitou-zw}, the number and position of clusters are acquired, which can support the more accurate computation of the initial value of OMP iterative algorithm and the number of iterations for precise channel estimation, as shown in Fig.~\ref{Communication System Transceiver Design}(a). Consequently, based on the MMICM with the multi-modal information, the acquisition of cluster information, which cannot be accurately obtained via the conventional channel modeling and RF-only intelligent channel modeling with the uni-modal information,  can 
further improve the accuracy of the developed methods in \cite{CE12,CE13}, which can essentially support the channel estimation.

 
\subsubsection{Channel Prediction}
Channel prediction refers to the process of forecasting the future channel behavior/characteristics. In general, channel prediction is based on historical channel data, statistical models, or ML algorithms to anticipate how channels change over time, which can aid in optimizing communication system performance and reliability \cite{CP11,CP12}. Due to the high cost of the orthogonal pilot based method in massive MIMO, the authors in \cite{CP13} combined channel prediction and channel estimation to obtain the future CSI at a lower cost by considering auto-correlation across CSI series under 10 MHz communication bandwidth. As a result, by observing the current and historical CSI, with the help of channel aging effect and channel temporal correlation, the future CSI is predicted \cite{CP13}. Based on the MMICM, the acquisition of velocity information related to the MPC, which cannot be obtained via the conventional channel modeling and RF-only intelligent channel modeling, can assist in exploring channel temporal correlation and further enhance the accuracy of the developed framework in \cite{CP13}, which can support channel prediction, as shown in Fig.~\ref{Communication System Transceiver Design}(b). 
Furthermore, the authors in \cite{CP14} leveraged transfer learning and meta-learning approaches to address channel prediction in a new propagation environment under the sub-6 GHz frequency band, thus reducing the demand for the volume of training data. 
 By exploiting the MMICM, the geometrical information of the new propagation environment, e.g., the number and angle of MPCs, can be obtained based on the multi-modal information. However, the geometrical information cannot  be  sufficiently acquired via  the conventional channel modeling and RF-only intelligent channel modeling by solely utilizing uni-modal information. The obtained geometrical information can further improve the accuracy of channel prediction according to the developed framework in \cite{CP14}.  As a result, based upon the developed frameworks in \cite{CP13,CP14}, the MPC and velocity information can be acquired through the MMICM, which can  facilitate the channel prediction.


\subsubsection{Beam Prediction}
Beam prediction refers to the process of forecasting the optimal beamforming parameters, which can be exploited in communication systems to maximize the signal quality at the Rx \cite{BP11}. Generally, channel prediction is based upon channel conditions, user mobility, and environmental changes, aiming to adapt the beamforming strategy proactively to enhance the communication performance \cite{BP12}. In \cite{BP13}, the authors proposed an image based beam prediction architecture under the sub-6 GHz frequency band. Specifically, beamforming vectors divided the scenario into multiple sectors, and then residual network (ResNet) \cite{resnet} was utilized to identify which sector the user belongs to, thus achieving beam prediction. Nonetheless, if there is beam blocking, the beam is aligned towards the cluster with the strongest power in the propagation environment. In this case, based on the MMICM, the cluster position and power can be obtained, which can further improve the accuracy of the beam prediction, as shown in Fig.~\ref{Communication System Transceiver Design}(c). To further conducted beam prediction under mmWave frequency band, the authors in \cite{BP14} proposed an environment semantics assisted network architecture via ResNet. In the proposed architecture, based on the street camera, the environment semantics was acquired, encoded, and transmitted to BS, thus obtaining the predicted beam index. Building upon the proposed architecture in \cite{BP14}, we can further incorporate cluster information, e.g., number, position, and power, to enrich the environment semantics via the MMICM. Since the inadequate modeling of cluster information based on the uni-modal information, the conventional channel modeling and RF-only intelligent channel modeling cannot support the beam prediction. Conversely, based on the proposed architectures in \cite{BP13,BP14}, the beam prediction performance can be further improved by fusing the precise cluster information through the MMICM with the multi-modal information.


\subsection{Network Optimization}
In addition to communication system transceiver design, the MMICM can also support applications of network optimization, including network planning and resource allocation, cell handover, and multi-hop networking.

\begin{figure*}[!t]
    \centering
    \subfigure[]{
    \includegraphics[width=0.32\textwidth]{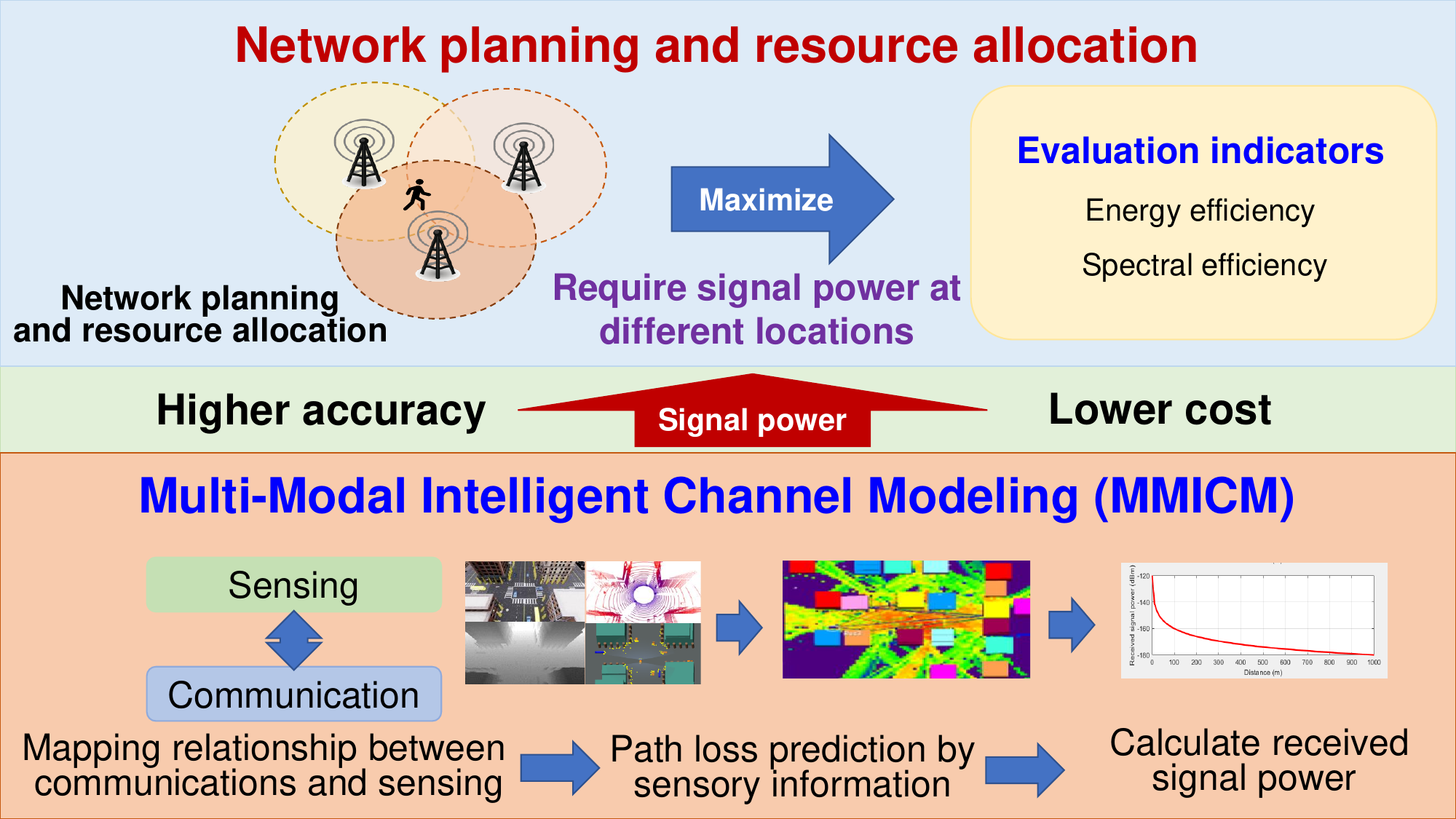}}
    \hfill
     \subfigure[]{
    \includegraphics[width=0.32\textwidth]{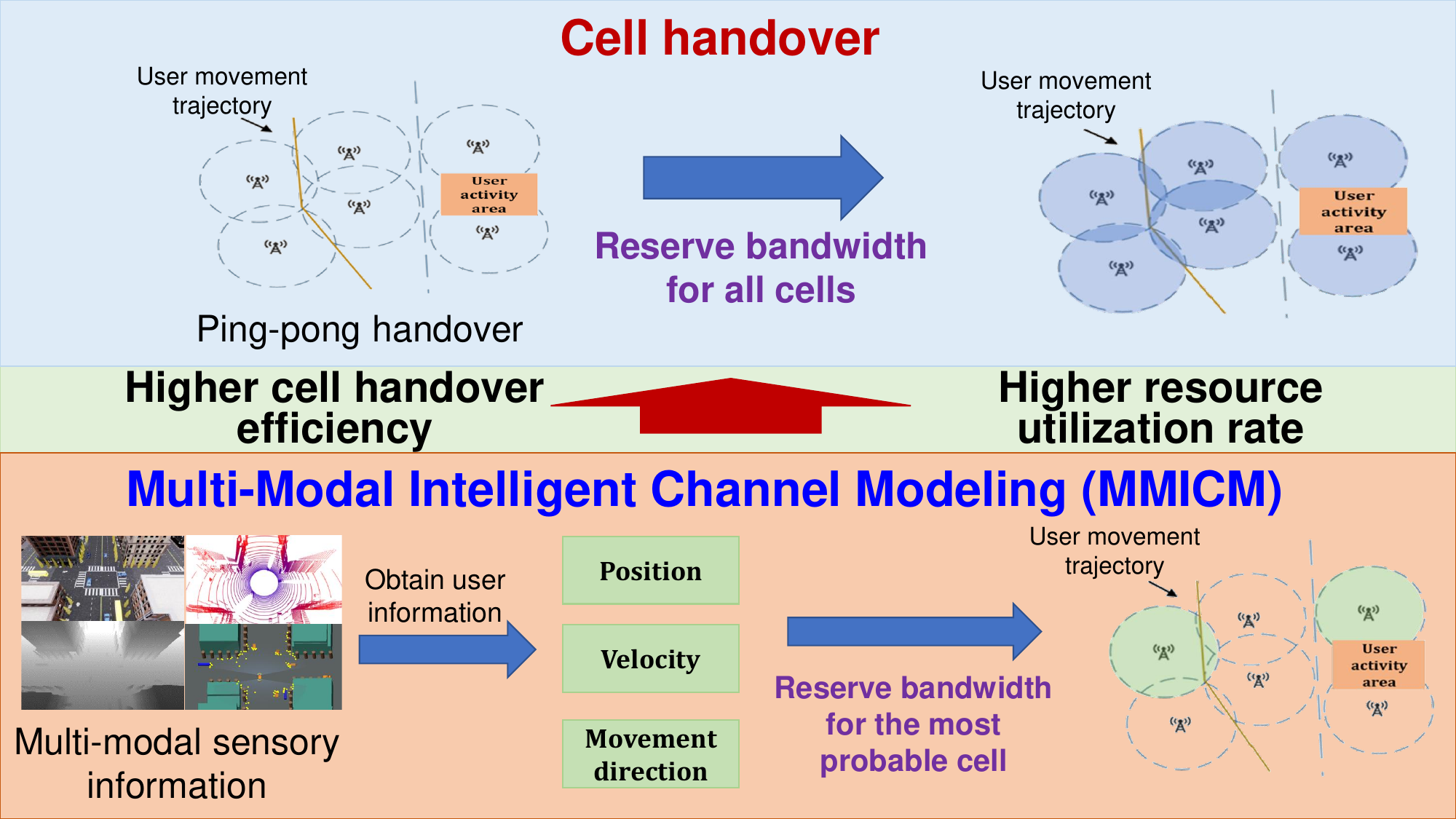}}
    \hfill
    \subfigure[]{
    \includegraphics[width=0.32\textwidth]{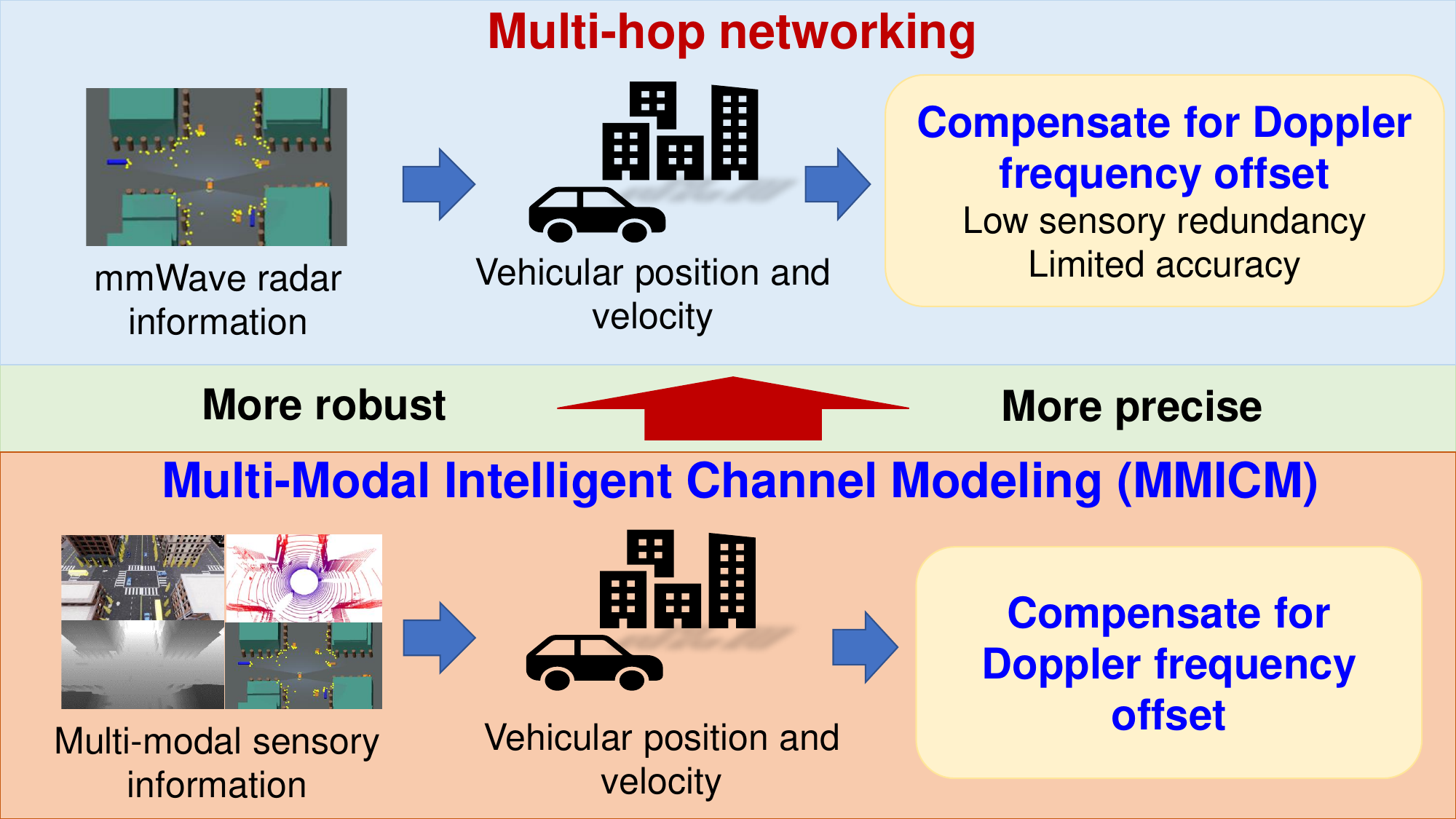}}
    \caption{Applications related to network optimization through the MMICM. (a) Network planning and resource allocation. (b) Cell handover. (c) Multi-hop networking.}
    \label{Network Optimization}
\end{figure*}

\subsubsection{Network Planning and Resource Allocation} 
In the network planning, the received signal power at different locations within the cell coverage area is usually obtained by the measurement and estimation \cite{NP11,NP12}. This can  be further exploited to plan the deployment of macro/micro BS nodes and optimize network configurations, including antenna downtilt, orientation, height, etc. As a result, cell capacity, coverage effectiveness, and energy savings can be improved. The authors in  \cite{NP13} proposed a factor, i.e., effective energy efficiency, in cognitive satellite terrestrial networks with 1 GHz system bandwidth, and further developed  a power allocation scheme to maximize effective energy efficiency while meeting  the interference constraint. The authors in \cite{NP14} extended the work in \cite{NP13} and proposed a metric in combination of energy efficiency (EE) and spectral efficiency (SE). According to the proposed metric, a power allocation scheme was developed in consideration of the EE–SE trade-off while meeting interference power constraints in the satellite communication. However, the received signal power, which is significantly related to path loss, at different locations needs to be obtained in the aforementioned schemes in \cite{NP13,NP14}. To further facilitate the network planning and resource allocation, by exploring the mapping relationship between communications and sensing, path loss is obtained through sensory information to calculate the received signal power in a low-complexity manner, as shown in Fig.~\ref{Network Optimization}(a). On one hand, the exploration of mapping relationship with the help of multi-modal information via the MMICM is more efficient than the measurement/estimation based approach. On the other hand, conventional channel modeling and RF-only intelligent channel modeling with the uni-modal information cannot accurately model path loss information, thus insufficiently supporting network planning and resource allocation.

 
\subsubsection{Cell Handover} 
Cell handover is a process in mobile networks and  allows for the seamless transfer of a device's ongoing call or data session from one cell to another \cite{CH11,CH12}. However, there is a typical issue in cell handover, named ``ping-pong" handover, i.e., the situation where a mobile device repeatedly switches back and forth between two base stations as their signal strengths is significantly similar \cite{CH13,CH14}. The continuous switching resembling the back-and-forth motion of a ping-pong ball, i.e., ``ping-pong" handover, has adverse effects on both the network and the user equipment. As stated in \cite{CH15}, seamless mobility demands proper resource reservation together with context transfer processes during handover, which is not sensitive to the random nature of the user mobility pattern. In such a condition, aiming at guaranteeing the service continuity for the dynamic user, the traditional proactive resource reservation scheme reserves bandwidth for all cells that mobile hosts may access during their active connection period. By characterizing the position, velocity, and motion direction information of users in the MMICM to predict motion states, the cell handover efficiency and resource utilization can be improved, as shown in Fig~\ref{Network Optimization}(b). On the one hand, we can anticipate the positions of users along with CSI in advance to develop proactive handover mechanisms, thus avoiding ``ping-pong'' handover. On the other hand, by obtaining the mobility pattern of users via sensory data, it is possible to prepare the bandwidth resource in the most probable cell for  the mobile user, thus saving resources reserved for other cells. By leveraging the MMICM, the position, velocity, and motion direction information can be characterized and predicted based on the multi-modal information, which can achieve more efficient cell handover. However, the aforementioned information cannot be accurately modeled by the conventional channel modeling and RF-only intelligent channel modeling with the uni-modal information, and thus cannot facilitate cell handover.

 
\subsubsection{Multi-Hop Networking} 
A multi-hop network refers to a network where nodes communicate through multiple hops to extend coverage beyond the range of a single-hop wireless radio with decent transmission power \cite{NM11}. The multi-hop network has been widely utilized owing to its low cost and high flexibility in deployment \cite{NM12}. Currently, the authors in \cite{NM13} developed a method to maximize the sum capacity of users by optimizing positions of flying BSs under UAV-aided multi-hop networks. To further consider the advantage of sensory information, the authors in \cite{NM14} leveraged the mmWave radar information to the multi-hop V2V network, which aims at augmenting end-to-end information deliver. Specifically, mmWave radar information was utilized to obtain the vehicular position and velocity to establish multi-hop V2V links and compensate for Doppler frequency offset to combat channel time selectivity \cite{NM14}. Nevertheless, utilizing single-modal sensory information to acquire the vehicular position and velocity has the drawback of low sensory redundancy. Based on the MMICM, with the help of multi-modal sensory data, the  position and velocity of dynamic objects can be accurately acquired. In this case, Doppler frequency offset can be accurately compensated, thus establishing more reliable  multi-hop links based on the multi-modal information, as shown in Fig.~\ref{Network Optimization}(c). Different from the MMICM, since the conventional channel modeling and RF-only intelligent channel modeling solely utilize the uni-modal communication information and ignore the sensory data, the position and velocity of dynamic objects cannot be obtained, and thus cannot support multi-hop networking.


\subsection{Localization Sensing}
Apart from the aforementioned application associated with communication tasks, the application of localization sensing, e.g., 3D reconstruction and mobile terminal positioning, can also be supported by the MMICM.


\begin{figure*}[!t]
    \centering
    \subfigure[]{
    \includegraphics[width=0.45\textwidth]{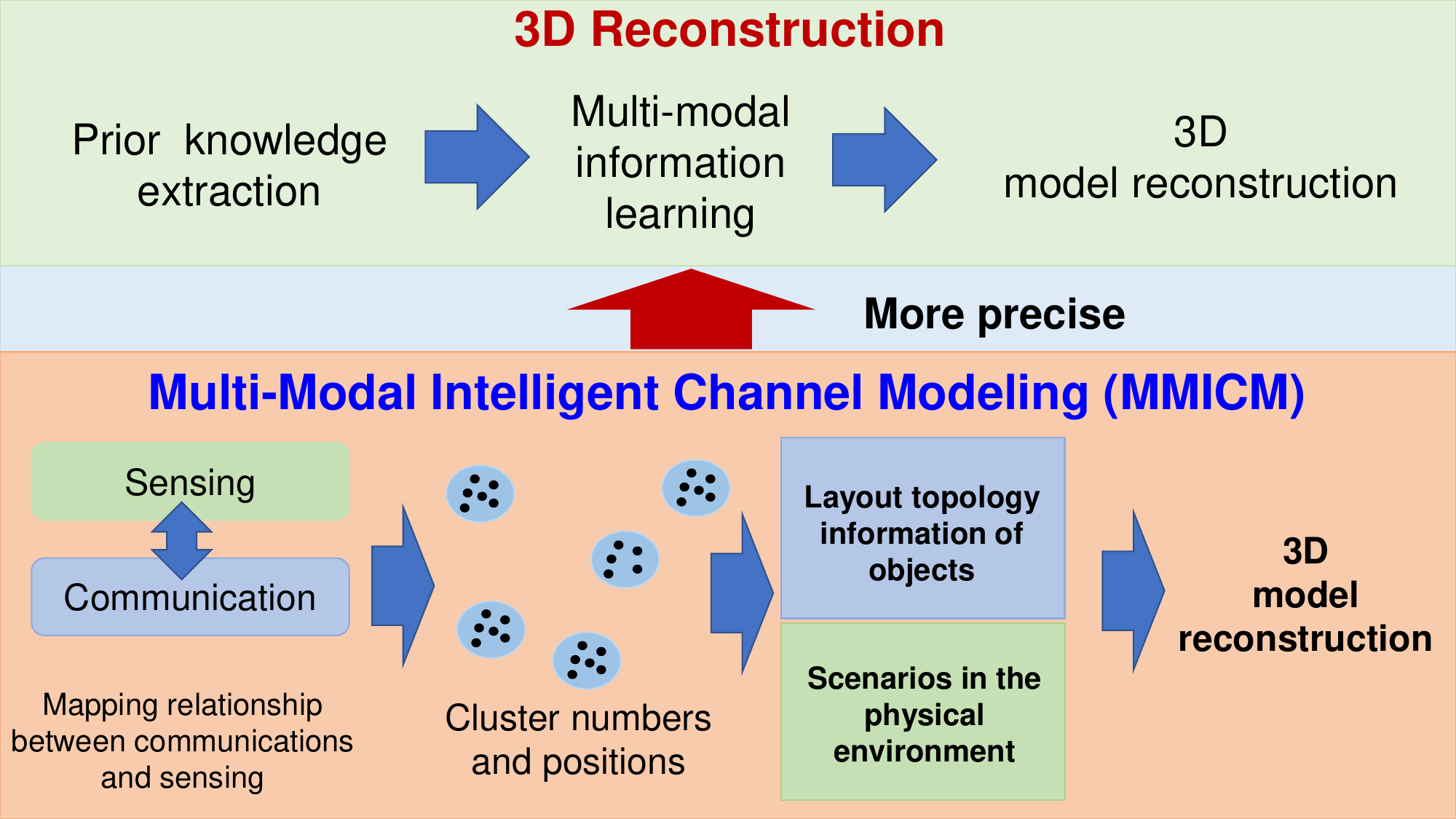}}
    \hfill
    \subfigure[]{
    \includegraphics[width=0.45\textwidth]{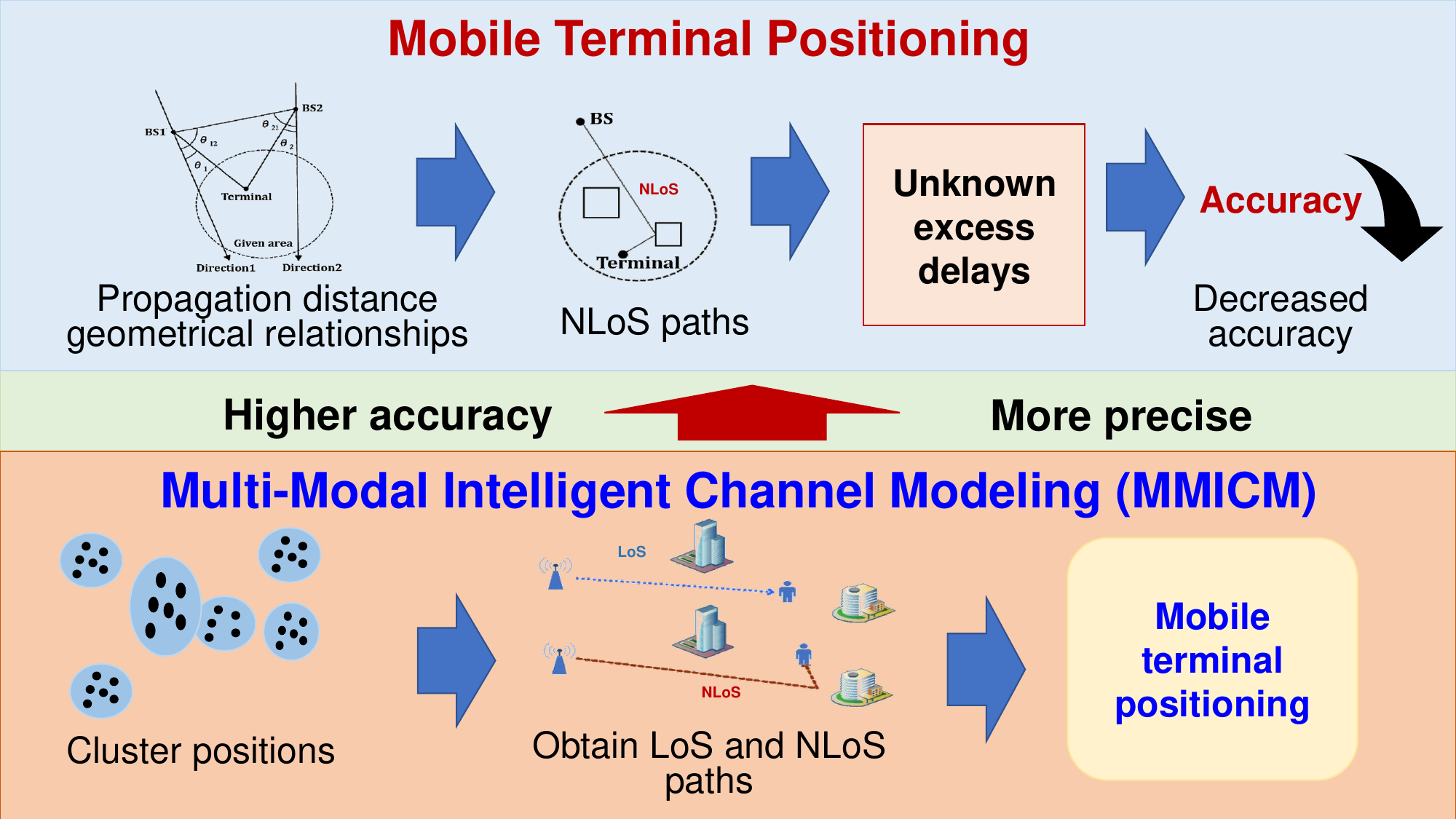}}
    \caption{Applications related to localization sensing through the MMICM. (a) 3D reconstruction. (b) Mobile terminal positioning.}
    \label{Localization Sensing}
\end{figure*}

\subsubsection{3D Reconstruction} 
3D reconstruction is the process of characterizing the shape and appearance of the real-world object or scenario to create a digital representation in 3D. The 3D reconstruction technique involves using sensory information from different viewpoints to reconstruct the geometry and texture of the subject \cite{3R11,3R12}. Aiming at reconstructing the high-fidelity textured 3D model under the single image, the authors in \cite{3R13} proposed a adversarial learning pipeline, which increases the realism of the textured 3D prediction. To further utilize the multi-modal information for the 3D reconstruction, the authors in \cite{3R14} proposed a method that contained three components, i.e., 3D shape retrieval module,  multi-modal information fusion module, and 3D reconstruction
module. Although the aforementioned work in \cite{3R14} considered multi-modal information, communication data was ignored in the 3D reconstruction. Note that clusters has the ability to mimic interactions between radio waves and objects, thus intuitively characterizing the transmission process in the propagation environment \cite{Cluster11}. By analyzing the number and position of clusters via the MMICM, we can obtain the layout topology information of objects and scenarios in the physical environment \cite{huawei11}, and further exploit it as auxiliary information can enhance the accuracy of 3D reconstruction, as shown in Fig.~\ref{Localization Sensing}(a). Therefore, with the help of the MMICM, the layout topology information of objects and scenarios can be obtained based on the mutli-modal information to facilitate more accurate 3D reconstruction. This cannot be achieved by the conventional
channel modeling and RF-only intelligent channel modeling, whose ability to extract environmental features is limited based on the uni-modal information.


\subsubsection{Mobile Terminal Positioning} 
Mobile terminal positioning refers to the process of determining the geographical location of a mobile device within a given area \cite{MTP11}. Mobile terminal positioning is crucial for various location-based services, navigation applications, as well as emergency services. Conventional mobile terminal positioning relies solely on RF channel information. To be specific, by utilizing the propagation distance, which can be further exploited to calculate delay, between the mobile terminal and multiple BSs, the position of the mobile terminal can be determined through geometrical relationships \cite{MTP12}. The AoA and AoD of LoS paths can improve the accuracy of mobile terminal positioning. However, NLoS paths can introduce unknown excess delays, which severely effect the accuracy of mobile terminal positioning \cite{MTP13,MTP14}. In this case, by leveraging various channel statistical properties, AI algorithms are employed to identify and eliminate NLoS conditions to enhance positioning accuracy, resulting in high complexity. To overcome this limitation and further improve the accuracy of mobile terminal positioning, by utilizing the MMICM, it is possible to obtain LoS and NLoS paths in the propagation environment based on the position of clusters, as shown in Fig.~\ref{Localization Sensing}(b). As a result, based on the multi-modal information, the acquirement of LoS and NLoS paths via the MMICM results in more precise mobile terminal positioning. Nonetheless, due to the limited understanding of the propagation environment by the uni-modal information, the conventional channel modeling and RF-only intelligent channel modeling cannot adequately support the mobile terminal positioning.


\subsection{Intelligent Agent Cognitive Intelligence}
In the future 6G era, AIoT is a typical application and deserves the extensive research. As an important component in AIoT, the intelligent agent cognitive intelligence has received widespread attention. Fortunately,  the MMICM can essentially support applications related to intelligent agent cognitive intelligence, e.g., automated guided vehicle (AGV) path planning and UAV path planning. 

\begin{figure*}[!t]
    \centering
    \subfigure[]{
    \includegraphics[width=0.45\textwidth]{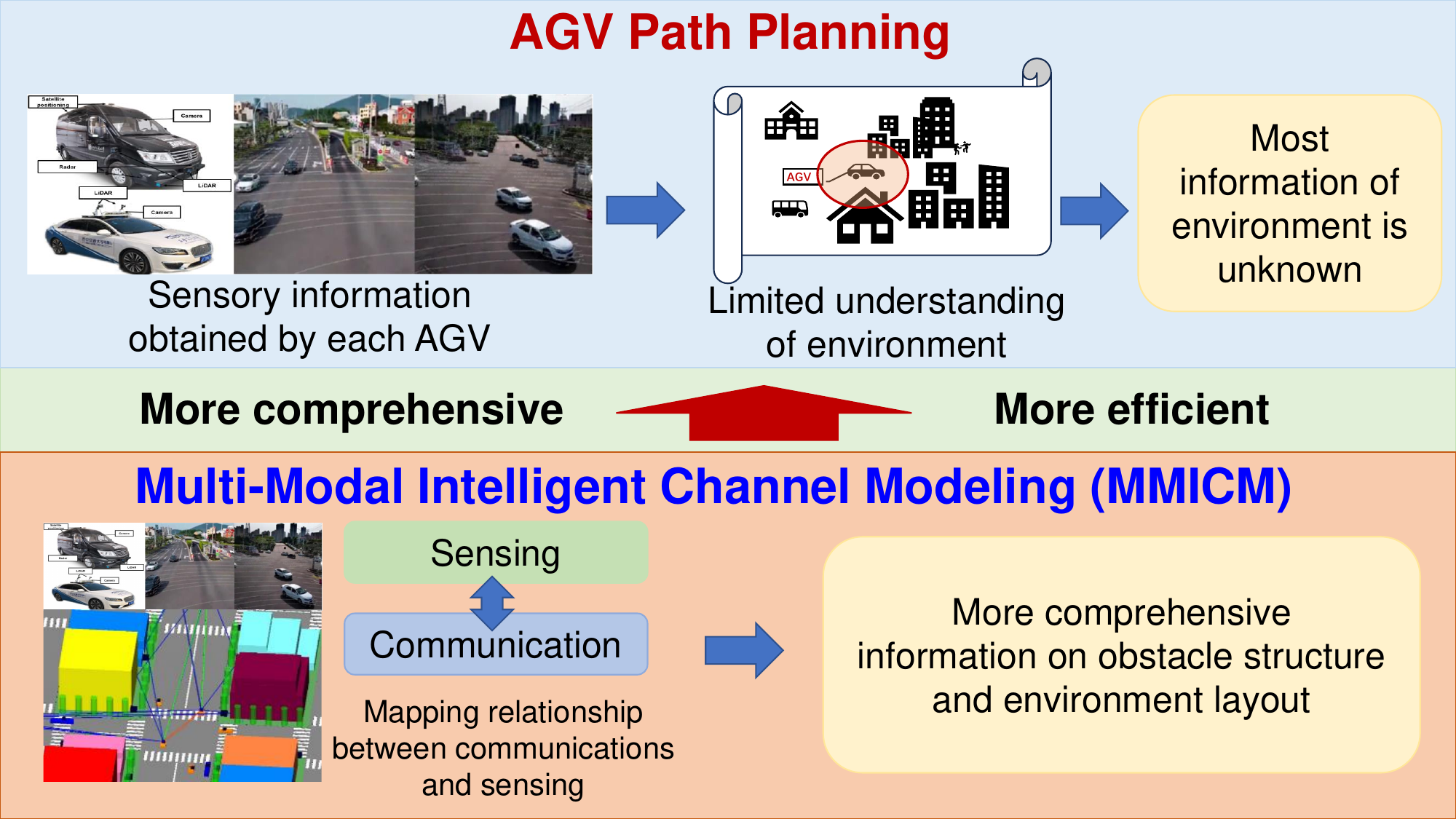}}
    \hfill
    \subfigure[]{
    \includegraphics[width=0.45\textwidth]{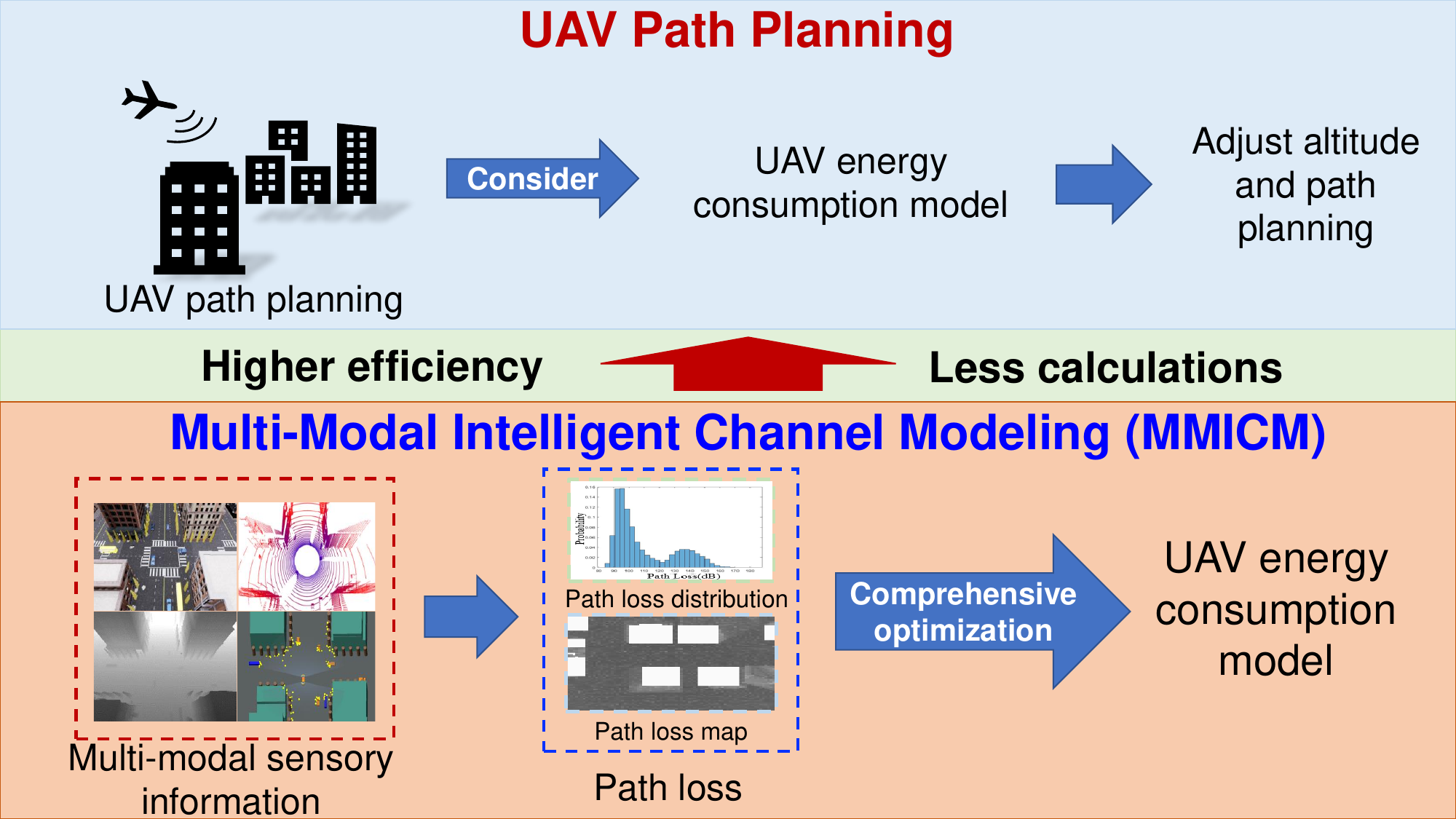}}
    \caption{Applications related to intelligent agent cognitive intelligence through the MMICM. (a) AGV path planning. (b) UAV path planning.}
    \label{Intelligent Agent Cognitive Intelligence}
\end{figure*}


\begin{table*}[!t]
\centering
    \begin{footnotesize}
\caption{Applications Enhanced through the MMICM.}
\renewcommand\arraystretch{2}
\begin{tabular}{|c|c|c|c|c|}
    \hline			
    \multicolumn{2}{|c|}{\textbf{Applications}}&\textbf{\makecell[c]{Conventional channel modeling and \\RF-only intelligent channel modeling \\ with uni-modal information }} &\textbf{\makecell[c]{Assistance of \\multi-modal intelligent \\channel modeling\\ with multi-modal information}} 
    &\textbf{\makecell[c]{Advantage of \\multi-modal intelligent \\channel modeling}}  \\
    \hline
\multirow{5}{*}{	\makecell[c]{Communication \\system  \\transceiver design}}
& Channel estimation & \makecell[c]{Inadequate modeling of \\ cluster number and position}  	& \makecell[c]{Accurate acquisition of \\cluster number and \\position information} 	&  \makecell[c]{More accurate setting \\of initial values \\of iterative algorithms}   \\					\cline{2-5} 
 & Channel prediction  &	\makecell[c]{Inadequate modeling of \\MPC number and angle}  & \makecell[c]{Accurate acquisition of \\velocity information \\related to MPCs}  & \makecell[c]{More accurate exploration \\of channel \\temporal correlation}   \\					\cline{2-5} 
& Beam prediction     & \makecell[c]{Inadequate modeling of \\cluster position and power}& \makecell[c]{Accurate acquisition of \\cluster position and \\power information} & \makecell[c]{More accurate beam \\alignment to the \\strongest  power component \\in beam blocking}\\			
    \hline
    \multirow{4}{*}{	\makecell[c]{Network   \\optimization}}
& \makecell[c]{Network planning \\and \\resource allocation} 	& \makecell[c]{High-complexity acquisition \\of path loss via the RF \\communication information} &\makecell[c]{Efficient acquisition of \\path loss via \\the sensory information}	& \makecell[c]{Lower complexity \\calculation  of \\received signal power}   \\					\cline{2-5} 
 & Cell handover  &	\makecell[c]{High-complexity acquisition \\of user information} &\makecell[c]{Efficient acquisition of \\positions and  movement \\information of users}  & \makecell[c]{More efficient \\cell handover and \\resource utilization}  \\					\cline{2-5} 
& \makecell[c]{Multi-hop \\networking }    & \makecell[c]{High-complexity  acquisition \\of dynamic object information}&\makecell[c]{Efficient acquisition of \\positions and velocity \\of dynamic objects}  & \makecell[c]{More precise \\compensation of \\Doppler frequency offset}\\	
    \hline 
      \multirow{3}{*}{	\makecell[c]{Localization    \\sensing}}
& 3D reconstruction &\makecell[c]{Inadequate modeling of  \\cluster number and position}& \makecell[c]{Accurate acquisition of     \\ number and position \\of clusters}	& \makecell[c]{More precise introduction \\of layout topology  \\information of \\objects and scenarios} \\					\cline{2-5} 
 & \makecell[c]{Mobile terminal \\positioning}  &\makecell[c]{High-complexity distinction \\between LoS and NLoS paths}	 & \makecell[c]{Precise acquisition of   \\ LoS and NLoS paths} & \makecell[c]{Lower complexity \\distinction between \\LoS and NLoS paths} \\					\cline{2-5} 
     \hline 
           \multirow{3}{*}{	\makecell[c]{Intelligent agent    \\cognitive \\intelligence}}
& AGV path planning &\makecell[c]{Imprecise characterization of \\environment information}& \makecell[c]{Precise characterization \\of interactions between \\radio wave and objects}	& \makecell[c]{More accurate \\introduction of obstacle \\structure and environment \\layout information } \\					\cline{2-5} 
 &  UAV path planning  &\makecell[c]{Non-real-time \\path loss acquisition}&	\makecell[c]{Efficient acquisition of  \\ path loss in real time }  & \makecell[c]{Lower complexity \\calculation of \\received signal power} \\					\cline{2-5} 
     \hline 
\end{tabular} 
\label{applications}
    \end{footnotesize}
\end{table*} 

\subsubsection{AGV Path Planning}
AGV path planning refers to the process of determining the optimal path for the AGV to navigate from its current location to a designated target while avoiding obstacles and adhering to specific constraints \cite{APP11,APP12}. Efficient path planning for AGVs is essential for optimizing operations in industrial settings, warehouses, and other environments where these AGVs are deployed for material handling and transportation tasks. According to whether all the information of the
environment is accessible or not, the AGV path planning can be classified into global path planning and local path planning \cite{APP13}.  For the global path planning, all the information of the environment is known to AGVs before starting \cite{APP14}. Different from global path planning, for the local path planning, almost all the information of the environment is unknown to AGVs before starting \cite{APP15}. However, relying solely on the sensory information obtained by each AGV, the understanding of obstacle structures and environmental layouts is limited. By further introducing the communication information and obtaining interactions between radio waves and objects via the MMICM with the multi-modal information, more comprehensive information on obstacle structure and environment layout can be acquired to facilitate more efficient AGV path planning, as shown in Fig.~\ref{Intelligent Agent Cognitive Intelligence}(a). However, the conventional channel modeling and RF-only intelligent channel modeling solely exploit uni-modal communication information and ignore the utilization of sensory data, and thus cannot support the AGV path planning.


\subsubsection{UAV Path Planning}
The communication quality of UAVs has received the extensive attention, which is significantly affected by other users together with complicated topography \cite{UPP11}--\cite{TAO33}. In this case, the proper UAV path planning is of paramount importance. UAV path planning aims to select the optimal route based on the task being performed by analyzing the position of UAVs relative to the entire environment and employing path optimization algorithms. The proper UAV path planning can improve energy efficiency, reduce UAV energy consumption, and decrease operation time. In \cite{UPP12,UPP13}, by considering the channel model and UAV energy consumption model, joint optimization of communication performance and UAV energy efficiency metrics can be achieved through adjusting altitude and path planning of UAVs. However, the iterative convergence of UAV path planning optimization algorithms requires extensive calculations and is of high complexity. Based on the MMICM, the multi-modal sensory information collected from the physical environment can obtain the path loss in real time, and thus comprehensive optimization of communication system performance and UAV energy consumption can be achieved, as shown in Fig.~\ref{Intelligent Agent Cognitive Intelligence}(b). This can assist more efficient UAV path planning and significantly reduce the computational overhead caused by the optimization algorithm based on the MMICM with multi-modal information. Nevertheless,  the conventional channel modeling and RF-only intelligent channel modeling cannot sufficiently support the UAV path planning attributed to the inability to provide real-time path loss by the uni-modal information.

\subsection{Summary for Supported Applications}
The applications, which can be supported by the MMICM, are summarized in Table~\ref{applications}. It can be observed from Table~\ref{applications} that the application and performance related to communication network and networked intelligence, including communication system transceiver design, network optimization, localization sensing, and intelligent agent cognitive intelligence, can be essentially supported and enhanced via the MMICM. 
For the communication system transceiver design, the accuracy of channel estimation, channel prediction, and beam prediction can be improved by obtaining the number/position of clusters, the velocity information related to MPCs, and the position/power of clusters, respectively. For the  network optimization, the efficiency of the network planning and resource allocation, cell handover, and multi-hop networking can be enhanced by obtaining the path loss via the sensory information, the position/movement information of users, and the position/velocity of dynamic objects, respectively. For the localization sensing, more precise 3D reconstruction and mobile terminal positioning can be achieved by obtaining the number/position of clusters and the LoS/NLoS path, respectively. For the intelligent agent cognitive intelligence, more efficient AGV path planning and UAV path planning can be conducted by characterizing the interaction
between radio waves and objects and obtaining path loss in real time via the sensory information, respectively. Therefore, various applications can be supported by the MMICM, which deserves more investigations. 

In summary, compared to the conventional channel modeling and RF-only intelligent channel modeling with the uni-modal information, the MMICM  can bring significant benefits to the aforementioned application by utilizing the multi-modal information and further exploring the mapping relationship between communications and sensing. On one hand, it significantly advances the development of channel modeling theory, which can increase the involvement of channel modeling in system design and applications. On the other hand, the MMICM enhances the accuracy and efficiency of the aforementioned application, which can facilitate the development of 6G communication networks and networked intelligence.

\section{Simulation Results and Analysis}
In this section, different channel modeling approaches are compared by simulation.  Through comparison, the necessity of conducting MMICM via SoM is shown.  Furthermore, some interesting observations and valuable analyses are  presented, which are summarized. 

\begin{table*}[!t]
\renewcommand\arraystretch{1.25}
\caption{Key Parameters Utilized in the Simulation.\label{tab:table1}}
\centering
\begin{tabular}{|c|c|c|c|c|c|c|c|}
\hline
\textbf{Setting}&\textbf{Fig.~\ref{TACF-zr}}&\textbf{Fig.~\ref{DPSD-zr}}&\textbf{Fig.~\ref{PDP}}&\textbf{Fig.~\ref{pl-distribution-my}} & \textbf{Fig.~\ref{pl-distribution-mean-value-my}}&\textbf{Fig.~\ref{pl-distribution-mr}}&\textbf{Fig.~\ref{pl-map-mr}} \\ 
\hline
$f_\mathrm{c}$ (GHz) & 28 & 28 & 28 & 28 & 28 & 28 & 28 \\
\hline
$BW$ (GHz) & 2 & 2 & 2  & 2 & 2 & 2 & 2\\
\hline
Batch size & $-$ & $-$ & 16  & 5 & 5 & 8 & 8 \\
\hline
Optimizer & $-$ & $-$ & Adam & Adam  & Adam & Adam & Adam \\
\hline
Learning rate & $-$ & $-$ & 1${\times}$10${^{-3}}$ & 1.2${\times}$10${^{-3}}$ & 1.2${\times}$10${^{-3}}$ & 1${\times}$10${^{-4}}$ & 2${\times}$10${^{-3}}$  \\
\hline
Epoch & $-$  & $-$ & 200 & 50 & 50& 120 & 100 \\
\hline
Loss function & $-$ & $-$ & MSE loss & MSE loss  & MSE loss &  MSE loss & GAN loss \\
\hline
\multicolumn{8}{|l|}{\small	{$-$: There is no information in the simulation} } \\
\hline
\end{tabular}
\end{table*}


 \subsection{Channel Statistical Properties}
\begin{figure}[!t]
    \centering	\includegraphics[width=0.49\textwidth]{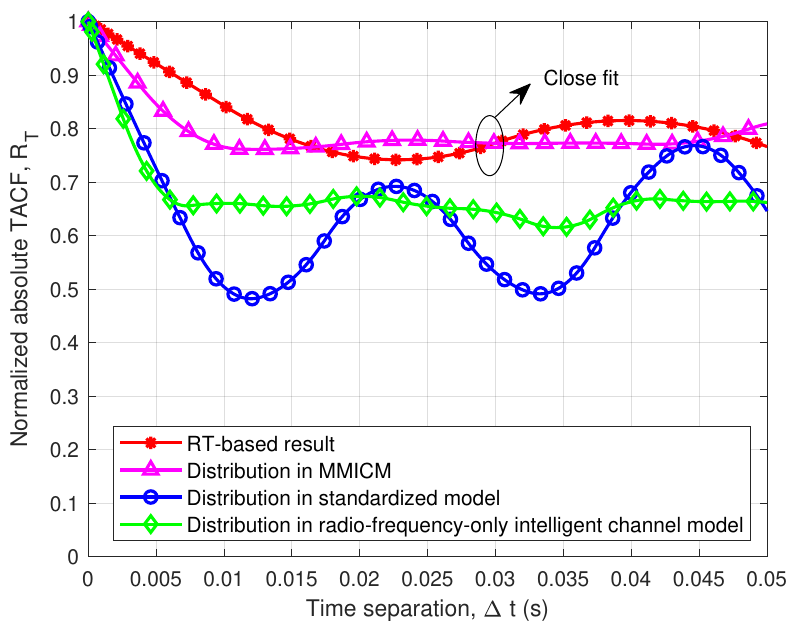}
\caption{Comparisons of TACFs with different channel modeling approaches.}
\label{TACF-zr}
\end{figure}

The TACF measures the correlation between a signal and a delayed version of itself over time, which can provide insights into the signal periodicity and temporal dependencies \cite{UAV-B-1,TACF2-zr}. 
Fig.~\ref{TACF-zr} shows that the RT-based TACF is compared with the TACFs based on MMICM \cite{daitou-zr}, the standardized channel model \cite{standardized-zr}, and the RF-only intelligent channel model \cite{huangchenscatterer-zr}. 
It can be seen that the TACF based on MMICM fits well with the RT-based TACF. 
In contrast, the TACFs based on the standardized channel model \cite{standardized-zr} and RF-only intelligent channel model \cite{huangchenscatterer-zr} cannot align well with the RT-based TACF. This is because that the TACFs based on the conventional channel modeling and RF-only intelligent channel modeling are obtained via uni-modal communication information.
Due to the lack of sensory data, the standardized channel model \cite{standardized-zr} and the RF-only intelligent channel model \cite{huangchenscatterer-zr} fail to achieve an in-depth  understanding of the environment. 
This phenomenon can be explained that the lack of sensory data results in the inaccurate modeling of cluster variation over time, and thus temporal correlation cannot be precisely characterized.
As a result, inadequate understanding and modeling of physical and electromagnetic space lead to lower temporal correlation.
Therefore, Fig.~\ref{TACF-zr} demonstrates the necessity of utilizing the multi-modal information in the channel modeling and the accuracy of the MMICM via SoM through the comparison of TACFs.

\begin{figure}[!t]
    \centering	\includegraphics[width=0.49\textwidth]{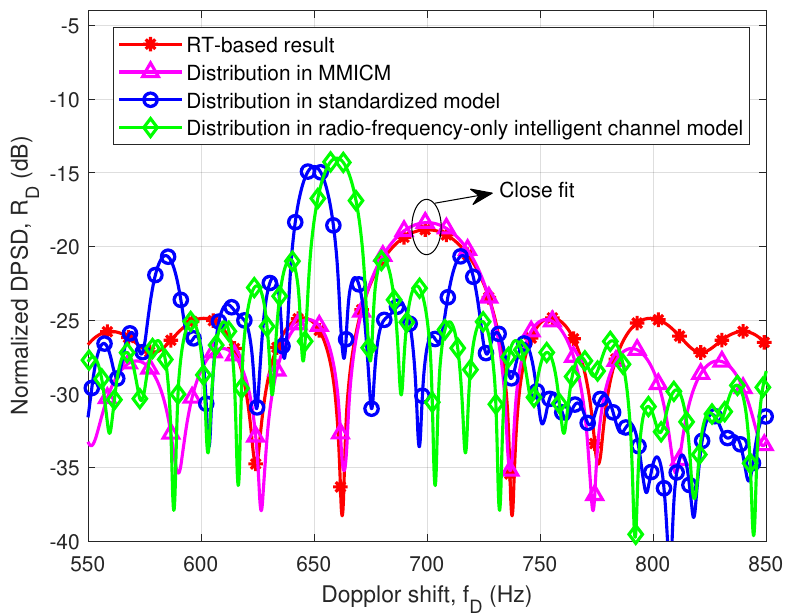}
\caption{Comparison of DPSDs in different channel modeling approaches.}
\label{DPSD-zr}
\end{figure}

DPSD quantifies the distribution of power in a signal as a function of Doppler frequency shifts and is caused by the relative movement between the transceiver \cite{general-zw,DPSD2-zr}. 
Fig.~\ref{DPSD-zr} demonstrates that the RT-based DPSD is compared with the DPSD of MMICM \cite{daitou-zr}, the standardized channel model \cite{standardized-zr}, and the RF-only intelligent channel model \cite{huangchenscatterer-zr}. 
It can be seen that the DPSD based on MMICM fits well with the RT-based DPSD. 
In contrast, the DPSD based on the standardized channel model \cite{standardized-zr} and RF-only intelligent channel model \cite{huangchenscatterer-zr} cannot align well with the RT-based DPSD. 
The standardized channel model \cite{standardized-zr} and RF-only intelligent channel model \cite{huangchenscatterer-zr} ignored the utilization of sensory data and cannot achieve an in-depth understanding of the environment. In this case, the lack of sensory data leads to the inaccurate modeling of cluster distribution in the propagation environment, where distributions of DPSDs of the standardized channel model \cite{standardized-zr} and RF-only intelligent channel model \cite{huangchenscatterer-zr} are steeper than those of RT-based DPSDs. Therefore, Fig.~\ref{DPSD-zr} shows the necessity of utilizing the multi-modal information in the channel modeling and the accuracy of the MMICM via SoM through the comparison of DPSDs.

\begin{figure*}[!t]
	\centering
      \subfigure[]{\includegraphics[width=0.49\textwidth]{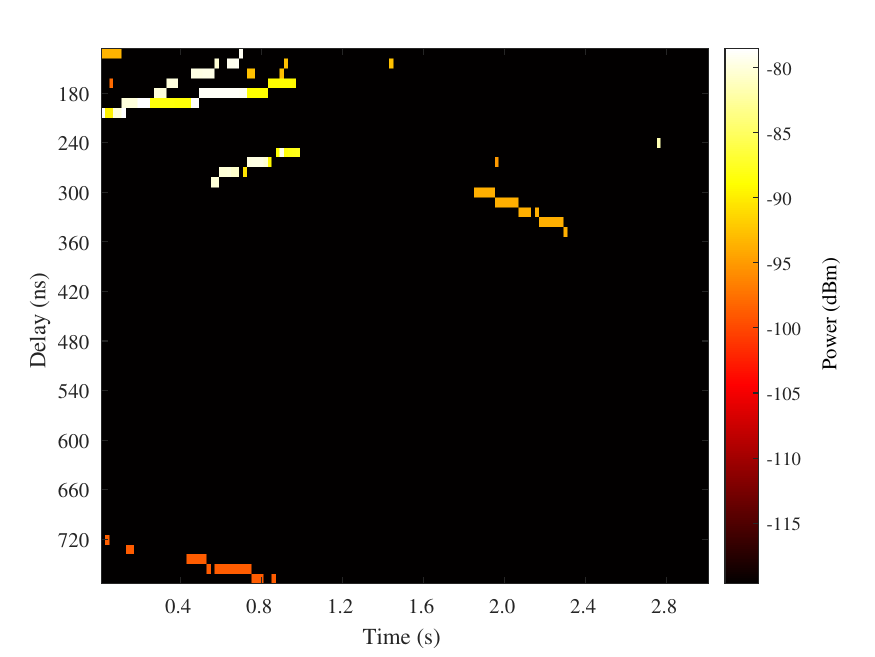}}
 \subfigure[]{\includegraphics[width=0.49\textwidth]{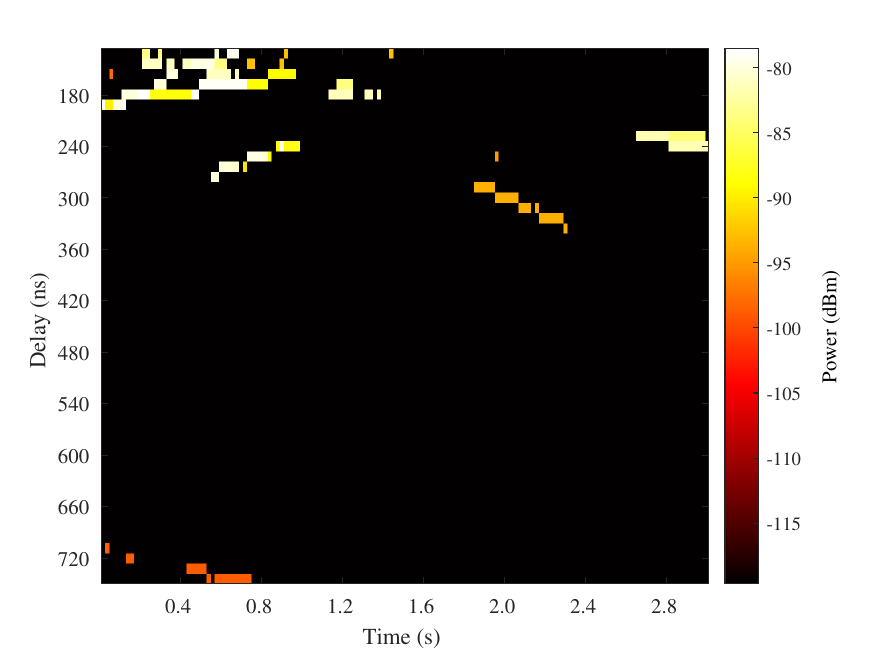}}
	\subfigure[]{\includegraphics[width=0.49\textwidth]{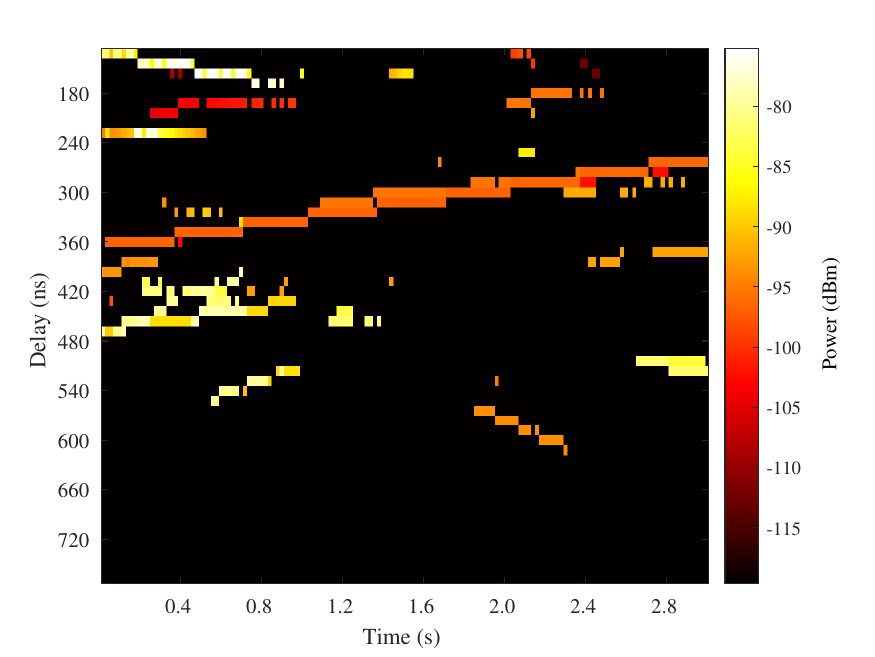}} 
   \subfigure[]{\includegraphics[width=0.49\textwidth]{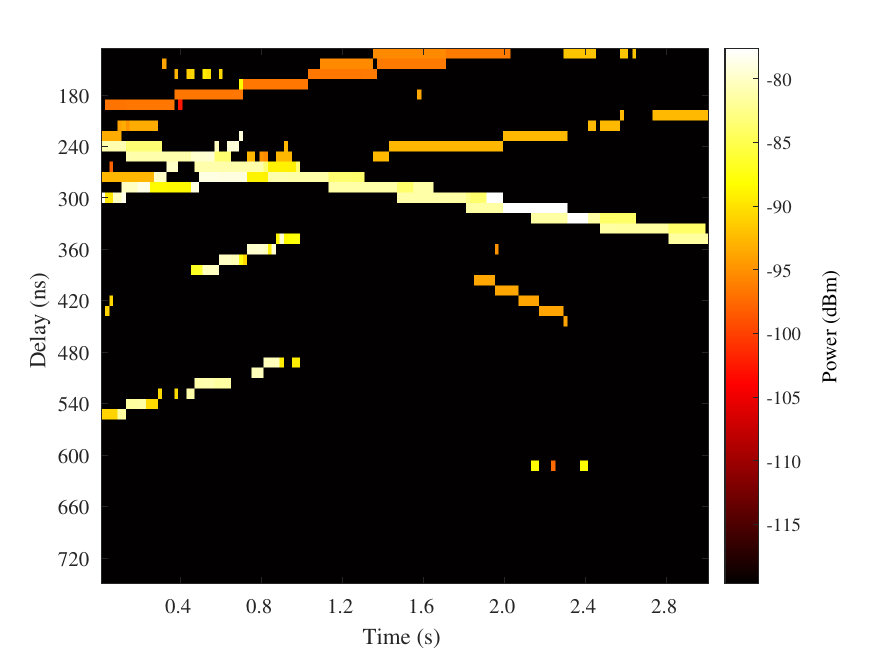}}
	\caption{Comparisons of PDPs. (a) RT-based result.(b) MMICM with mapping relationship exploration \cite{Daitou-zw}. (c) Standardized channel model with uni-modal communication information in \cite{CCC-zw}. (d) MMICM without mapping relationship exploration in \cite{LA-GBSM-zw}. }
 \label{PDP}
\end{figure*}

PDP can represent the power of received MPCs, which can be properly characterized  by scatterers/clusters, with different propagation delays \cite{general-zw,general2-zw}. To demonstrate the necessities of utilizing multi-modal information and modeling the mapping relationship, PDPs of RT-based results, the MMICM with mapping relationship exploration in \cite{Daitou-zw}, the standardized channel model with uni-modal communication information  in \cite{CCC-zw}, and the MMICM without mapping relationship exploration in \cite{LA-GBSM-zw}, are compared in Fig.~\ref{PDP}. To conduct a fair comparison, the main model parameters, including carrier frequency, communication bandwidth, antenna number, as well as vehicular speed, are set to the same. Since the mapping relationship is precisely explored, the simulation result of the MMICM in \cite{Daitou-zw} fits well with RT-based results. As a consequence, the accuracy of the MMICM in \cite{Daitou-zw} can be validated. More importantly, by modeling the mapping relationship via ANNs, PDPs smoothly and continuously vary over time and fit well with the RT-based PDP, which shows the precise capturing of  environment-channel non-stationarity and consistency in the MMICM  \cite{Daitou-zw}.  Nonetheless, the standardized channel model in \cite{CCC-zw} solely utilized uni-modal communication information to capture channel non-stationarity and consistency by modeling
spatial relationship via geometry, and thus cannot fit the RT-based PDP. Although the MMICM in \cite{LA-GBSM-zw} exploited multi-modal information, i.e., LiDAR point clouds and communication data, to capture channel non-stationarity and consistency by characterizing the coupling relationship via statistical learning, the exploration of complex mapping relationship between LiDAR point clouds and scatterers/clusters were neglected. In such a condition, environment-channel non-stationarity and consistency cannot be properly captured. Consequently,  the simulated PDP of the MMICM in \cite{LA-GBSM-zw} cannot fit the RT-based PDP. Through the comparison of PDPs in Fig.~\ref{PDP}, the necessities of utilizing multi-modal information and modeling the mapping relationship are adequately shown.  

  \subsection{Path Loss}

\begin{figure}[!t]
    \centering	\includegraphics[width=0.49\textwidth]{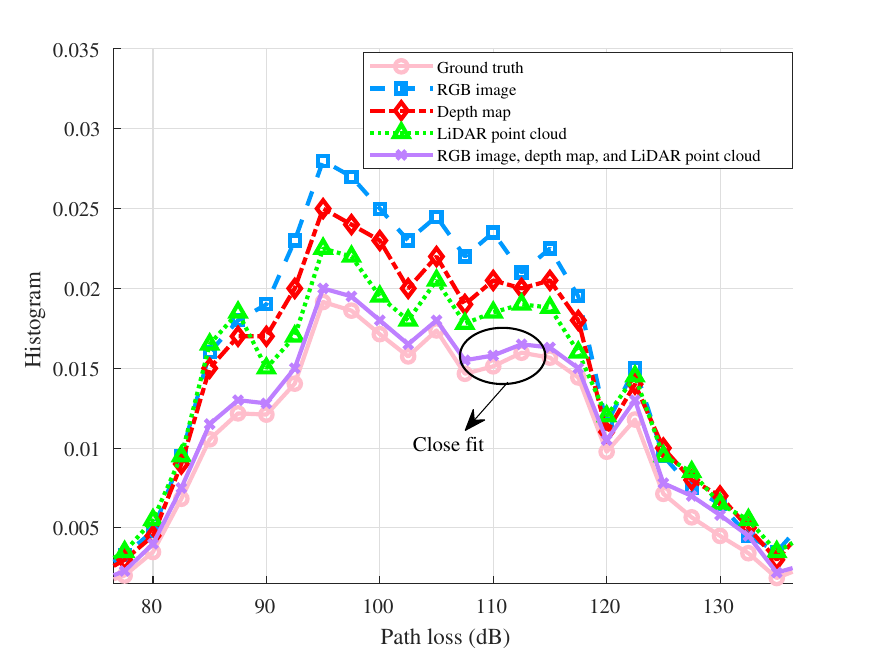}
\caption{Path loss distribution prediction in the vehicular scenario.}
\label{pl-distribution-my}
\end{figure}

\begin{figure}[!t]
    \centering	\includegraphics[width=0.49\textwidth]{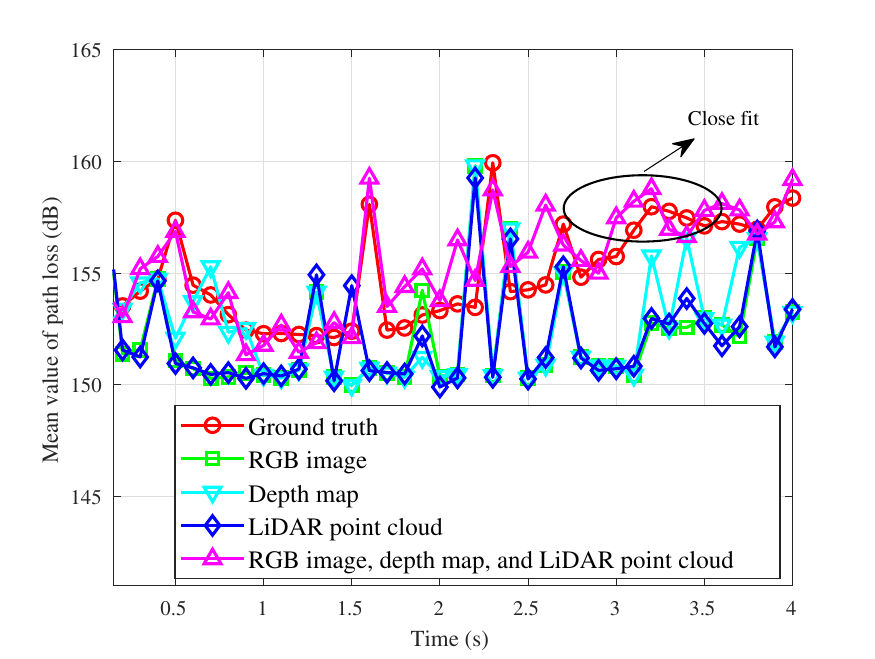}
\caption{Mean value of path loss distribution prediction results for one vehicle under different moments in the vehicular scenario.}
\label{pl-distribution-mean-value-my}
\end{figure}

Path loss is the phenomenon in which the strength of a radio signal between a Tx and a Rx is reduced as it propagates through propagation environment \cite{pl-discription1-my,pl-discription2-my}. Path loss is one of the most significant factors that affect the localization of base stations in cellular networks. 
To demonstrate the gain of multi-modal sensory information, we carry out path loss distribution prediction work in vehicular scenarios and simulation results are given in Figs.~\ref{pl-distribution-my} and \ref{pl-distribution-mean-value-my}.
Specifically, Fig.~\ref{pl-distribution-my} presents path loss distribution prediction results with uni-modal and multi-modal sensory information. Furthermore, Fig.~\ref{pl-distribution-mean-value-my} presents the variation of mean values of the path loss prediction result for a certain vehicle under different  moments.
In the prediction of  path loss distribution,
the uni-modal sensory information includes RGB images, depth maps, or LiDAR point clouds, while the multi-modal sensory information simultaneously contains RGB images, depth maps, and LiDAR point clouds. According to  different modalities of sensory information, different ANNs are trained  to explore the mapping relationship between sensory information and path loss.
The utilization of uni-modal sensory information as input solely requires different pre-processing of the data to input network and map to the path loss distribution. Unlike uni-modal sensory information, the prediction of path loss distribution with multi-modal sensory information requires the fusion of sensory information from different modalities. The feature-level fusion is carried out and processed with the attention mechanism to obtain the environmental feature map to predict the path loss distribution. Furthermore, Fig.~\ref{pl-distribution-mean-value-my} shows the mean value of prediction results for a certain vehicle with different sensory information as input.
The comparison shows that the distribution of the path loss can be approximately fitted. Compared to the utilization of uni-modal sensory information, i.e., RGB images, depth images, or LiDAR point clouds, the fusion of multi-modal sensory information  shows that the accuracy of path loss distribution prediction can be significantly enhanced. 
The physical reason underlying this phenomenon is that the RGB image can provide the texture information of the scenario, the depth map can provide the depth information of the object distribution, and the LiDAR point cloud can provide the detailed coordinate information of objects. Note that RGB images, depth maps, and LiDAR point clouds can be adequately combined to obtain the comprehensive environment information, which is closely associated with path loss. As a result, the facilitation of exploiting multi-modal sensory information in the path loss prediction under vehicular scenarios can be demonstrated.

\begin{figure}[!t]
    \centering	\includegraphics[width=0.49\textwidth]{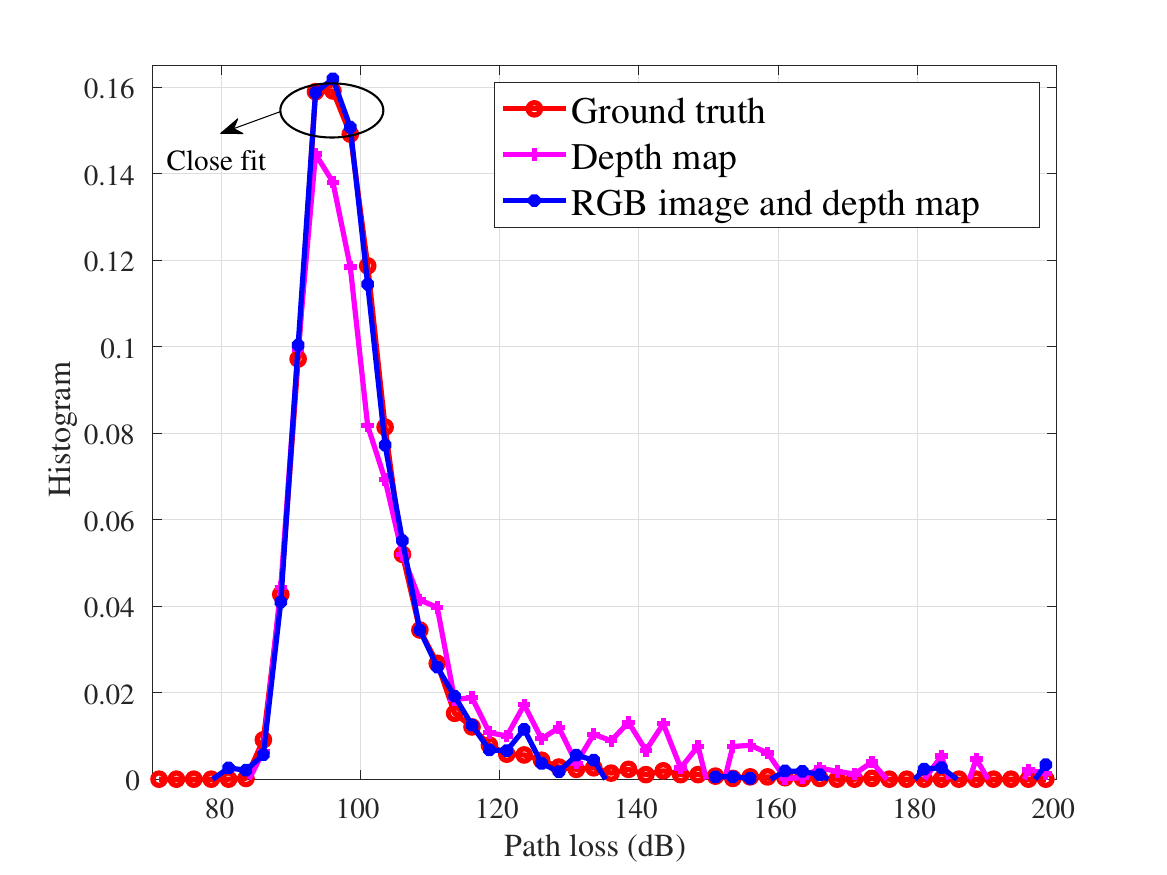}
\caption{Path loss distribution prediction in the UAV-to-ground scenario.}
\label{pl-distribution-mr}
\end{figure}

In addition to path loss prediction in the vehicular scenario, we also carry out path loss prediction in the UAV-to-ground scenario. To predict path loss distribution  of UAV-to-ground communications, we utilize the MLP to explore the mapping relationship between sensory information in the physical environment and path loss distribution in electromagnetic space. In Fig.~\ref{pl-distribution-mr}, the uni-modal sensory information includes depth maps, while the multi-modal information simultaneously contains  RGB images and depth maps. Fig.~\ref{pl-distribution-mr} demonstrates that the path loss distribution prediction model with multi-modal sensory information outperforms the path loss distribution prediction model with the uni-modal sensory information. As a consequence, the facilitation of exploiting multi-modal sensory information in the path loss prediction under  UAV-to-ground scenarios is validated.

\begin{figure}[!t]
    \centering	\includegraphics[width=0.49\textwidth]{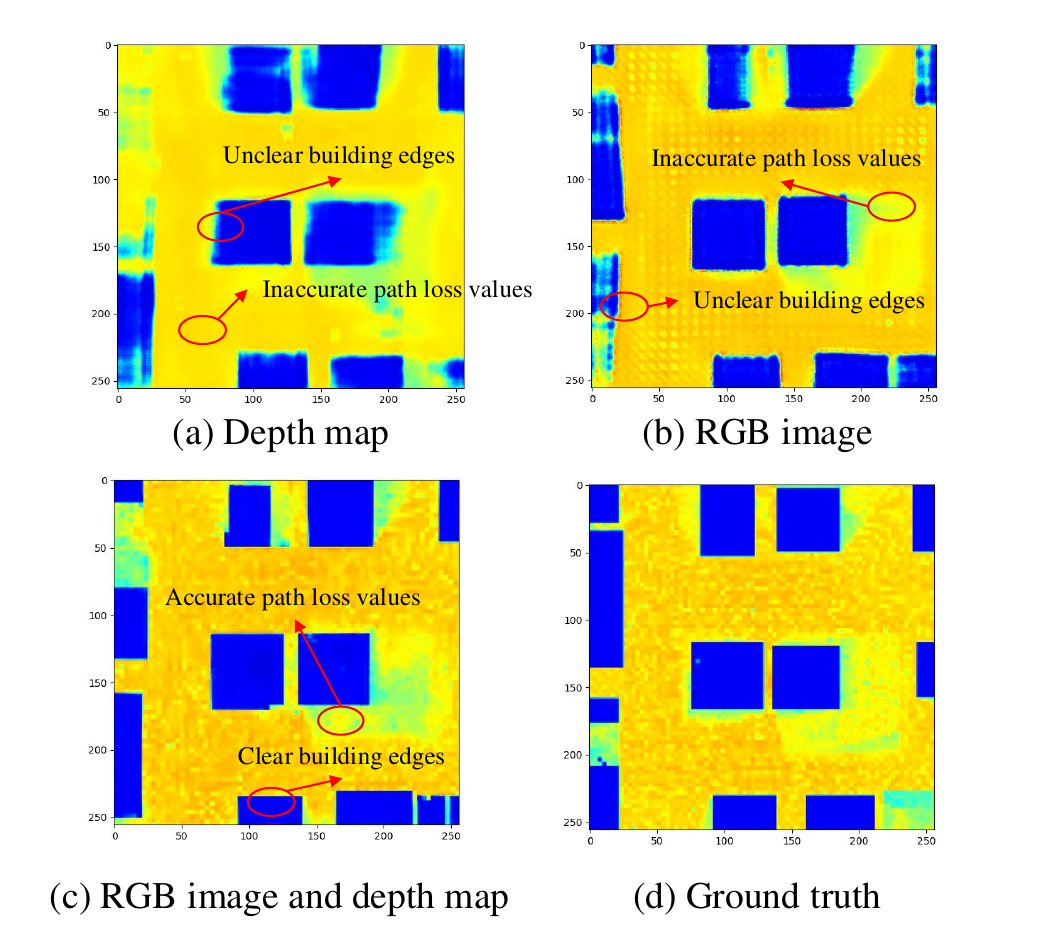}
\caption{Path loss map prediction in the UAV-to-ground scenario.}
\label{pl-map-mr}
\end{figure}

To further explore the mapping relationship between sensory information and path loss, we predict more comprehensive path loss information in the electromagnetic space, i.e., path loss map, based on multi-modal sensory information and uni-modal sensory information. The uni-modal sensory information includes RGB images or depth maps, while the multi-modal information simultaneously contains RGB images and depth maps.
Note that the data format of path loss map and sensing data, including RGB images and depth maps, is consistent. As a consequence, we utilize GAN, which is suitable for solving the image-to-image translation task, to predict path loss map based on uni-modal sensory information and multi-modal sensory information. Similar to the conclusion in Fig.~\ref{pl-distribution-mr}, Fig.~\ref{pl-map-mr} shows that the accuracy of the path loss map prediction model with the multi-modal sensory information is higher than that of the path loss map prediction model with the uni-modal sensory information. By intelligently processing the multi-modal sensory information, environmental features are extracted more intensively compared to the uni-modal sensory information. As path loss significantly depends on the surrounding environment, more comprehensive environmental information based on the multi-modal sensory information leads to higher prediction accuracy of path loss.
Therefore, the facilitation of utilizing multi-modal sensory information in the channel modeling under UAV-to-ground scenarios is further verified.

  \subsection{Summary for Simulation Results}

In this section, different channel modeling approaches are
compared by simulation. Through simulation,  the necessities of utilizing the multi-modal information and exploring the mapping relationship have been shown. For clarity, Table~\ref{sim_compare11} is presented to summarize the conclusion drawn from the simulation result. 

\begin{table*}[!t]
\renewcommand\arraystretch{2.5}
\centering
    \begin{small}
\caption{Conclusions Drawn from Simulation Results.}
\begin{tabular}{|c|c|c|}
    \hline			
    \multicolumn{2}{|c|}{\textbf{Simulation}} & \textbf{\makecell[c]{Conclusion}}  \\
    \hline
\multirow{6}{*}{\makecell[c]{Channel \\statistical properties}}
& \multirow{2}{*}{TACF} & \multirow{1.1}{*}{\makecell[c]{TACF with multi-modal information fits well with RT-based TACF}} \\				
\cline{3-3}
 &  & \multirow{1.1}{*}{\makecell[c]{TACF with uni-modal communication information \\is lower than RT-based TACF}} \\
\cline{2-3} 
 & \multirow{2}{*}{DPSD} & \multirow{1.1}{*}{\makecell[c]{DPSD with multi-modal information fits well with RT-based DPSD}} \\				
\cline{3-3}
 &  & \multirow{1.1}{*}{\makecell[c]{DPSD with uni-modal communication information exhibits a steeper \\distribution than RT-based DPSD}} \\				
\cline{2-3} 
& \multirow{2}{*}{PDP} & \makecell[c]{PDP with multi-modal information and mapping relationship \\exploration fits well with RT-based PDP by modeling \\environment-channel non-stationarity and consistency} \\			
\cline{3-3}
 &  & \multirow{1.1}{*}{\makecell[c]{PDP without multi-modal information and mapping relationship \\exploration cannot fit well with RT-based PDP}} \\
    \hline
    \multirow{4.3}{*}{\makecell[c]{Path loss}}
& \multirow{1.1}{*}{\makecell[c]{Vehicular path loss \\distribution prediction}} & \multirow{1.1}{*}{\makecell[c]{Increasing the number of sensory modalities enhances the accuracy \\of path loss distribution prediction in vehicular scenarios}} \\				
\cline{2-3} 
& \makecell[c]{Mean value of vehicular path loss \\distribution  prediction results\\ for one vehicle under different moments}  & \makecell[c]{Increasing the number of sensory modalities \\ enhances the accuracy of  mean value of  path loss\\ distribution prediction in vehicular scenarios} \\				
 \cline{2-3} 
 & \multirow{1.1}{*}{\makecell[c]{UAV-to-ground path loss \\distribution prediction}}  & \multirow{1.1}{*}{\makecell[c]{Increasing the number of sensory modalities enhances the accuracy \\of path loss distribution prediction in UAV-to-ground scenarios}} \\					
 \cline{2-3} 
& \multirow{1.1}{*}{\makecell[c]{UAV-to-ground path loss \\map prediction }}  & \multirow{1.1}{*}{\makecell[c]{Increasing the number of sensory modalities enhances the accuracy \\of path loss map prediction in UAV-to-ground scenarios}} \\	
    \hline 
\end{tabular}
\label{sim_compare11}
\end{small}
\end{table*} 

Fig.~\ref{TACF-zr} compares TACFs of the MMICM in \cite{daitou-zr}, the standardized channel model in \cite{standardized-zr}, and the RF-only intelligent channel model in \cite{huangchenscatterer-zr}. In Fig.~\ref{TACF-zr}, it can be seen that the TACF with the multi-modal information based on the MMICM via SoM has a close agreement with the RT-based TACF. Nonetheless, TACFs with the uni-modal information based on the conventional channel modeling and RF-only intelligent channel modeling are lower than the RT-based TACF. Therefore, the necessity of utilizing the multi-modal information and the accuracy of the MMICM via SoM are validated through the comparison of TACFs.

Fig.~\ref{DPSD-zr} compares DPSDs of the MMICM in \cite{daitou-zr}, the standardized channel model in \cite{standardized-zr}, and the RF-only intelligent channel model in \cite{huangchenscatterer-zr}. It can be seen from Fig.~\ref{DPSD-zr} that the multi-modal information based on the MMICM via SoM can fit well with the RT-based DPSD. However, DPSDs with the uni-modal information based on the conventional channel modeling and RF-only intelligent channel modeling exhibit the steeper distribution than the RT-based DPSD. Therefore, the necessity of utilizing the multi-modal information and the accuracy of the MMICM via SoM are verified through the comparison of DPSDs.

Fig.~\ref{PDP} compares PDPs of RT-based results, the MMICM with mapping relationship exploration in \cite{Daitou-zw}, the standardized channel model with uni-modal communication information in \cite{CCC-zw}, and the MMICM without mapping relationship exploration in \cite{LA-GBSM-zw}. It can be observed from Fig.~\ref{PDP} that, by utilizing the multi-modal information and exploring the mapping relationship, the  environment-channel non-stationarity and consistency can be modeled, and thus the PDP can fit well with the RT-based PDP. However, PDP without the multi-modal information and the mapping relationship exploration cannot capture environment-channel non-stationarity and consistency, and thus the PDP cannot fit well with the RT-based PDP. Therefore, the necessities of utilizing the multi-modal information and exploring the mapping relationship are demonstrated through the comparison of PDPs.

Figs.~\ref{pl-distribution-my} and \ref{pl-distribution-mean-value-my} compare the prediction results of path loss distribution between uni-modal and multi-modal sensory information under vehicular scenarios.  Figs.~\ref{pl-distribution-my} and \ref{pl-distribution-mean-value-my}  indicate that, for path loss distribution, the prediction model, which incorporates multi-modal sensory information, outperforms the model that relies solely on uni-modal sensory information. Therefore, Figs.~\ref{pl-distribution-my} and \ref{pl-distribution-mean-value-my} show that the utilization of multi-modal sensory information can significantly enhance the performance of path loss predictions in vehicular scenarios.

Fig.~\ref{pl-distribution-mr} compares the performance of path loss distribution prediction related to RT-based results, uni-modal sensory information, and multi-modal sensory information in UAV-to-ground scenarios. Fig.~\ref{pl-distribution-mr} presents that the path loss distribution prediction model with the multi-modal sensory information outperforms the path loss distribution prediction model with the uni-modal sensory information. Therefore, the necessity of utilizing multi-modal sensory information to enhance the performance of path loss distribution prediction in UAV-to-ground communications is verified in Fig.~\ref{pl-distribution-mr}.

Fig.~\ref{pl-map-mr} compares the performance of path loss map prediction related to RT-based results, uni-modal sensory information, and multi-modal sensory information in UAV-to-ground scenarios. Similar to Fig.~\ref{pl-distribution-mr}, Fig.~\ref{pl-map-mr} shows that the path loss map prediction model with multi-modal sensory information outperforms the path loss map prediction model with uni-modal sensory information. Therefore, Fig.~\ref{pl-distribution-mr} validates the necessity of utilizing multi-modal sensory information to enhance the performance of path loss map prediction in UAV-to-ground communications.

\section{Future Research Directions for Multi-Modal Intelligent Channel Modeling}
Open issues and future research directions, which are regarded as the guideline of conducting MMICM, are given from three perspectives, including measurement perspective, modeling perspective, and application perspective. 

\subsection{Measurement Perspective}
\subsubsection{Measurement Platform Establishment} 
As the cornerstone of the MMICM via SoM, a comprehensive real-world communication and multi-modal sensing dataset, including RF communication information, i.e., channel
matrices and path loss, RF sensory information, i.e., mmWave radar point clouds, and non-RF sensory information, RGB images, depth maps, and LiDAR point clouds, is essential. However, compared to the real-world dataset for ISAC \cite{Huawei1} that focuses on the integration of RF communications and RF sensing, constructing a real-world dataset for intelligent multi-modal sensing-communication integration, i.e., SoM, requires a more complex and diverse set of hardware devices. These hardware devices need to achieve the precise spatio-temporal synchronization, which further increases the challenge of the real-world dataset construction. Currently, the authors in \cite{DeepSense1} attempted to overcome the aforementioned challenge and constructed a real-world communication and multi-modal sensing  dataset, named DeepSense 6G. To construct the DeepSense 6G dataset \cite{DeepSense1}, the mmWave communication device and mutli-modal sensors, such as radar, RGB camera, and LiDAR, were exploited and synchronized. However, the DeepSense 6G dataset in \cite{DeepSense1} ignored the collection of depth maps and lacked a typical weather condition, i.e., snowy day. 

In the future, to support the research on MMICM via SoM, a proper hardware measurement platform, which has the capability to collect comprehensive real-world communication and multi-modal sensory data, needs to be established.

\subsubsection{Simulation Platform Establishment} Although the real-world dataset is of high accuracy and can facilitate the approach verification, it is difficult to flexibly customize desired scenarios owing to the labor and cost concerns. Unlike the real-world dataset, the synthetic dataset achieves a proper trade-off between complexity and quality via efficient software measurement platforms. However, due to the huge difference between communication data and multi-modal sensory data, there is currently no software measurement platform available for the simultaneous collection of communication and multi-modal sensory data. By using software measurement platforms, i.e., Wireless InSite \cite{WI} and Blender \cite{Blender}, the authors in \cite{ViWi} proposed a synthetic communication and multi-modal sensing  dataset, named  Vision-Wireless (ViWi), including RGB images, depths, and LiDAR information. To include mmWave radar information, our previous work in \cite{M3SC} proposed a new synthetic communication and multi-modal sensing  dataset, named M$^3$SC. The M$^3$SC dataset utilized three high-fidelity software measurement platforms, i.e., Wireless InSite, AirSim \cite{AirSim}, and WaveFarer \cite{WF}, and further achieved in-depth integration and precise alignment of them. Nevertheless, the M$^3$SC dataset lacked channel large-scale fading, i.e., path loss, 
and neglected a typical high-mobility scenario, i.e., UAV scenario. 

In the future, a new software measurement platform, which can construct a more  comprehensive synthetic communication and multi-modal sensing dataset, is urgently required to facilitate the MMICM via SoM.

\subsubsection{Self-Generated and Self-Evolving Dataset Construction} 
Artificial intelligence generated content (AIGC) exploits the advanced generative AI (GAI) technique to
produce content corresponding to human instructions through deciphering the intent and generating
appropriate content in response \cite{AIGC1, AIGC111}. Because of its capability to enhance creativity and accelerate the design process, AIGC has been extensively utilized in the wireless communication research \cite{AIGC2,AIGC3}. To further increase the amount of communication data and multi-modal sensing data and support the MMICM via SoM, AIGC and transfer learning can be leveraged to achieve self-generation and self-evolution of data. Specifically, considering various communication scenarios, such as urban, suburban, and rural areas, and  different conditions, such as frequency bands, antenna types, and weather conditions, the research on self-generating data needs to be conducted to explore continuously evolving and enriching data generation algorithms. As a consequence, massive accurate communication data and multi-modal sensory data provide a solid foundation for the investigation of the MMICM via SoM.

In the future, based upon the real-world and
synthetic communication and multi-modal sensing data, there is an urgent need to construct a self-generated and self-evolving dataset that combines real-world data with synthetic data under various scenarios and conditions.

\subsection{Modeling Perspective}
\subsubsection{Large Language Model (LLM)-Driven Mapping Relationship Exploration} The large language model (LLM) is the AI system, which is trained on billions of words obtained from the article, book, as well as
other internet-based content \cite{natureme}. In general, the LLM exploits the neural network architecture to explore the complex associative relationship between in the word in text-based training datasets \cite{natureme1}. Considering the powerful capability of LLMs in solving nonlinear problems, our previous work in \cite{boxun} leveraged the LLM to empower wireless communication physical layer tasks for the first time, and further propose a novel channel prediction scheme based on the pretrained LLM. As previously mentioned, there are three significant differences between communications and multi-modal sensing, including   data structures, frequency bands, and applicability. As a result, the mapping relationship between communications and multi-modal sensing is complicated and nonlinear. It can be predictable that the LLM is an enabling technology for the mapping relationship exploration. To be specific, first, the constructed multi-modal sensing-communication measurement and simulation dataset can be intelligently processed via ML algorithms. Then, in combination with electromagnetic propagation mechanisms, the complex and nonlinear mapping relationship between  entities in the physical environment and path loss and multipath parameters in the electromagnetic propagation can be explored via LLMs. In addition, mapping relationships under different scenarios  and conditions can be explored by leveraging transfer learning and incremental learning, and inherent rules between mapping relationships in different  scenarios and conditions are further investigated. 

In the future, based on its powerful ability in solving nonlinear problems, the LLM can be adopted to explore the complex and nonlinear mapping relationship between communications and multi-modal sensing under different scenarios  and conditions.

\subsubsection{Model-Enhanced Data-Driven Channel Modeling} Generally, for the stochastic channel modeling \cite{Bailu-1,Ziwei-1}, it is general and has low complexity, while the drawback is the inability to accurately predict channels in real-time. For the intelligent prediction channel modeling, it has the ability to forecast future CSI from an external information perspective \cite{huangchen1,yanngmi11}. However, the drawback is the lack of generality in predicting channel models due to the limitation of data-driven approaches. To enhance the accuracy, intelligence, and involvement in system design of channel modeling, establishing a model-enhanced data-driven MMICM in consideration of multi-modal sensory data is extremely important. Model-enhanced data-driven approaches can significantly improve the modeling accuracy, inclusiveness, and generalizability of channel models with low complexity.
However, the parameter representation and calculation methods of stochastic channel models and intelligent prediction channel models are different, resulting in the fact that the establishment of a consistent integration approach and mechanism is significantly challenging. The significant difference between multi-modal sensory data and wireless channel data further increases the challenge of establishing the model-enhanced data-driven MMICM. To overcome the aforementioned challenge, utilizing the data-driven framework of intelligent prediction channel models as the main architecture, based on the statistical properties of stochastic channel models, conceptualize the network structure of ML algorithms and initialize parameters of the statistical model. Meanwhile, multi-modal sensory data is intelligently processed based on ML algorithms to extract physical environment features, which are further matched with the electromagnetic environment. As a result, the investigation of model-enhanced data-driven MMICM is significantly important whereas is of huge challenge. 

In the future, by combining stochastic channel modeling with intelligent prediction channel modeling, the model-enhanced data-driven MMICM in consideration of multi-modal sensory data is urgently required. 

\subsubsection{Multi-Modal Intelligent Cooperative Channel Modeling} Certainly, a cooperative multi-vehicle/UAV platoon has the ability to cooperatively transmit the data stream via the cooperative communication technology \cite{Cooperative-zw1, Cooperative-zw2}. In this case, the distributed MIMO system performance can be implemented to improve communication efficiency and reliability. As a result, cooperative multi-vehicle and multi-UAV communication technologies have been widely exploited in civil and military applications \cite{Cooperative-zw3,Cooperative-zw4}. To support the cooperative multi-vehicle/UAV intelligent sensing-communication system design, the mapping relationship between cooperative multi-vehicle/UAV communications and multi-modal sensing needs to be adequately modeled and investigated. However, different from point-to-point vehicular/UAV communications, there are many complex links/sub-channels in multi-vehicle/UAV communications. This significantly increases the difficulty of exploring mapping relationships. Fortunately, in the multi-vehicle/UAV communication, different links/sub-channels possess several same scatterers/clusters, thus exhibiting the strong space-time correlation between multiple links/sub-channels. In such a condition, there is also a strong correlation between the mapping relationships between multi-modal perception and multi-vehicle/UAV links/sub-channels, which can facilitate the exploration of mapping relationship in the cooperative multi-vehicle/UAV scenarios. 

In the future, to support the design of cooperative multi-vehicle/UAV intelligent sensing-communication systems, it is urgent to explore mapping relationships between cooperative multi-vehicle/UAV communications and multi-modal sensing.
Furthermore, by considering the correlation between multiple links/sub-channels in cooperative multi-vehicle/UAV scenarios, the complexity of mapping relationship exploration can be significantly reduced.

\subsection{Application Perspective}

\subsubsection{Multi-Modal Intelligent Channel Modeling for Digital Twin} 
With the current advance in the software platform, AI/ML together with efficient computing accelerated through the graphics processing unit, the DT has received extensive attention in diverse areas, including the smart city, manufacturing, as well as retail \cite{DT-zw1}--\cite{DT-zw3}. DT is the digital representation of the physical entity and the system. As stated in \cite{DT-zw4}, the real-time DT technology can be leveraged to make the real-time or near real-time decision for the physical-world sensing-communication system. On one hand, the BS can utilize real-time DTs to conduct the downlink channel prediction. On the other hand, the mobile user can utilize real-time DTs to conduct the blockage prediction. Consequently, the real-time nature of the DT can enhance the performance of sensing and communications, and further provide a more comprehensive awareness related to the surrounding environment in making the real-time decision.
To achieve the  real-time DT  for physical-world sensing-communication systems, it is necessary to explore the nonlinear mapping relationship between communications and multi-modal sensing  and further conduct MMICM. Towards this object, there are extremely high demands for the real-time and accuracy requirements of the mapping relationship exploration together with MMICM, which poses significant challenges.

In the future, the MMICM needs to be conducted in a real-time and accurate manner to support making  the real-time or near real-time decision for the physical-world sensing-communication system via the DT technology.

\subsubsection{Multi-Modal Intelligent Channel Modeling for Embodied Intelligence} 
A paradigm shift is occurring from the era of ``Internet AI'' to the era of ``Embodied AI," where the AI algorithm and agent no longer primarily learn from the image, video, or text dataset sourced from the internet \cite{EI-zw1,EI-zw2}. Conversely, the AI algorithm and agent learn through the interaction with the environment in a self-centered perceptual manner similar to humans. As a result, there is a significant increase in demand for embodied AI frameworks to facilitate diverse research tasks in the area of embodied AI \cite{EI-zw3}. In 6G embodied AI scenarios, embodied AI agents will deploy a large number of communication and multi-modal sensory devices to  achieve efficient interactions of embodied AI agents with the environment. Towards this objective, the intelligent multi-modal sensing-communication integration, i.e., SoM, urgently needs to be investigated, which can achieve a bidirectional deep integration of communications with multi-modal sensing and  support embodied AI applications.  As the foundation of SoM research, it is essential to conduct the proper MMICM by exploring the mapping relationship between communications and multi-modal sensing. Nevertheless, a large number of embodied AI agents moving and interacting in the environment result in a complex and high-mobility scenarios, thus posing significant challenges of  together with mapping relationship exploration. 

To support future 6G embodied AI research tasks, the proper MMICMs  need to be conducted by exploring the mapping relationship between communications and multi-modal sensing  
in complex embodied AI scenarios.

\subsubsection{Multi-Modal Intelligent Channel Modeling for Networked intelligence} 

With the recent development in the communication and networking
technology in networked intelligent systems, connectivity can be provided to autonomous vehicles/UAVs to share information and cooperate with other vehicles/UAVs and infrastructures. Consequently, autonomous vehicles/UAVs have the ability to operate with a global view \cite{NI-zw1,NI-zw2}.
In the networked intelligent system, autonomous vehicles and UAVs are naturally equipped with diverse multi-modal sensors and communication devices. In such a condition, there is a vast amount of communication information and multi-modal sensory information, which can be intelligently processed to achieve robust environmental perception as well as efficient link establishment. Fortunately, by precisely exploring the complex mapping relationships between communications and multi-modal sensing, MMICM has the capability to enhance the performance of communication system transceiver design, network optimization, localization sensing, and intelligent agent cognitive intelligence. However, as the precision of mapping relationship exploration increases, the computational complexity of the MMICM also increases. Excessive computational time may fail to meet the real-time requirements of networked intelligent system under high-mobility scenarios. To reduce the computational time, while supporting the application related to networked intelligence and accurately exploring the mapping relationship, the requirement of computational complexity is further required to be taken into account.

In the future, the closer integration of applications associated with networked intelligence and channel modeling will bring huge challenges to the mapping relationship exploration and MMICM. More efforts need to be given to propose a proper mapping relationship exploration way and an efficient MMICM approach with a decent trade-off between the accuracy and complexity.

\subsection{Summary for Future Research Directions}
In this section, the open issue and potential research direction in the area of the MMCIM via SoM are given from three perspectives, including measurement perspective, modeling perspective, and application perspective. 

For the measurement perspective, a hardware measurement
platform that can collect real-world  communication and multi-modal sensory data needs to be established. To further enrich the communication and multi-modal sensing dataset, a new software measurement platform, which can  generate  high-fidelity synthetic communication and multi-modal sensory data, needs to be constructed. Based on the real-world and synthetic communication and multi-modal sensing data, a self-generated and self-evolving dataset containing diverse scenarios and conditions, needs to be developed based on the AIGC and transfer learning.

For the modeling perspective, considering the advantage of LLM in solving nonlinear problems, the LLM can be exploited to explore the mapping relationship between communications and multi-modal sensing in various scenarios and conditions to facilitate the MMICM. In addition, by combining stochastic channel modeling with intelligent prediction channel modeling, the model-enhanced data-driven MMICM with the help of multi-modal sensory data needs to be conducted. To further support the cooperative multi-vehicle/UAV intelligent sensing-communication system design, more efforts need to be given to explore more complicated mapping relationship between cooperative multi-vehicle/UAV communications and multi-modal sensing, where the correlation between multiple links/sub-channels needs to be considered to reduce the complexity. 

For the application perspective, based on the DT technology, the MMICM needs to be carried out in a real-time and accurate way to support making the real-time or near real-time decision for the physical-world sensing-communication system. Moreover, to support future 6G embodied AI research tasks, it is necessary to carry out the proper MMICM by exploring the mapping relationship between communications and multi-modal sensing under complex embodied AI scenarios. Finally, the tight integration of network intelligence applications with channel modeling poses significant challenges to the mapping relationship exploration and MMICM, where a
proper mapping relationship exploration way and an efficient
MMICM approach with a decent trade-off between the accuracy and complexity are urgently required.

\section{Conclusions}
In this paper, we have provided a comprehensive survey regarding MMICM via SoM. First, we have reviewed and compared the existing channel modeling approaches, including the channel modeling evolution from conventional channel modeling with uni-modal communication information to RF-only intelligent channel modeling with uni-modal communication information to MMICM with multi-modal information from communication devices and sensors. Furthermore, we have reviewed the recent advance in the topic of capturing channel non-stationarity and consistency based on the  conventional channel modeling, RF-only intelligent channel modeling, and MMICM by characterizing the mathematical, spatial, coupling, and mapping relationships. Additionally, we have analyzed the role of MMICM via SoM in facilitating the application related to communication network and networked intelligence, including communication system transceiver design, network optimization, localization sensing, and intelligent agent cognitive intelligence. Simulation results related to MMICM have been given, where the necessities of utilizing the multi-modal information and exploring the mapping relationship have been demonstrated. We have further given some open issues and potential directions   for the MMICM via SoM from three different perspectives, including measurement, modeling, and application perspectives. Finally, we expect that more accurate and efficient work related to MMICM via SoM will be conducted to facilitate more potential applications of communication network and networked intelligence in the future 6G era, and we also hope this survey will be of use in that regard.

\section*{APPENDIX}
	\subsection*{List of Abbreviations}
	\begin{tabbing}
		\hspace*{80bp}\=article\quad \=ÎÄÕÂÀà\kill
1D  \>One-Dimensional \\
2D \> Two-Dimensional \\
2G \> Second-Generation \\
3D \> Three-Dimensional \\
3G \> Third-Generation \\
3GPP \>  Third Generation Partnership Project \\
4G \> Fourth-Generation \\
5G \> Fifth-Generation \\
5GCM \> 5G Channel Model \\
6G  \> Sixth-Generation	\\
A2A \> Air-to-Air \\
AGV  \> Automated Guided Vehicle \\
AI \> Artificial Intelligence \\
AIGC \>Artificial Intelligence Generated Content\\ 
AIoT \> Artificial Intelligence of Things \\
ANN  \> Artificial Neural Network \\
AoA  \> Angle of Arrival \\
AoD  \> Angle of Departure \\
BS \> Base Station \\
CFR \>Channel Frequency Response\\
CIR \> Channel Impulse Response\\
CNN \> Convolution Neural Network\\
COST \>	European COoperation in  the field of \\  \>Scientific and Technical research \\
CSI \> Channel State Information \\
D2D \> Device-to-Device \\
DBSCAN \> Density-Based Spatial Clustering \\ \> of Applications with Noise\\
DoA \> Direction of Arrival \\
DPSD \> Doppler Power Spectral Density \\
DT \> Digital Twin \\
EE \> Energy Efficiency\\
eMBB \> Enhanced Mobile Broadband \\
FDTD \> Finite-Difference Time-Domain \\
GAI \> Generative AI \\ 
GAN \> Generative Adversarial Network \\
GBDM \> Geometry-Based Deterministic Modeling\\
GBSM \> 	Geometry-Based Stochastic Modeling \\
GHz \>   Gigahertz\\
HST \> High-Speed Train \\
ISAC \>  Integrated Sensing and Communication \\
IS-GBSM \> Irregular-Shaped Geometry-Based  \\
\> Stochastic Modeling	\\
ITU-R \> International Telecommunication Union \\ \> Radiocommunication Sector\\
K-M \> Kuhn-Munkres\\
LASSO  \> Least Absolute Shrinkage and Selection Operator\\
LiDAR \> Light Detection and Ranging \\
LLM   \> Large Language Model \\
LoS \> Line-of-Sight \\
LSTM \> Long Short-Term Memory \\
METIS \>Mobile and wireless communications  \\ 
\>Enablers for the Twenty-twenty \\ \>  Information Society \\
MIMO \> Multiple-Input Multiple-Output \\ 
ML \> Machine Learning \\
MLP \> Multilayer Perceptron \\
mmMAGIC \>mm-wave based Mobile radio Access \\ \>	network 
for fifth Generation    Integrated \\ \> Communications  \\
mMTC 	\> massive Machine Type Communications	\\
MMICM \> Multi-Modal Intelligent Channel Modeling\\
mMTC \>Massive Machine Type Communications \\
mmWave  \> Millimeter Wave\\
MPC \> Multipath Component \\
MS \> Mobile Station \\
MSE\>Mean Square Error \\
NGSM \>	Non-Geometry Stochastic Modeling	\\
NLoS \> Non-Line-of-Sight \\  
OMP  \>	 Orthogonal Matching Pursuit \\
PAS \> Power-Angle-Spectrum \\
PDF  \> Probability Density Function \\
PDP \> Power Delay Profile\\
QuaDRiGa \> QUAsi Deterministic RadIo channel \\ \>GernerAtor \\
ResNet \> Residual Network \\
RF  \> Radio Frequency	\\
RS-GBSM \>Regular-Shaped Geometry-Based Stochastic \\
\>Modeling	\\
RT \> Ray-Tracing\\
Rx \>  Receiver \\
SAGE \>Space-Alternating Generalized \\ \>	Expectation- Maximization \\
SAGSIN \> Space-Air-Ground-Sea Integrated Network \\
SCCF\> Space Cross-Correlation Function\\
SE \> Spectral Efficiency \\
SIMO \> Single-Input Multiple-Output \\
SISO \> Single-Input Single-Output\\
SoM  \> Synesthesia of Machines	\\
ST-CF\>  Space-Time Correlation Function\\
STF-CF\>	Space-Time-Frequency Correlation Function	\\
SVM  \> Support Vector Machine	\\
TACF\> Time Auto-Correlation Function \\
TDL  \> Tapped Delay Line \\
TF-CF   \> Time-Frequency Correlation Function \\
THz \> Terahertz \\
Tx \> Transmitter\\ 
UAV \>  Unmanned Aerial Vehicle \\
UPA \> Uniform Planar Antenna Array \\
uRLLC \> Ultra-Reliable and Low Latency \\ \>Communications\\
V2V  \>       Vehicle-to-Vehicle \\
VTD  \>       Vehicular Traffic Density\\
		
	\end{tabbing}

\ifCLASSOPTIONcaptionsoff
  \newpage
\fi

\end{document}